\documentclass{mn2e}
\usepackage{natbib}
\usepackage{booktabs}                           
\usepackage{tabularx}                           
\usepackage{longtable}                           
\usepackage{rotating}
\usepackage{psfig}
\usepackage{epsfig}
\usepackage{times}

\def\halpha{\mbox{H$\alpha$}}

\def\kms{\,km~s$^{-1}$}      
\def\lya{\mbox{Ly$\alpha$}}

\def\subsun{\mbox{$_{\normalsize\odot}$}}

\def\lesssim{\mathrel{\hbox{\rlap{\hbox{\lower4pt\hbox{$\sim$}}}\hbox{$<$}}}}
\def\gtrsim{\mathrel{\hbox{\rlap{\hbox{\lower4pt\hbox{$\sim$}}}\hbox{$>$}}}}
\def\la{\mathrel{\hbox{\rlap{\hbox{\lower4pt\hbox{$\sim$}}}\hbox{$<$}}}}
\def\ga{\mathrel{\hbox{\rlap{\hbox{\lower4pt\hbox{$\sim$}}}\hbox{$>$}}}}

\title[HII and H$_2$ in cooling flows]{HII and H$_2$ in the envelopes of cooling flow central galaxies}

\author[Jaffe, W., Bremer, M.N., Baker, K.]{W. Jaffe$^{1}$, M. N. Bremer$^{2}$, K. Baker$^{2}$\\
$^1$Leiden Observatory, P.B. 9513, Leiden 2300RA, Netherlands \\
$^2$ Dept. of Physics, Bristol Univ. H.H. Wills Lab, Tyndall Ave. Bristol, BS8 1TL, UK}

\begin{document}
\def\Ht {H$_2$\ }
\def\Hp {H$_2$}
\def\Ha{H$\alpha$\ }
\def\Hap{H$\alpha$}
\def\Pa{Pa$\alpha$\ }
\def\Pap{Pa$\alpha$}
\def\kms{km s$^{-1}$\ }
\def\kmsp{km s$^{-1}$}
\def\eg{{\it e.g.}\ }
\def\chandra{{\it Chandra} }

\pagerange{\pageref{firstpage}--\pageref{lastpage}} 
\pubyear{2004}

\maketitle

\label{firstpage}

\begin{abstract}
We report observations of ionized and warm molecular gas in the
extended regions of the central galaxies in several cooling flow
clusters. These show that both gas phases are present in these
clusters to large radii. We trace both \Pa and \Ht lines to radii in
excess of 20 kpc. The surface brightness profiles of the two phases
trace each other closely. Apart from very close to the central AGN,
line ratios in and between the phases vary only slowly with
position. The kinematics of the phases are indistinguishable and away
from the influence of the central AGN both the mean and dispersion in
velocity are low ($\leq 100$\kmsp) ruling out kinematic support of the
molecular gas. All of the above indicates that the mechanisms for
heating the molecular gas and ionizing the HII regions are highly
coupled. The highest surface brightness emission within a few kpc of
the central AGN is distinct, both kinematically and thermally from
that at larger radii. The relative strengths of the \Pa to the \Ht
lines indicates a source of UV excitation rich in EUV relative to FUV
photons, \eg a black body with a temperature  $\geq 10^5$K.

\end{abstract}  

\section{Introduction}

 The density of the hot ($10^{7}-10^{8}K$) intracluster medium (ICM)
at the centres of many galaxy clusters is high enough for it to cool
on timescales less than the age of the cluster, potentially leading to
the formation of a cooling flow (see \citealt{fab94} for a
review). Recent observations of galaxy clusters with Chandra and
XMM-Newton have shown that the situation at the centres of clusters
thought to contain cooling flows is considerably more complex than
previously thought. Chandra and XMM-Newton data indicates that the
size of a cooling region is typically smaller than was indicated by
earlier ROSAT data (\eg \citealt{peres98}), cooling may be intermittent,
and there is considerable interaction between the hot phase at the
centre of a cluster and any radio plasma emitted by the central radio
source (\citealp{fab01,bohringer01,mcnamara01}).
Chandra and XMM-Newton spectroscopy of the central regions of clusters
reveals a lack of line emission expected from gas at $\sim 2$ keV, implying
that the amount of gas cooling through these temperatures is
considerably less than the mass deposition rates determined from ROSAT
data ({\it e.g.}  see \citealt{peterson03}).

The existence of atomic and ionised gas emitting at $10^4$K in and around the dominant
galaxies in many of these cooling flow clusters has been known for two
decades ({\it e.g.} \citealp{Hu83, heckman89}). Despite clearly
being statistically correlated with the presence of the cooling flow
in the cluster, the emission-line gas discovered in these early
studies has too high a surface brightness and luminosity to be simple
recombination radiation from material cooling through 10$^4$K. It has
long been recognised that this relatively high surface brightness
emission must be excited by one or more mechanisms: hot stars, shocks,
turbulence or X-ray heating.

Cooler molecular material associated with the optical nebulosity has
recently been detected from its K-band \Ht line emission
(\citealp{jaffe97, falcke98, wilman00, donahue00}).  This material
is dense ($n_e \ge 10^5$ cm$^{-3}$) and its occurance
correlates with that of atomic nebular emission \citep{jaffe01}.
This, plus detection of large amounts of cold ($<$50K) CO in cooling
flows \cite{edge01} indicates that large amounts of cool material is
present at the centres of these clusters, even if the new X-ray data
indicates weaker cooling than was previously estimated.

With the exception of the CO observations whose spatial resolution is
so far quite poor, most previous studies of gas in brightest
cluster galaxies (BCGs) have only
detected gas within a few kpc of the nuclei, where the energetics of
the gas are without doubt dominated by the AGN activity, although
\cite{hatch05} have detected molecular line emission in several
star-forming regions at distances of $\gtrsim 25$ kpc from the center
of the Perseus cluster. Our
previous low resolution 4m $K-$band spectroscopy has shown that 
there is a clear link between the intensities of emission lines from 
ionized and warm ($\sim 2000$ K) molecular gas at the centres of these clusters.
We now extend this work by exploring the kinematic and
morphological link between these two phases to larger radii (and lower
surface brightness limits).This is essential to separate the role of
the AGN in these systems from that of the host galaxy and cooling flow
in the excitation and dynamics of this gas.  

With the goal of connecting the cooler IGM states to that of the hot
X-ray gas on scales of $> 10$ kpc, we combine very deep narrow-band imaging
of the BCGs of cooling flow clusters in
the light of H$\alpha\ \lambda$ 6563 and [NII] $\lambda$ 6584 lines
with deep 8m long slit K-band spectroscopy of a subset, obtained in order
to examine the distribution and thermal state of the molecular gas
through the 1-0S and 2-1S emission lines of \Hp.  The \Pa emission in
the $K-$band spectra allow us to link both data sets through
comparison of the H$\alpha$ and \Pa lines.
\section{Observations}

\subsection{Optical narrow-band imaging}

The H$\alpha$+[NII] observations were performed at the Anglo-
Australian Telescope (AAT) using the Taurus Tunable Filter (TTF) to
obtain images with a bandwidth of $\sim 12$\AA\ in all but one case.
This Fabry-Perot etalon instrument, described in \cite{bland98}, has
several advantages over conventional narrow band imaging: very narrow
bandpass, very short wavelength step between ``on" and ``off" bands
and \emph {shuffling} between on- and off- images on the CCD chip,
which allows rapid switching between the two bands without increase in
read-out noise.  Together, these techniques yield images with very
clean continuum subtraction, and independent maps of \Ha and [NII].

The galaxies observed are given in Table 1.  These include 5 strong
cooling flow galaxies (as defined by the ROSAT observations, before
{\it Chandra}/XMM reobservations) and the non-cooling flow radio
galaxy 3C445 (a known line-emitting source at a similar redshift to
the clusters) and the non-cooling flow cluster A3667.  Here and
further below, apparent quantities are converted to intrinsic
quantities assuming H$_o$=72 km/s/Mpc, $\Omega_m=0.3$ and $\Omega_\Lambda=0.7$.

\begin{table}
\label{table:1}
\centering
\begin{tabular}{lllrrrr}
\hline\\
Cluster & Redshift & D & \multicolumn{2}{c}{Flux} 
& \multicolumn{2}{c} {Luminosity}  \\
& &Mpc &H$\alpha$ &  N[II] & H$\alpha$ & N[II] \\
\hline\\
 3C445  & 0.056 &233&  14.0 & -  & 9.1 & - \\
 Abell 3667  & 0.056&232& $<$0.7 & $<$0.7  & $<$0.6 & $<$0.6 \\
 Sersic~159-03 & 0.056 &233& 12.7 & 12.7 & 8.4 & 8.3 \\
 Abell 2029  & 0.076 &316& $<2.9$ & $<1.4$ & $<4.8$ & $<2.3$ \\
 Abell 2204  & 0.152 &632& 48.5  & * & 232.7 & * \\
 Abell 2597  & 0.082 &343& 41.5  & 28.7 & 80.7 & 55.7 \\
 Abell 1795  & 0.063&264& $^*$51.7 & N.A. & $^*$58.1 & N.A. \\
\end{tabular}
\caption{Clusters observed with TTF, with observed total fluxes and 
luminosities, or 3 $\sigma$ upper limits.
Fluxes are in units of $10^{-15}$ erg s$^{-1}$ cm$^{-2}$. 
Luminosities in units of $10^{40}$ erg s$^{-1}$.  All values
are uncorrected for extinction.
$^{*}$ No independent [NII] measurement was taken for Abell~1795.
The flux quoted for H$\alpha$ includes the [NII] emission.  The
[NII] images of Abell~2204 were corrupted and could not be reduced.
}
\end{table}

All clusters were observed with a bandwidth of 12\AA\ (FWHP) except
Abell~1795, which was observed with a bandwidth of 60\AA.  Except for
this cluster, each was observed at the redshifted wavelengths
of \Ha\ and [NII] (rest wavelength 6584\AA), although the
[NII] image for Abell~2204 was corrupted and could not
be reduced. Abell~1795 was observed
with a single broader bandpass covering both lines. For our data, 
the etalon step between on- and off-line
images was equivalent to approximately 40\AA\ blueward of H$\alpha$
and 40\AA\ redward of [NII] (about twice this value for Abell~1795 due to
its wider bandpass). The off-line image was sufficiently far removed
in wavelength to avoid picking up any extended line emission present
in any high velocity wing of the emission-line, but sufficiently close
in wavelength that colour differences of the continuum between
on- and off- bands will have a negligible effect on
its removal from the on-line image.

\begin{figure*}
\centering
\resizebox{17cm}{!}{\vbox{
  \includegraphics[width=17cm,angle=90]{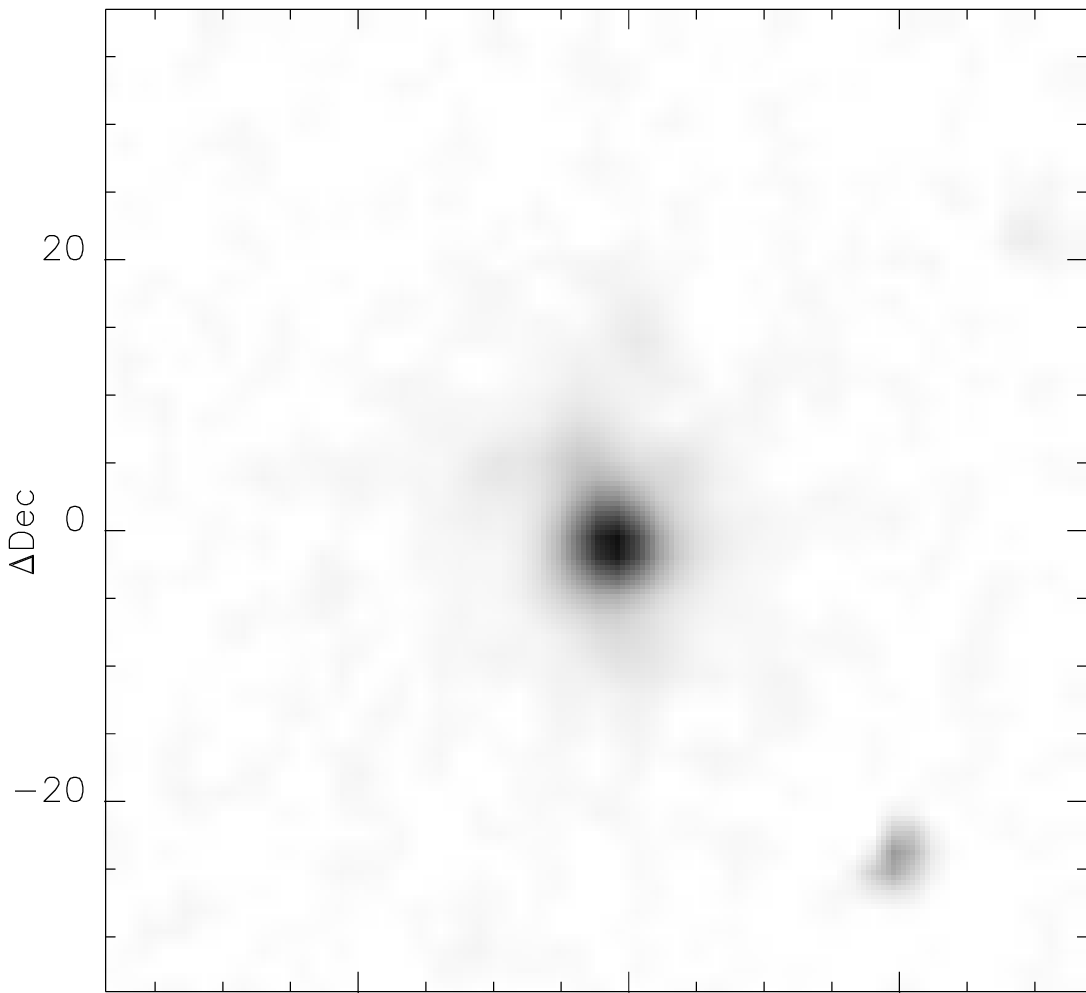}\vskip 3.truecm}
   \includegraphics{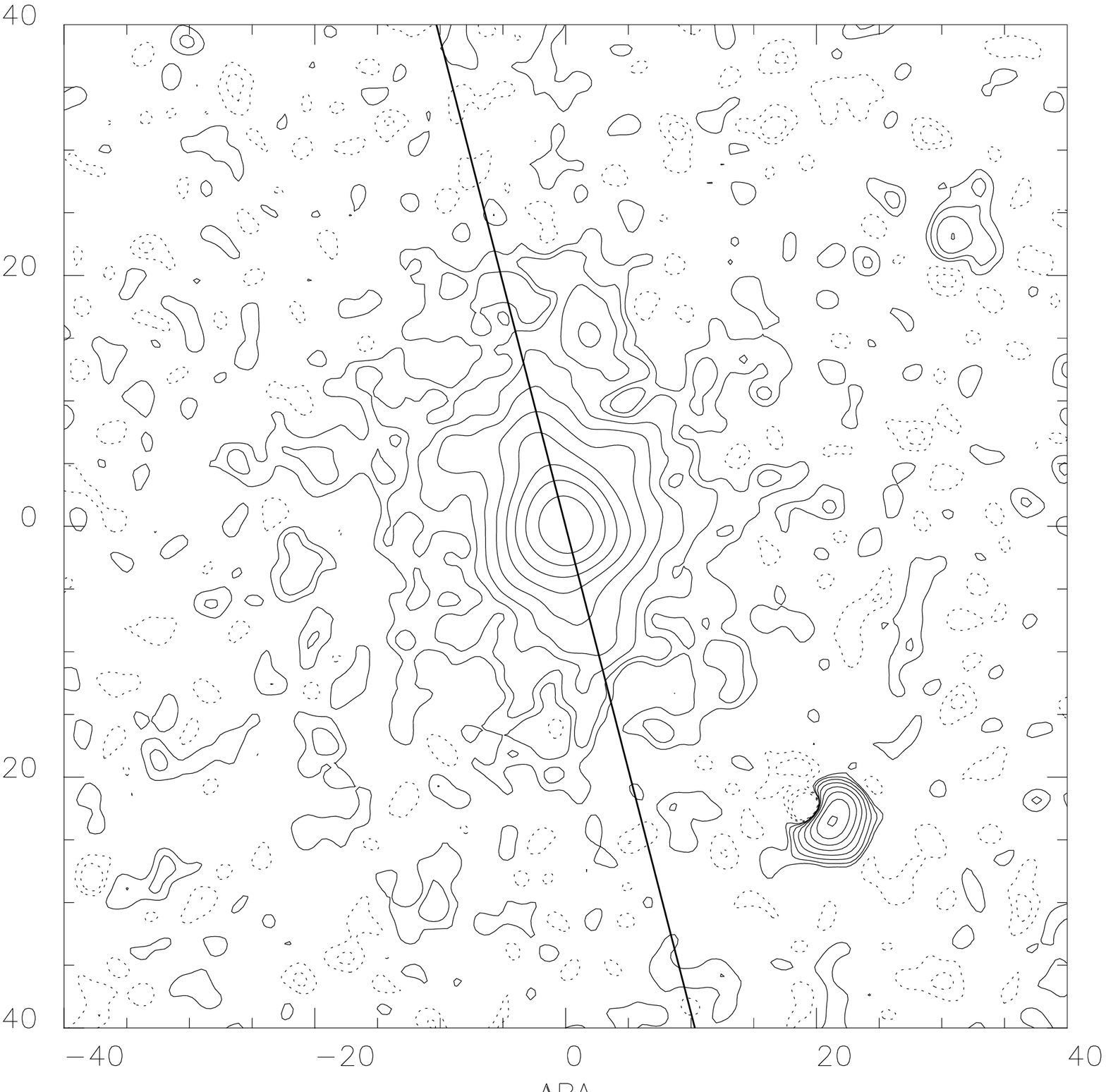}
   \includegraphics{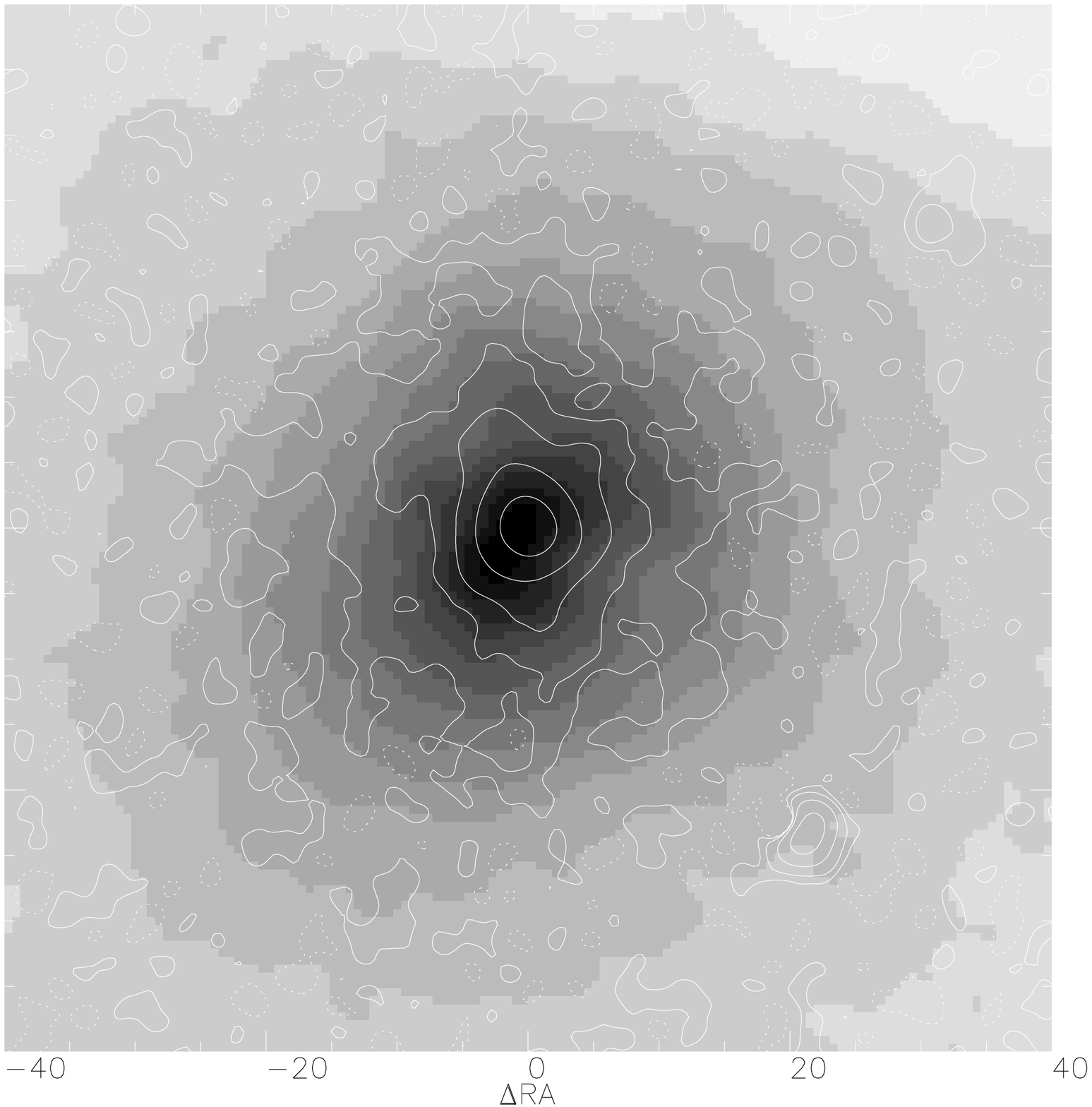}}
\caption{Images of the line emission around A2597. North is up and
East at left.  Coordinates give the offsets from the nucleus
in arcsec on the sky. At Abell~2597, 1 arcsec = 1.66 kpc.  The small feature
in the bottom right is the residual image of a bright star. Left: Greyscale and
Middle: contour plot of H$\alpha$+[NII]. Lowest solid contour is at
3.0$\times 10^{-18}$ergs s$^{-1}$cm$^{-2}$arcsec$^{-2}$. In this
and following figures, the subsequent
contours double in surface brightness, while dashed contours represent
negative (noise) values. Right: Contour plot of line
emission overlayed on greyscale of \chandra X-ray emission from cluster.
For clarity in the overlay every other contour has been omitted.} 
\label{figure:A2597optical}
\end{figure*}

\begin{figure*}
\centering
\resizebox{17cm}{!}{\vbox{
  \includegraphics[width=17cm,angle=90]{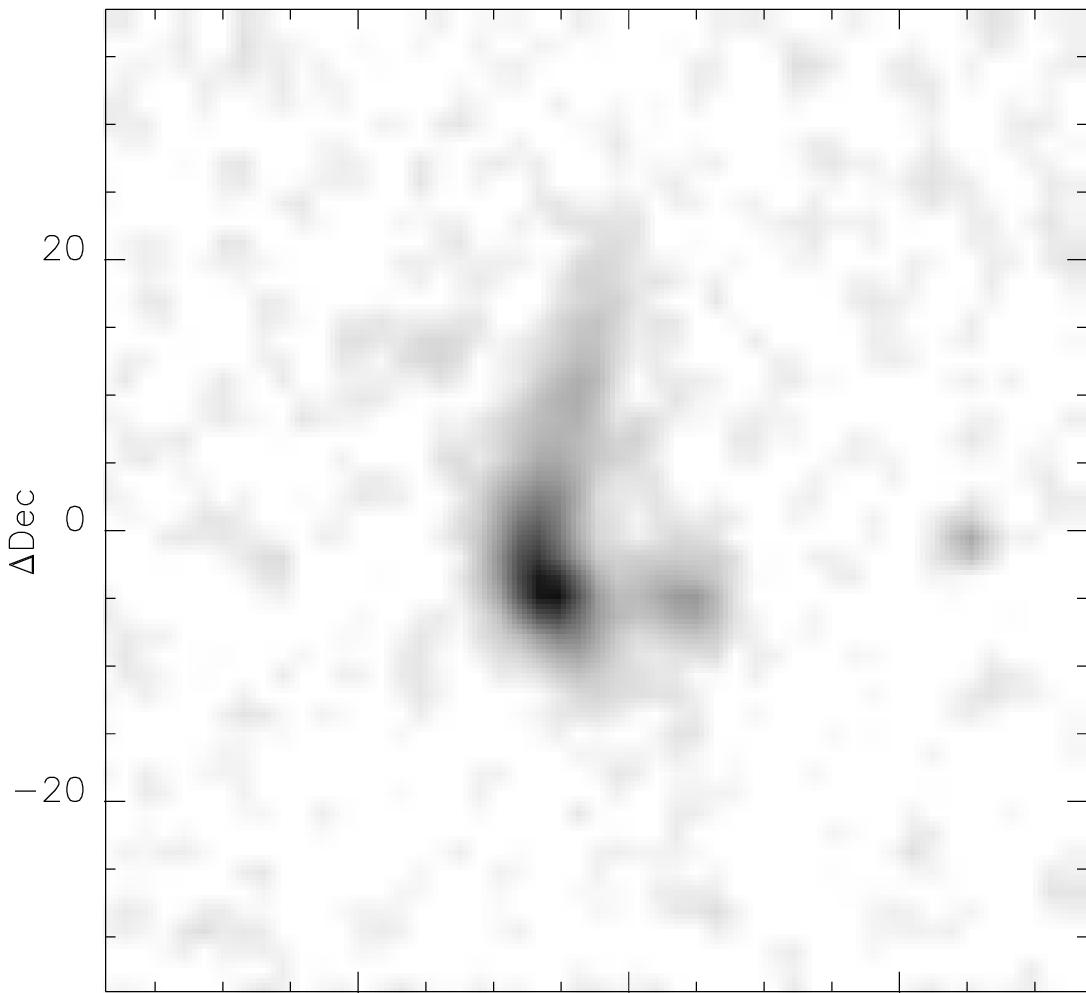}\vskip 3.truecm}
   \includegraphics{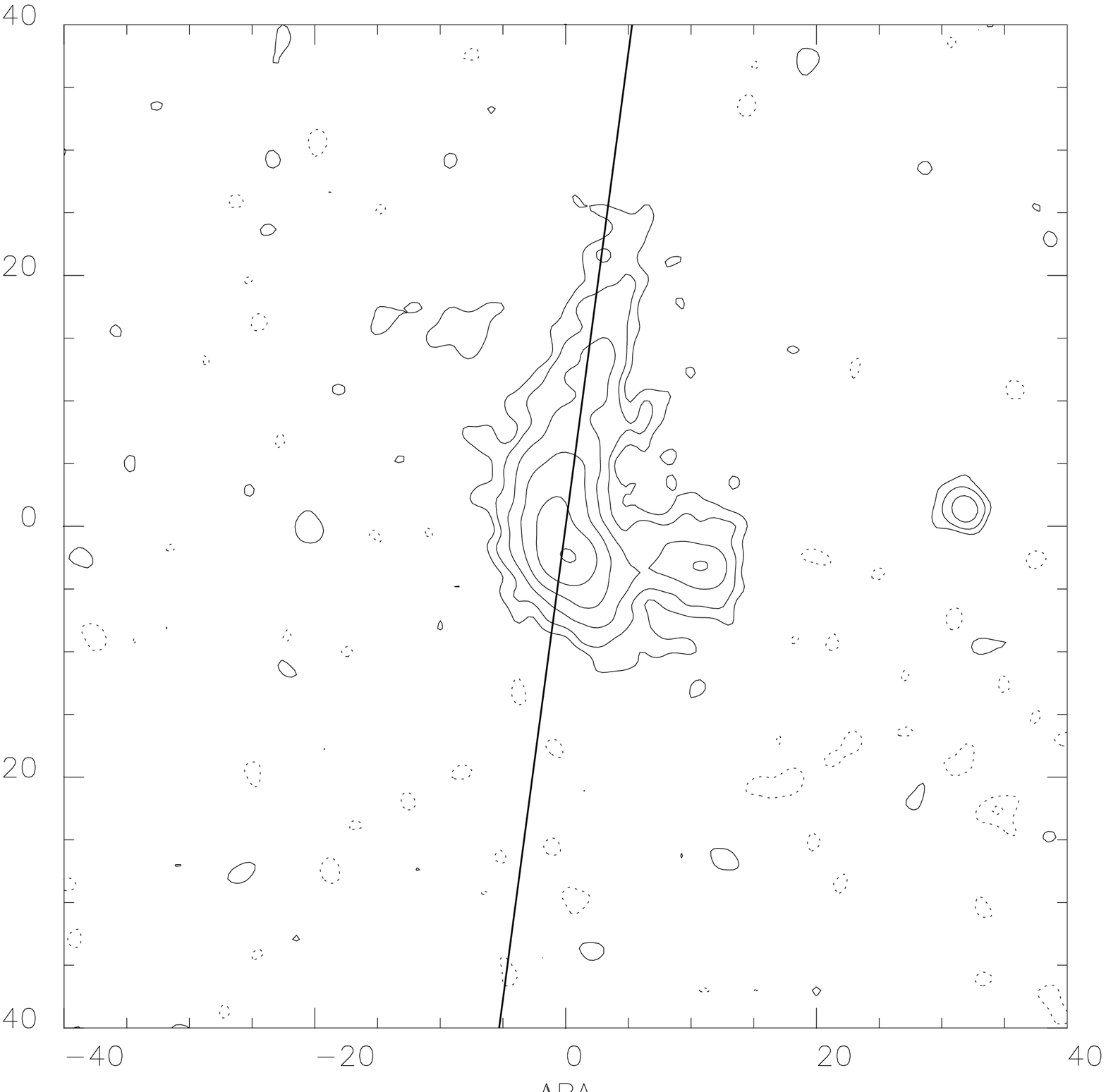}
   \includegraphics{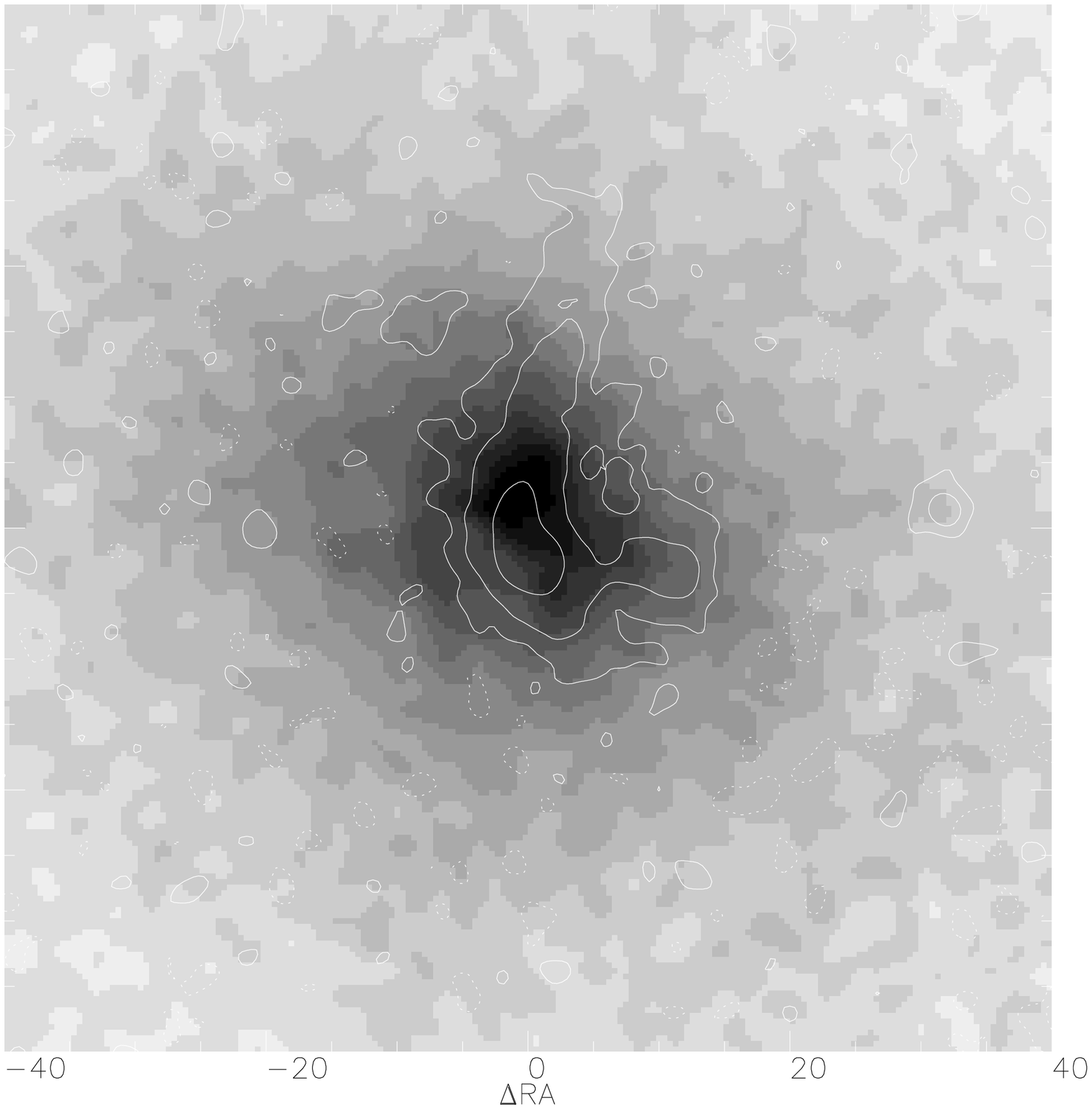}}
\caption{Images of the line emission around Sersic~159-03. Left: Greyscale and
Middle: contour plot of  H$\alpha$+[NII]. Lowest contour is at
3.0$\times 10^{-18}$ergs s$^{-1}$cm$^{-2}$arcsec$^{-2}$. 
Right: Contour plot overlayed on \chandra image.
1 arcsec = 1.13 kpc}
\label{figure:S159optical}
\end{figure*}

\begin{figure*}
\centering
\resizebox{17cm}{!}{\vbox{
  \includegraphics[width=17cm,angle=90]{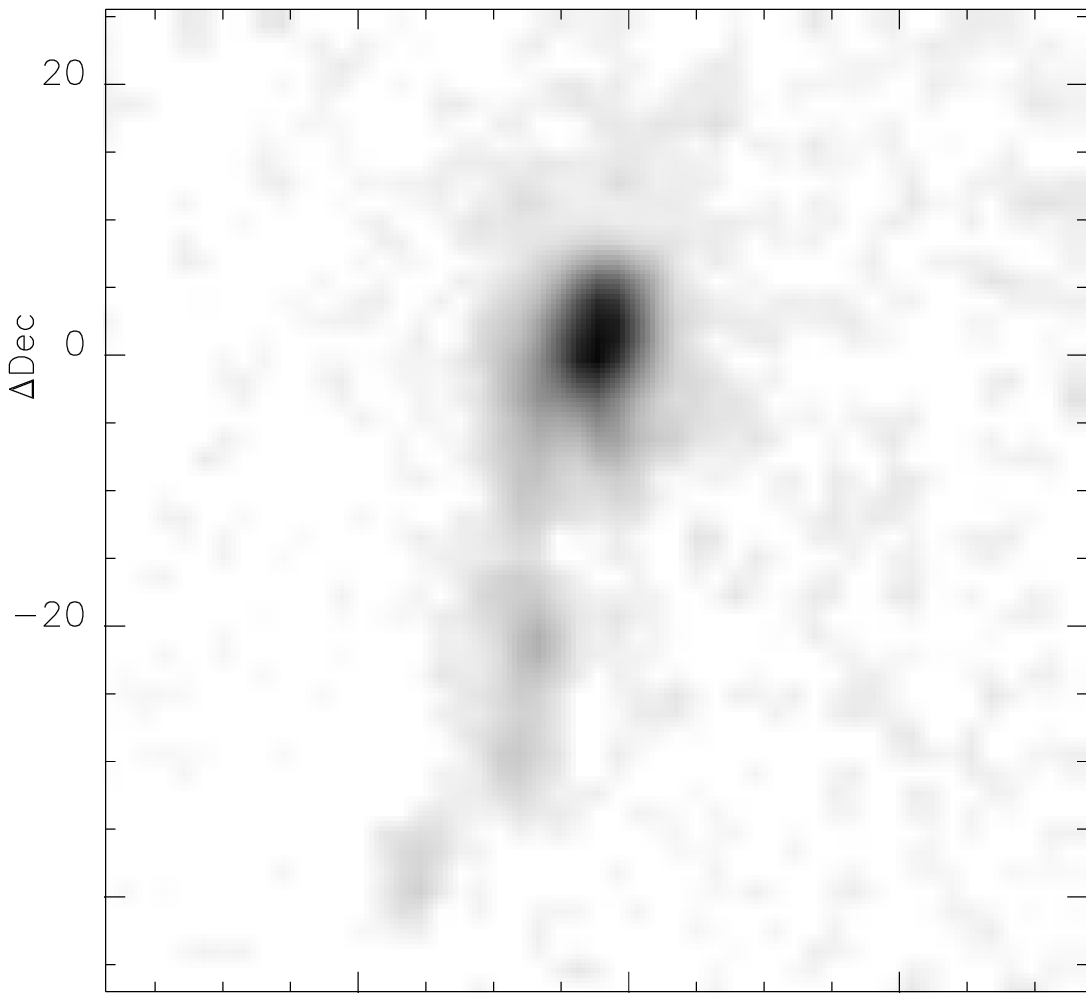}\vskip 3.truecm}
   \includegraphics{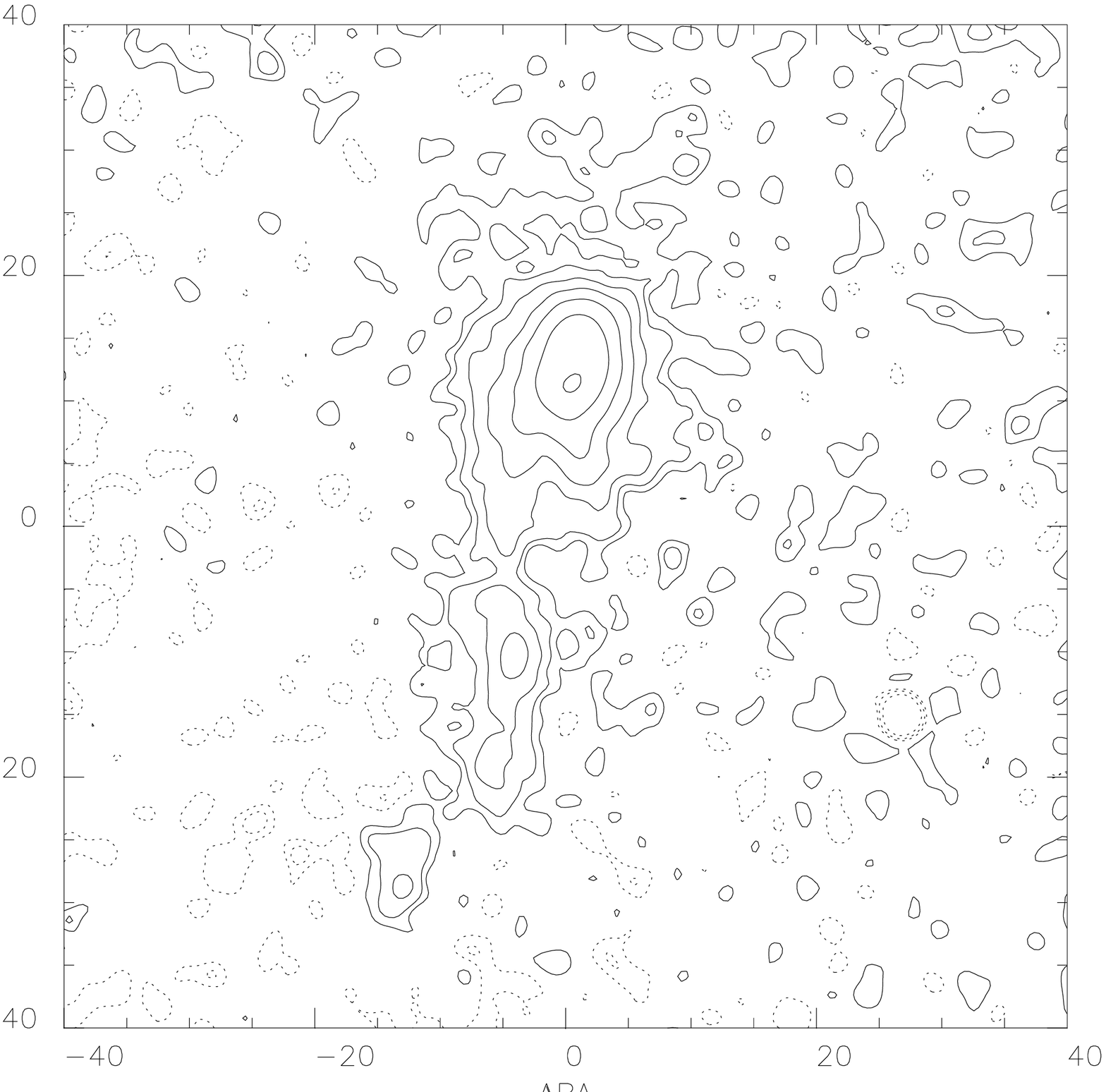}
   \includegraphics{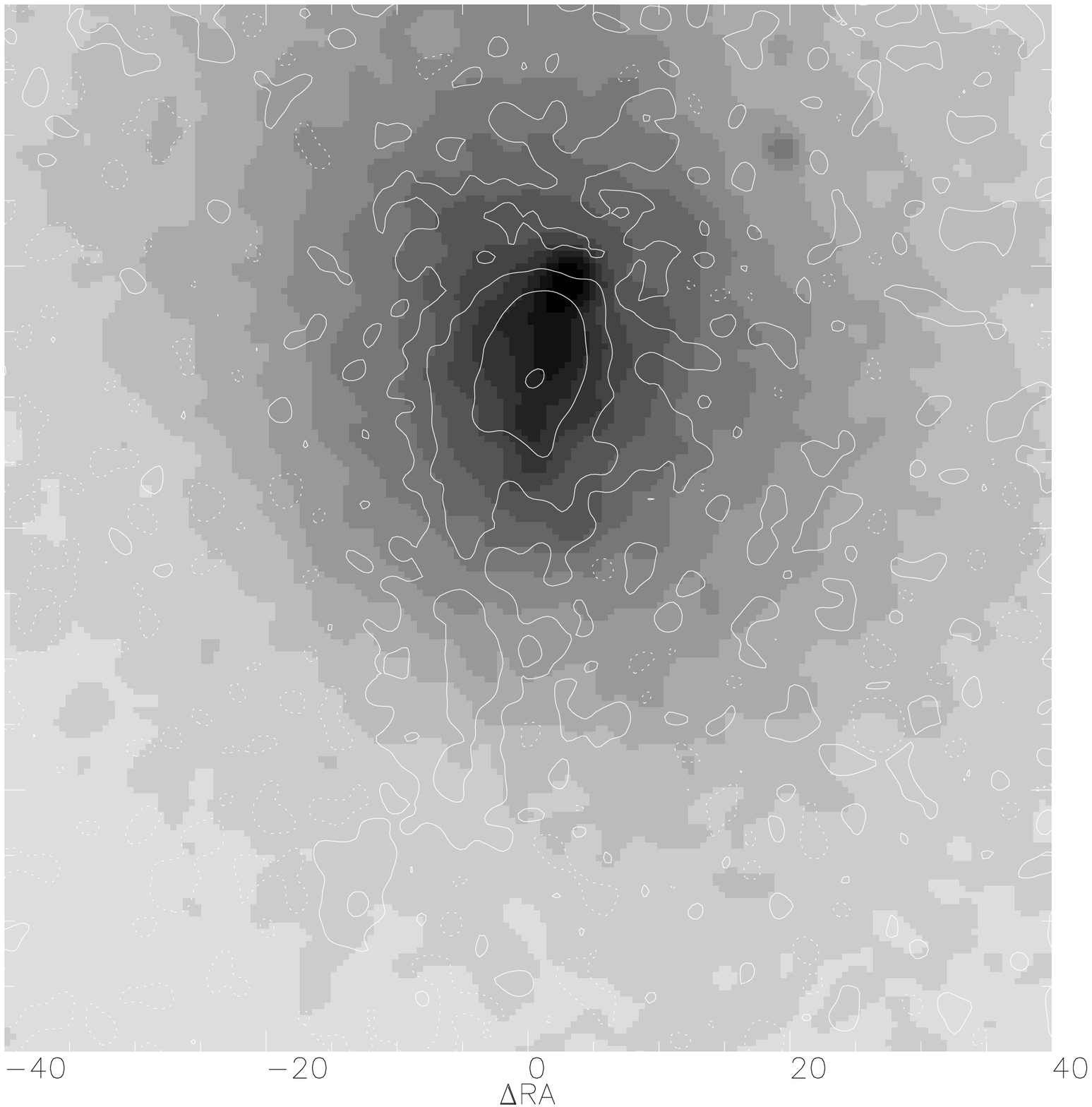}}
\caption{Images of the line emission around Abell~1795. Left: Greyscale and
Middle: contour plot of  H$\alpha$ only. Lowest contour is at
4.6$\times 10^{-18}$ergs s$^{-1}$cm$^{-2}$arcsec$^{-2}$. 
Right: Contour plot overlayed on \chandra image. 
1 arcsec = 1.27 kpc}
\label{figure:A1795optical}
\end{figure*}

\begin{figure*}
\centering
\resizebox{17cm}{!}{\vbox{
  \includegraphics[width=17cm,angle=90]{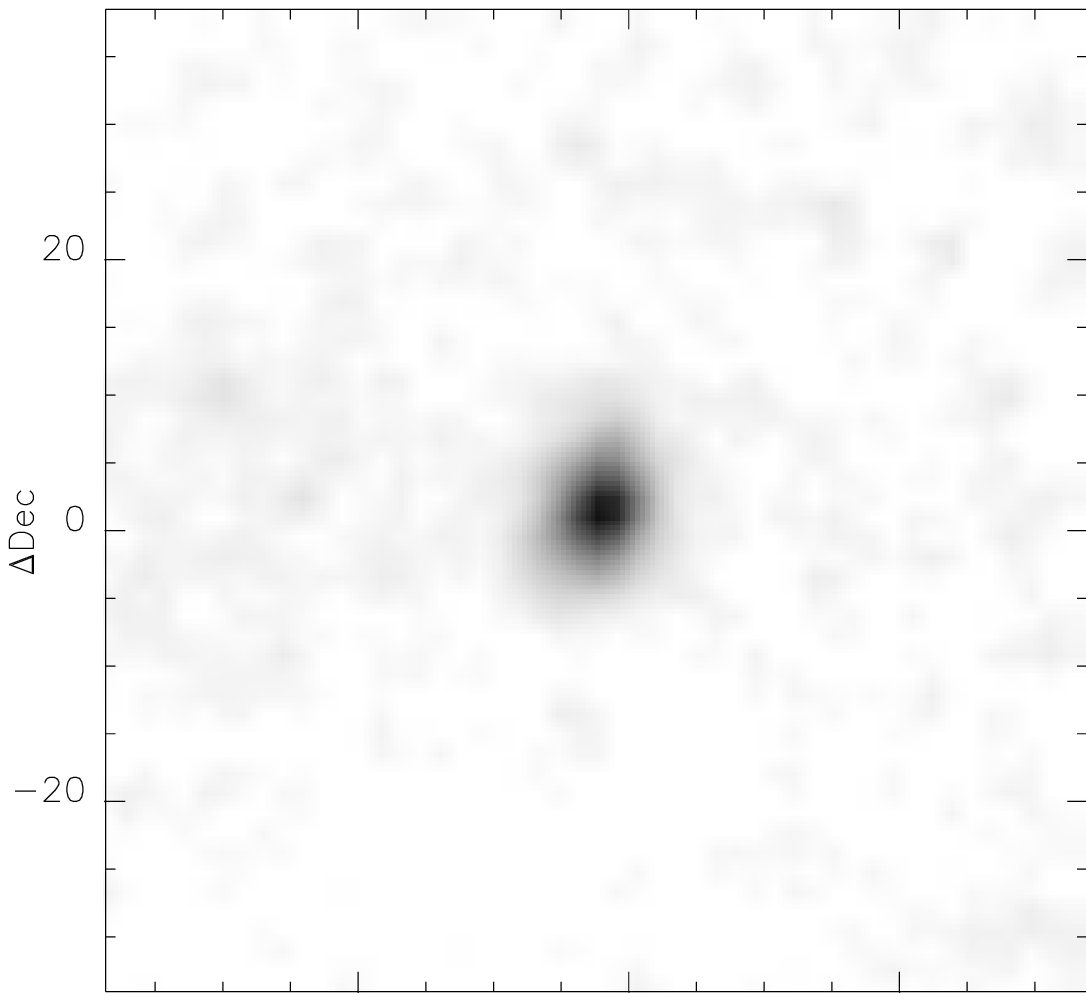}\vskip 3.truecm}
   \includegraphics{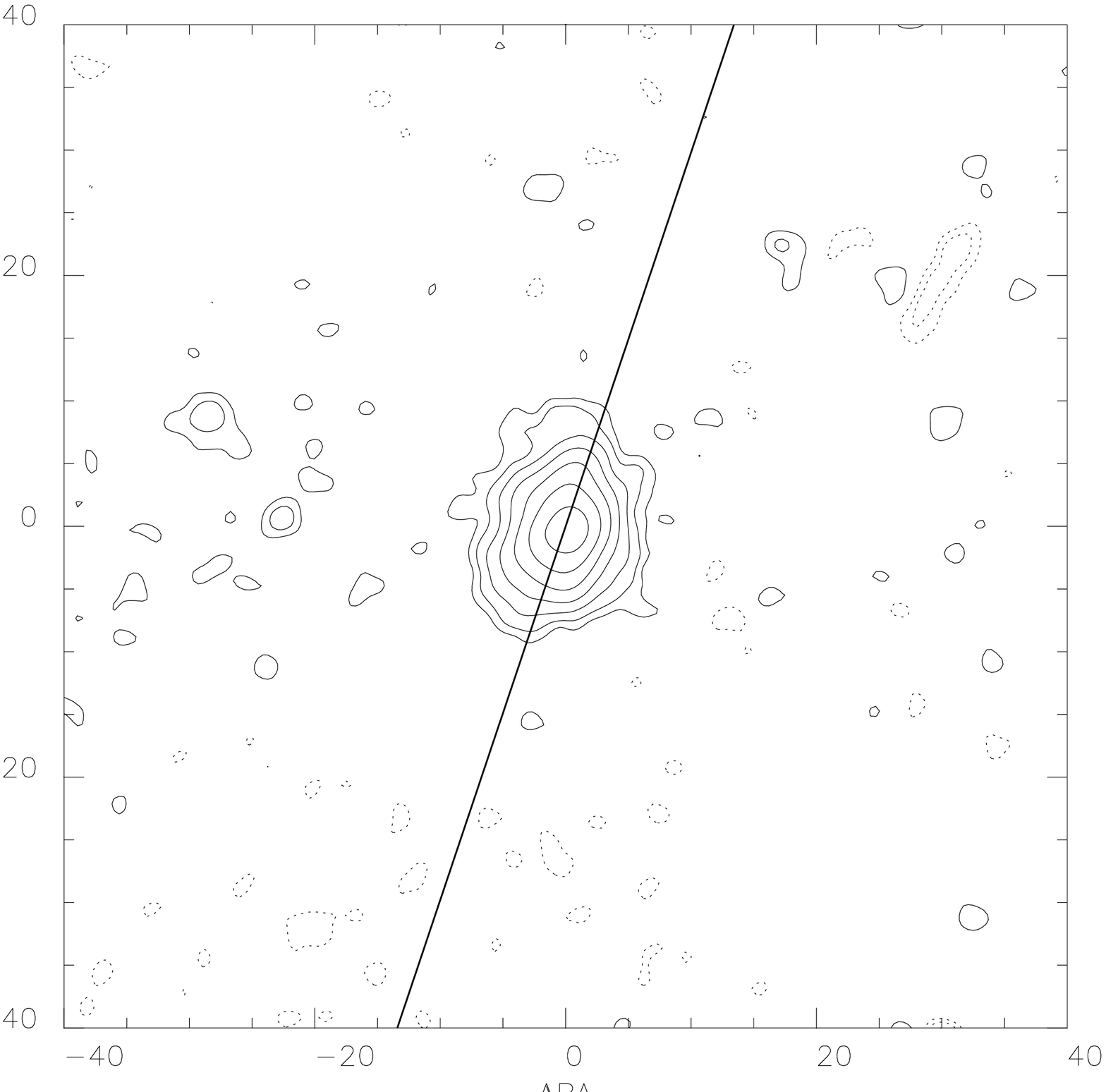}
   \includegraphics{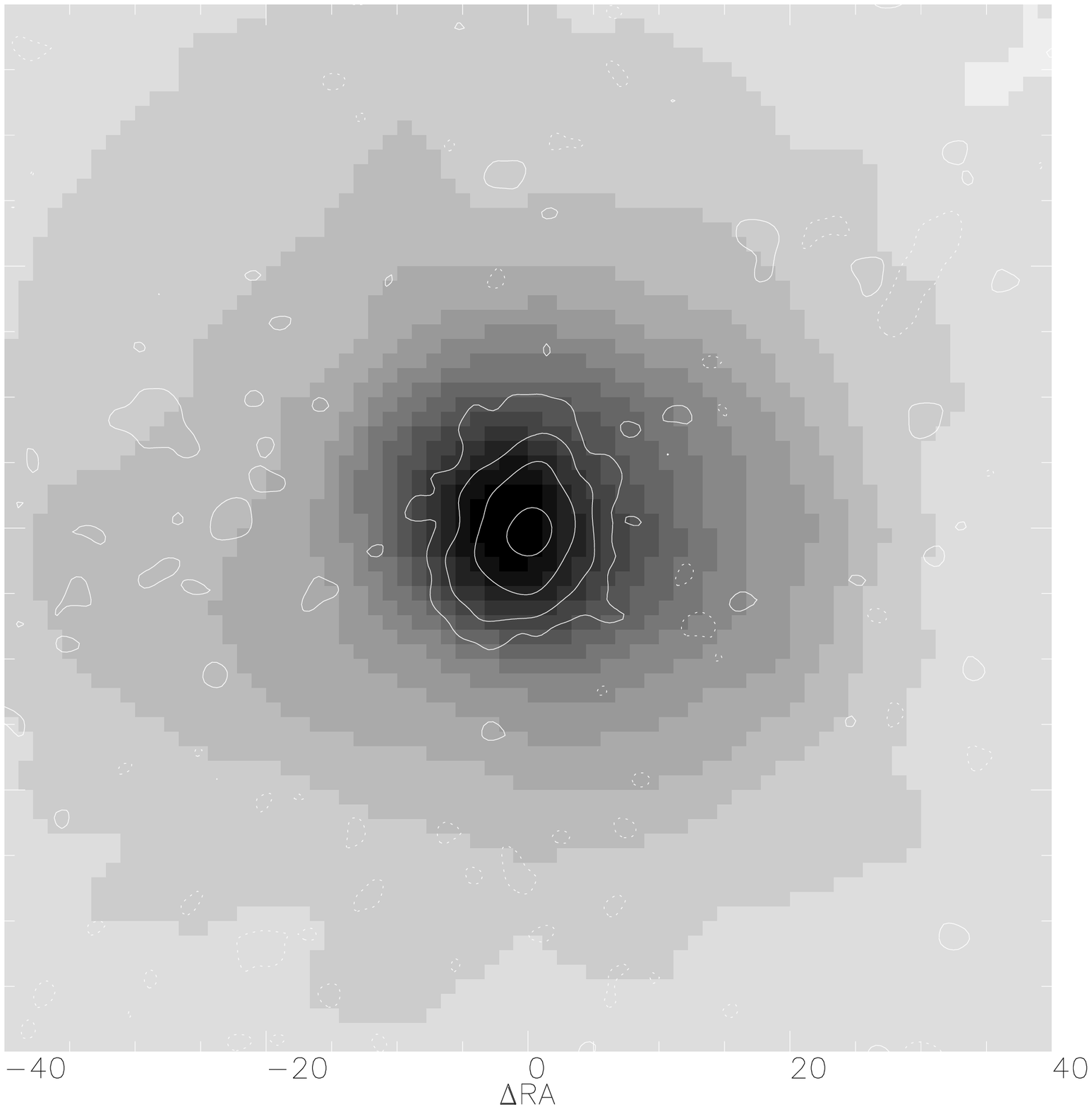}}
\caption{Images of the line emission around Abell~2204. Left: Greyscale and
Middle: contour plotof  H$\alpha$+[NII]. Lowest contours is at
1.3$\times 10^{-17}$ergs s$^{-1}$cm$^{-2}$arcsec$^{-2}$. 
Right: Contour plot overlayed on \chandra image.  1 arcsec = 3.2 kpc}
\label{figure:A2204optical}
\end{figure*}

\begin{figure*}
\centering
\resizebox{22cm}{!}{
\vbox{\includegraphics[height=9cm,angle=90]{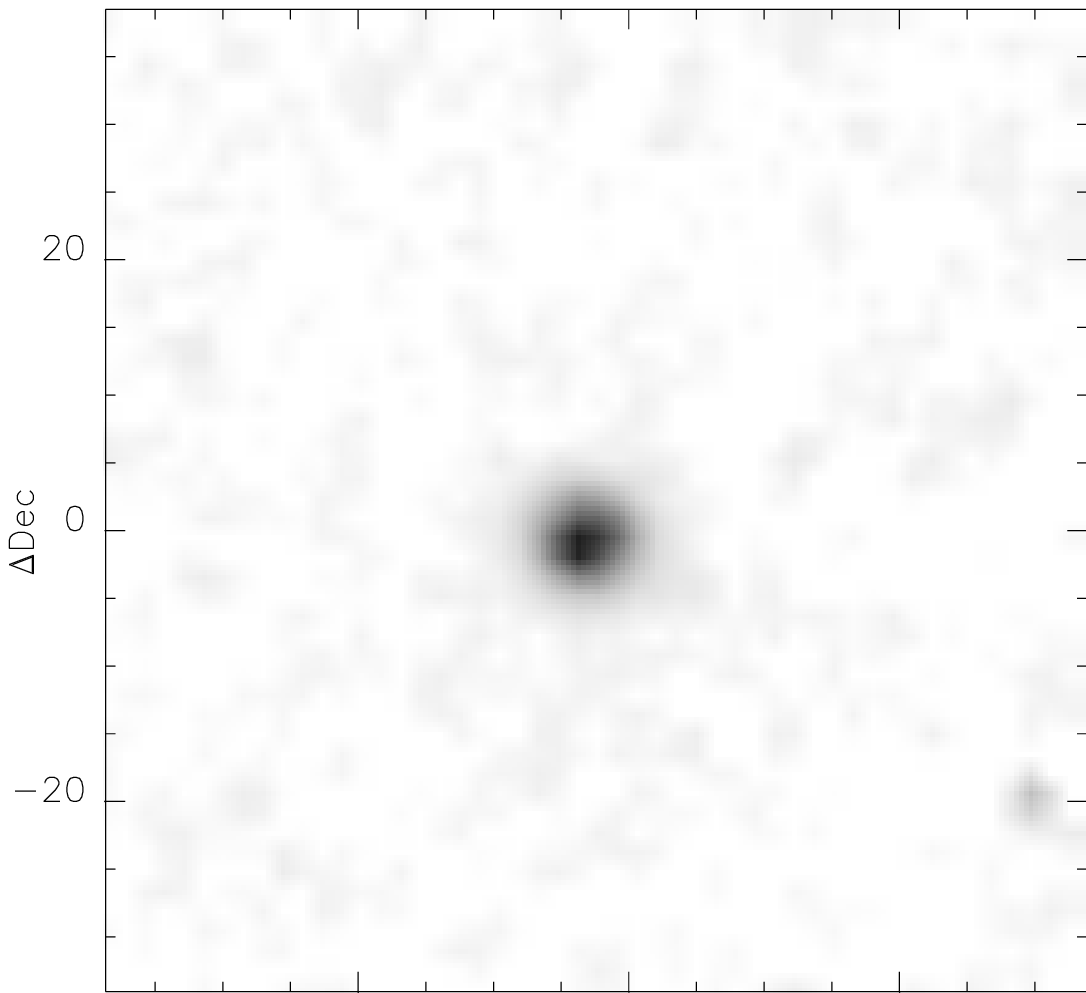}\vskip 0.5truecm}
\includegraphics[height=10cm]{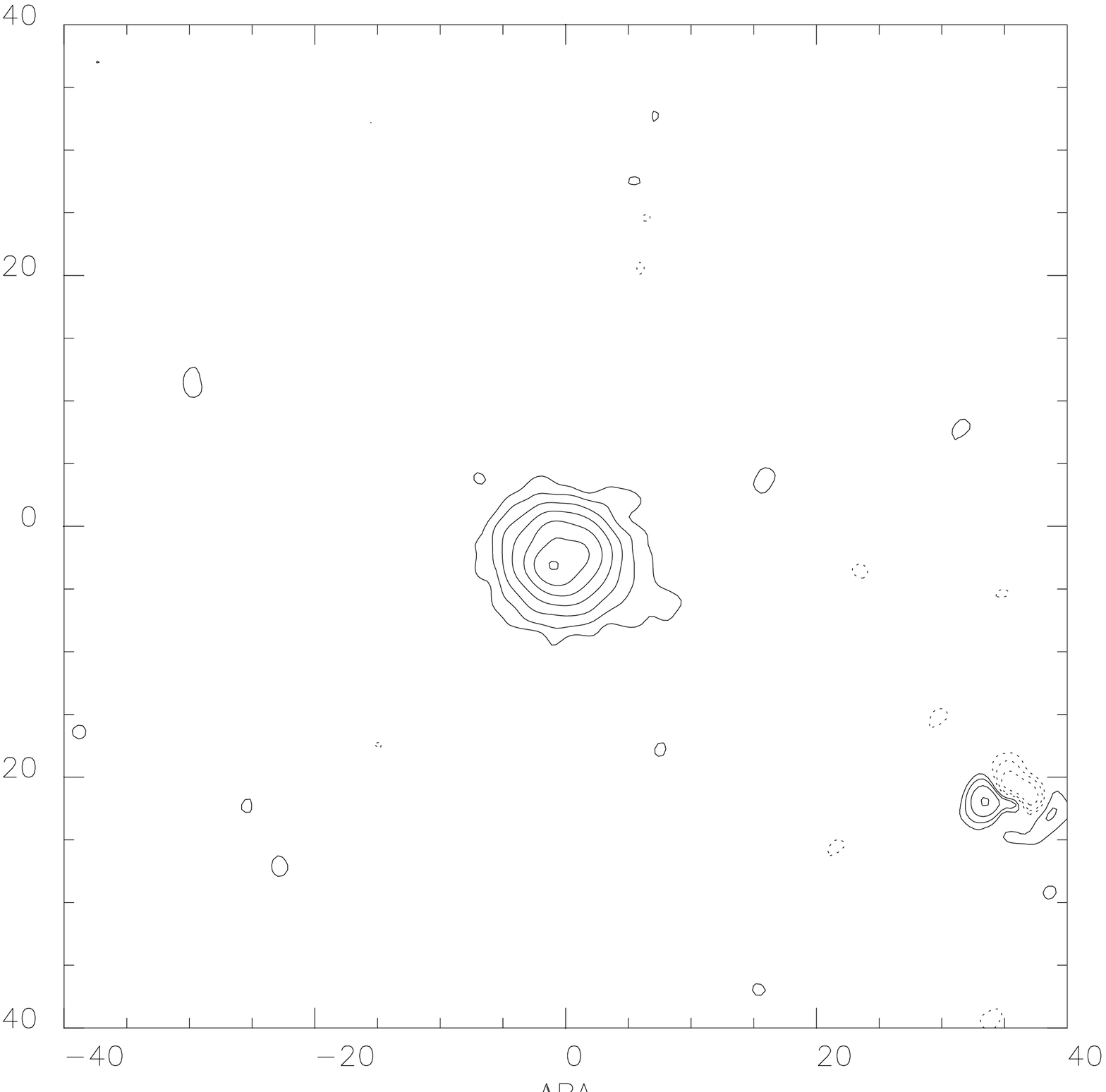}
\includegraphics[width=15cm]{blankx.eps}
}
\caption{Contour plot of 3C445 H$\alpha$ emission.  Lowest
contour is at 5.25$\times
10^{-18}$ergs s$^{-1}$cm$^{-2}$arcsec$^{-2}$.  1 arcsec = 1.13 kpc}
\label{figure:3C445optical}
\end{figure*}

As expected from previously-published less-sensitive observations,
optical emission lines were detected from five of the sources.  No
significant emission was detected from Abell~2029 or the control
cluster Abell~3667.  Grey scale and contour plots of the detected
clusters are shown the left and center panels of Figures
\ref{figure:A2597optical}-\ref{figure:3C445optical} respectively
and for four clusters the contour plots are overlayed on \chandra X-ray images
in the righthand panel.
These images were extracted from the \chandra science archive site
{\bf http://asc.harvard.edu/targets}.  They represent data from the
ACIS-S3 chip, summed over the 0.3-7.0 keV energy range, and reduced
with the {\tt CIAO 3.0} software package.  The photon counts were
smoothed with a 4.4 arcsec full-width at half maximum gaussian kernel.
The log of the Chandra observations is given in Table~2.

\begin{table}
\label{table:2}
\centering
\begin{tabular}{lcr}
\hline\\
Cluster & Exposure Time (ks)& Observation Date \\
\hline\\
 Abell 1795  & 20 &21 03 2000\\
 Abell 2204  & 30 &29 07 2000\\
 Abell 2597  & 40 &28 07 2000\\
 Sersic~159-03 & 10 &13 07 2001\\
\end{tabular}
\caption{Observing Log of Clusters observed with Chandra}
\end{table}

\begin{figure*}
\centering
\resizebox{17cm}{!}{\vbox{
\includegraphics[width=8cm]{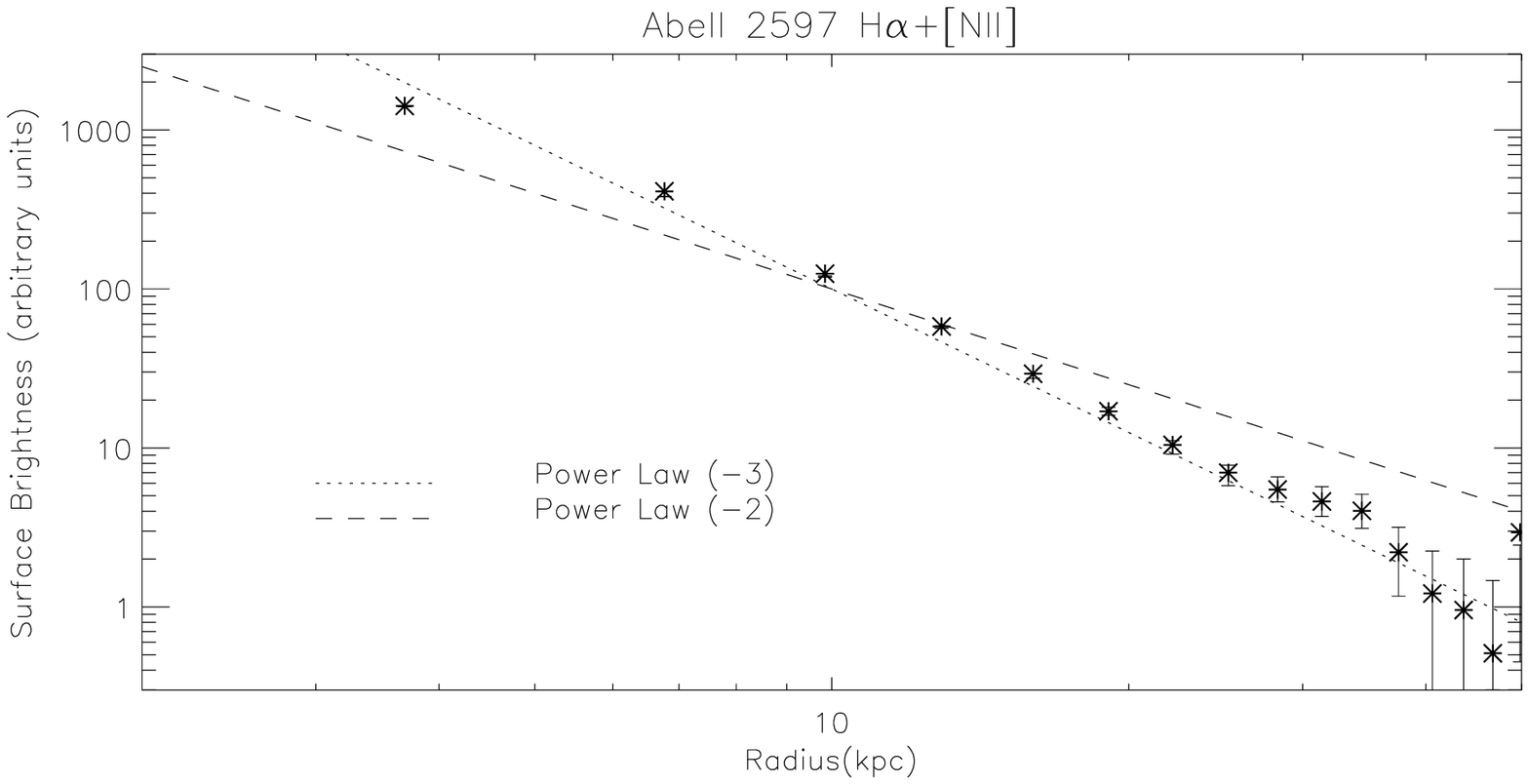}
\includegraphics[width=8cm]{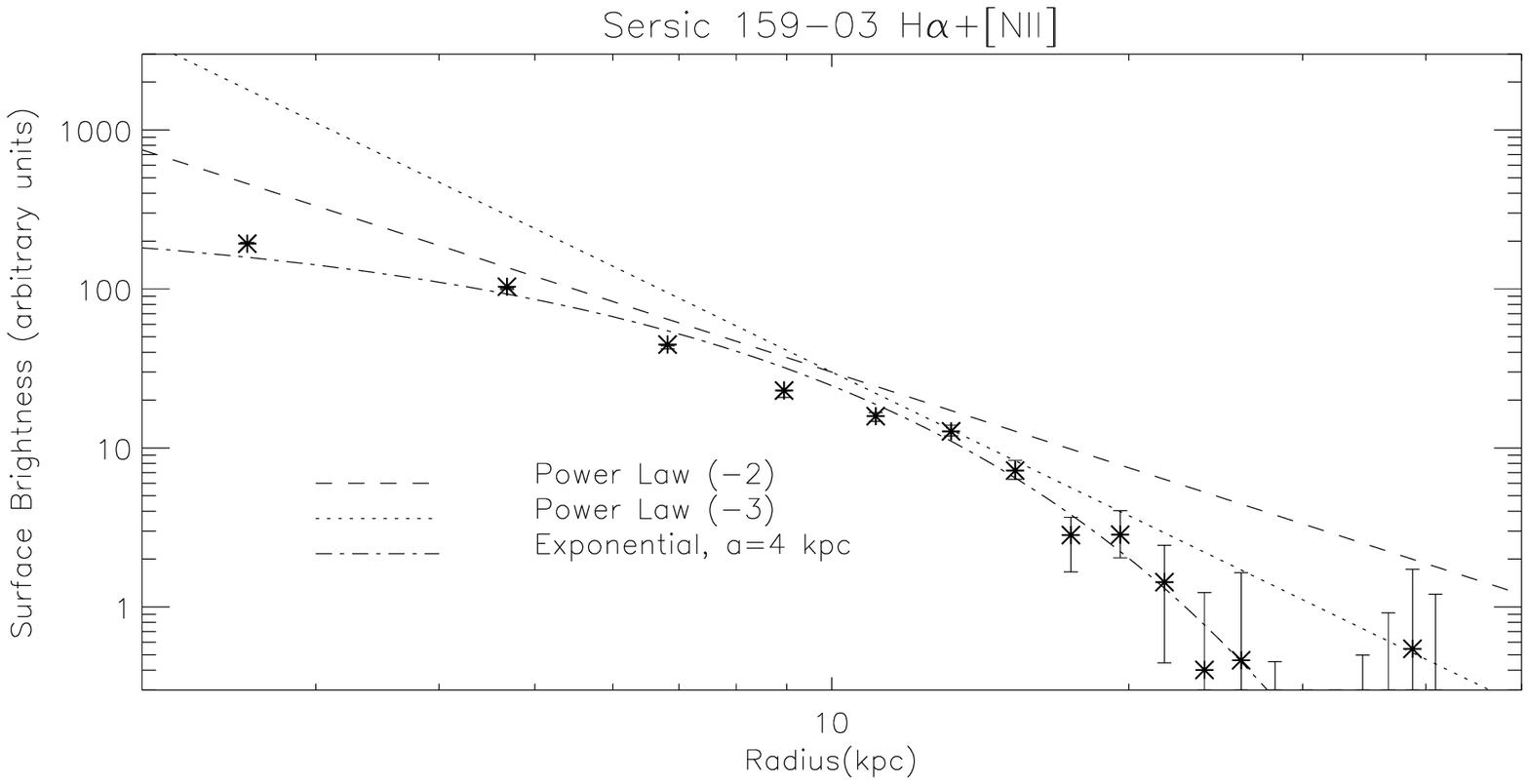}}}
\caption{Left: Averaged surface brightness in combined \Ha and [NII] emission
in circular annuli around the nucleus
of Abell~2597.  Right: Sersic~159-03.  The error bars include noise, 
intrinsic non-circularity, and estimates of the uncertainty in the 
background subtraction at large radii.  For radii less than $\sim 15$
kpc the error bars are smaller than the plot symbols.
Also shown are power law curves for $r^{-2}$ and $r^{-3}$ laws and for Sersic~159-03
an exponential curve $\Sigma\propto \exp(-r/a)$.}
\label{figure:A2597radial}
\end{figure*}

\begin{figure*}
\centering
\resizebox{17cm}{!}{\vbox{
\includegraphics[width=8cm]{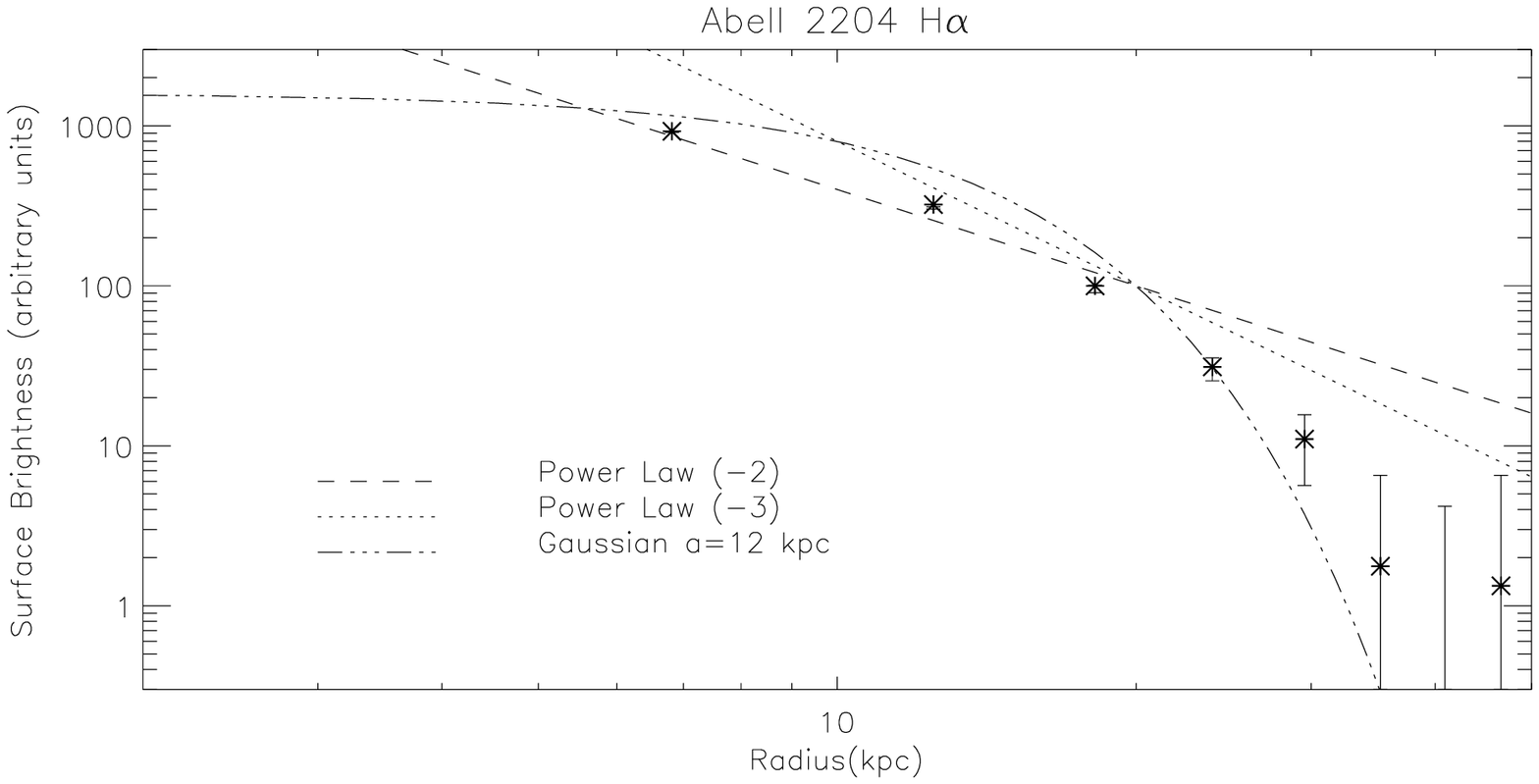}
\includegraphics[width=8cm]{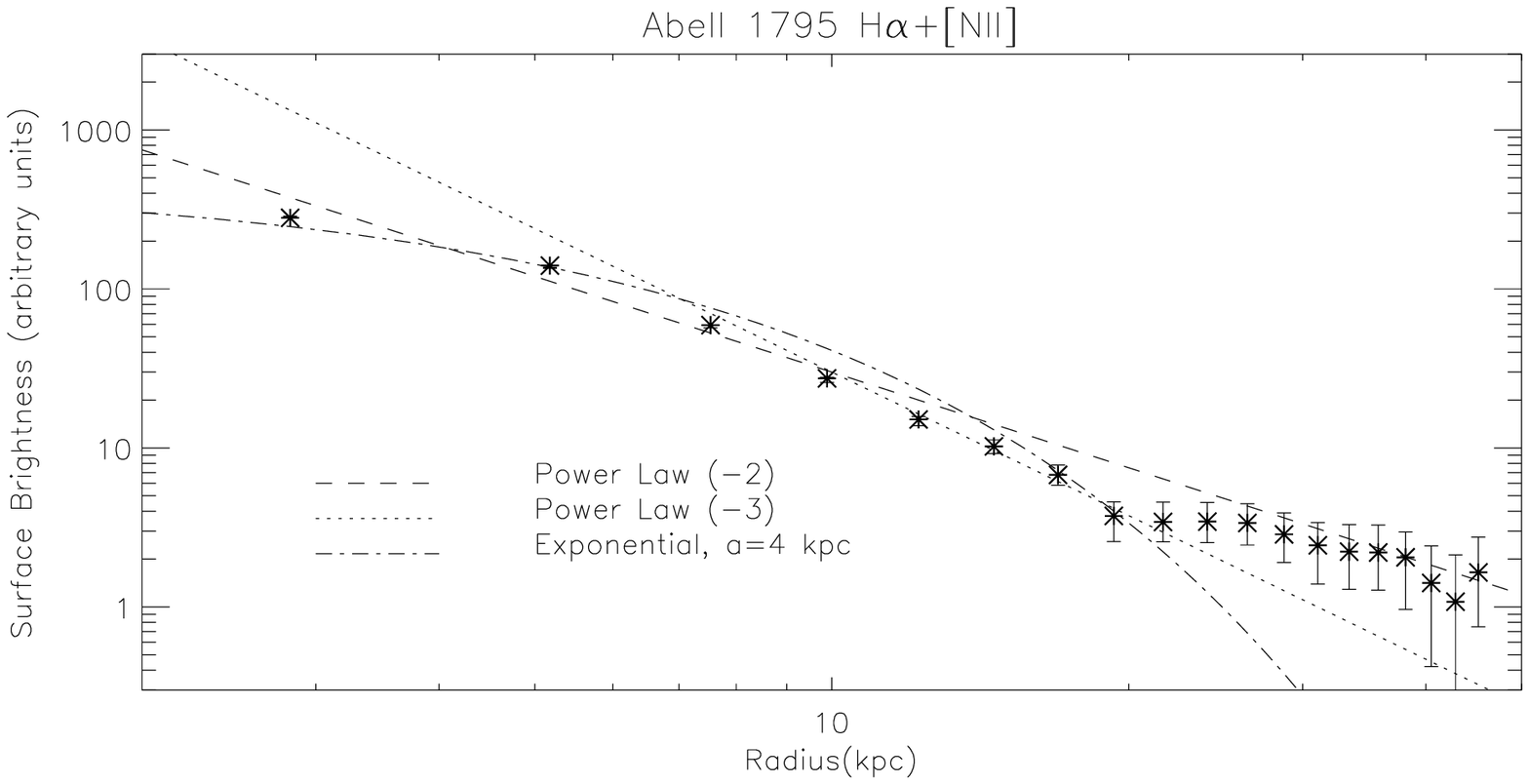}}}
\caption{Averaged surface brightness in circular annuli around the nucleus
of Left: \Ha in Abell~2204; Right: \Ha combined with [NII] in Abell~1795.  
Also shown are power laws, for Abell~2204 a gaussian relation:
$\Sigma\propto \exp(-r/a)^2$, and for Abell~1795 an exponential relation.}
\label{figure:A2204radial}
\end{figure*}

In Table 1 we also list the integrated fluxes and luminosities
contained within a perimeter drawn at a level equal to the rms noise in
the image. In Figures \ref{figure:A2597radial} and \ref{figure:A2204radial}
we plot the azmuthally averaged surface brightness of the emission
lines in each of the four well resolved sources as a function of
radius from the galactic nucleus.  

\subsection{$K-$band spectroscopy}

In order to probe in detail the connection between the atomic and
molecular phases of the line-emitting gas, we obtained
medium-resolution spectroscopy of a subset of the above cluster
sample.  These observations were obtained with ISAAC on the Very Large
Telescope (VLT).  The sample was that subset of the optical imaging
sample described above, showing emission at large radii and easily
visible from the Southern hemisphere.  
Figures \ref{figure:A2597optical},
\ref{figure:S159optical} and \ref{figure:A2204optical} show the
positions of the slits superimposed on the contour maps obtained in
the previous section.

Because of the extended nature of the \Ha\ emission we used a
relatively wide slit (1") for the IR spectroscopy, assuming
(correctly) that the IR lines would be similarly extended.  The slit
was generally oriented through the galaxy nucleus and at a position
angle at which we had previously observed extended \Ha\ emission.  We
used the ISAAC medium resolution (MR) grating, with a resolution of
$\sim$ 3000 after experience showed that the results with the low
resolution (LR) grating ($R\sim 500$) were inferior.  The main
advantage of the MR system is that the K-band contains several strong
telluric absorption systems near the 2-micron end of the band and with
the LR system the individual lines in these systems blend
together. This makes it difficult to detect weak extraterrestrial
emission in this region.  With the MR system the individual lines are
separated, and emission can easily be detected between the absorption
lines.

A disadvantage of this approach is that we cannot observe the entire
K-band in one setting of the spectroscope.  The observable band was
about 0.12$\mu$ at any one setting of the central wavelength.  We have
chosen to take one spectrum in the vicinity of \Pap, \Hp:1-0S(3)
(and occasionally other lines), and one in the vicinity of
\Hp:1-0S(1).  Because of variations of the exact bandpass with
redshift, and small errors in redshift, some of the less important
lines at the edges of the spectra (e.g. Br$\gamma$ were not observed
for all clusters).

Standard stars, usually B-stars of magnitude K$\sim$ 7, 
were observed to correct for telluric absorption and 
to provide calibratable fluxes.  These standard spectra were used 
to remove the continuum emission by a linear fit to the galactic 
emission in regions where no galactic emission or absorption lines are 
expected.  Additional smooth power-law components are also removed, 
which mostly reflect the difference in blackbody slope between the 
galactic stars (mostly K- and M- giants) and the standards.  Any 
absorption lines in the standard stars then remain as emission lines in 
the final spectra.  This is sometimes a problem near rest wavelength Br$\gamma$.  
The aforementioned spectral mismatches appear mostly as very weak
absorption or emission features, before correcting for the total
telluric absorption.

\begin{figure*}
\centering
\includegraphics[angle=90,width=17cm]{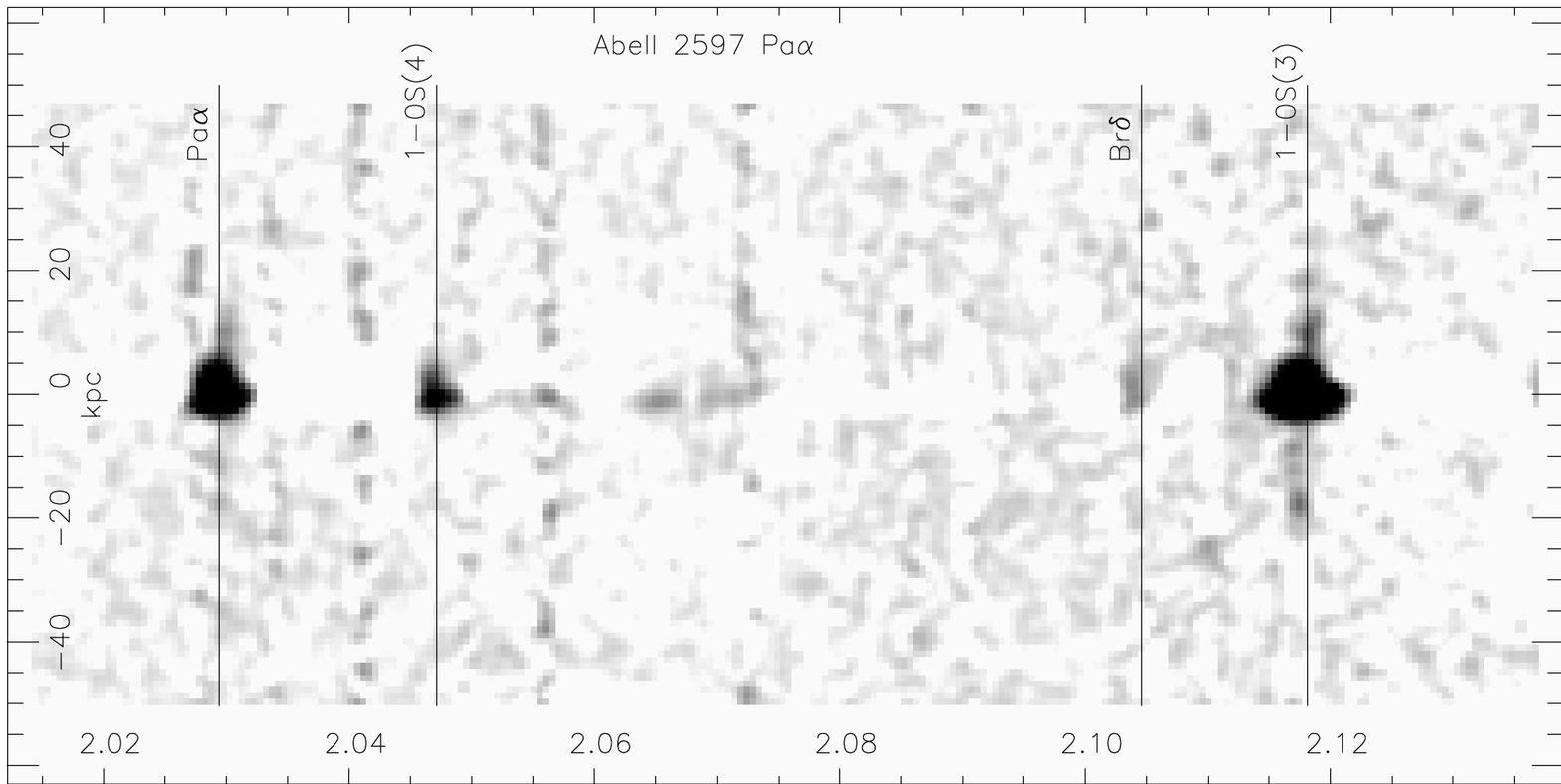}
\includegraphics[angle=90,width=17cm]{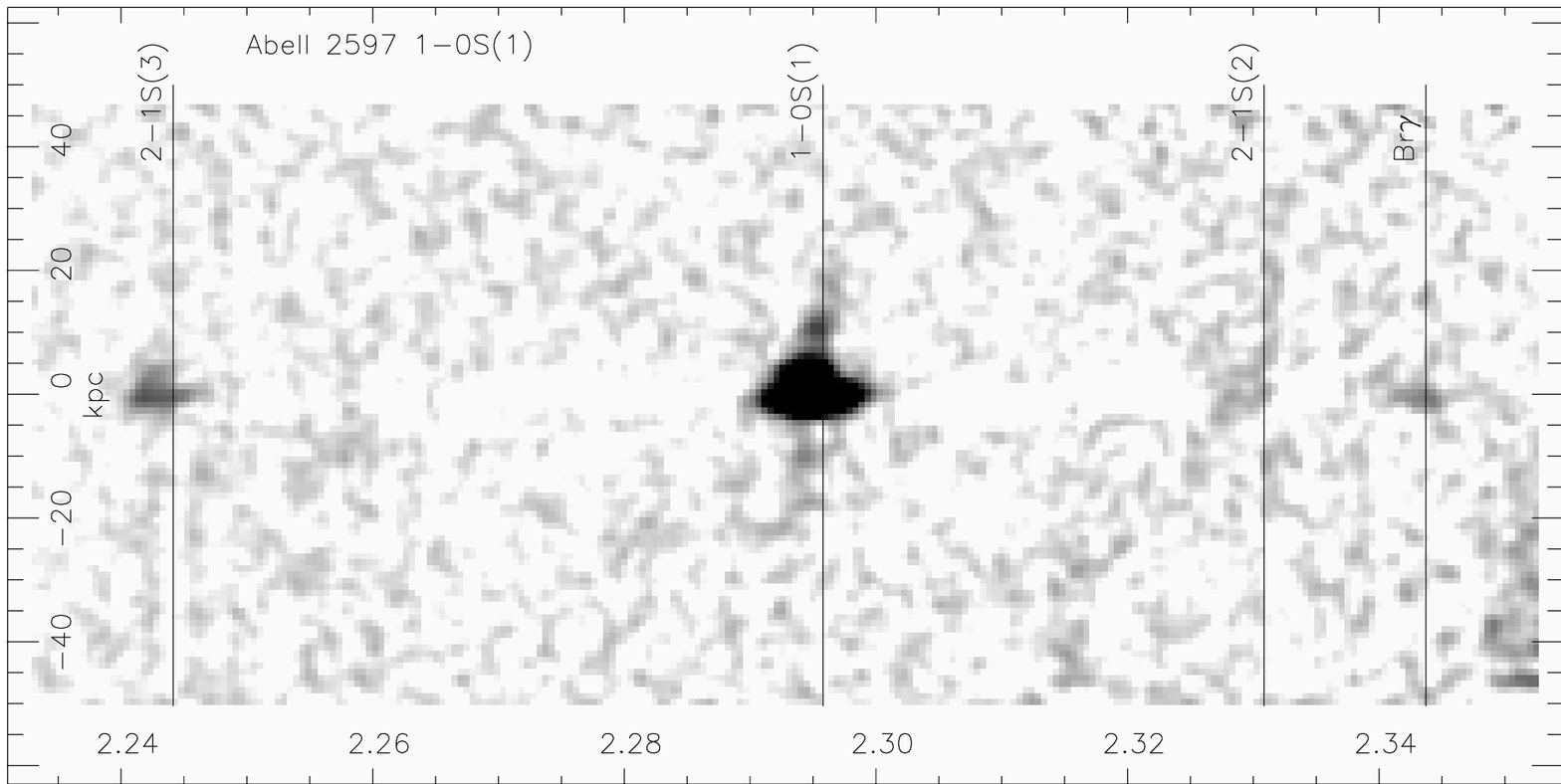}
\caption{Raw two-dimensional spectra of Abell~2597 in
the regions of Pa$\alpha$ and 1-0S(1) respectively.  The abscissa
gives the laboratory wavelength in microns.}
\label{figure:A2597IR}
\end{figure*}

\begin{figure*}
\centering
\includegraphics[angle=90,width=17cm]{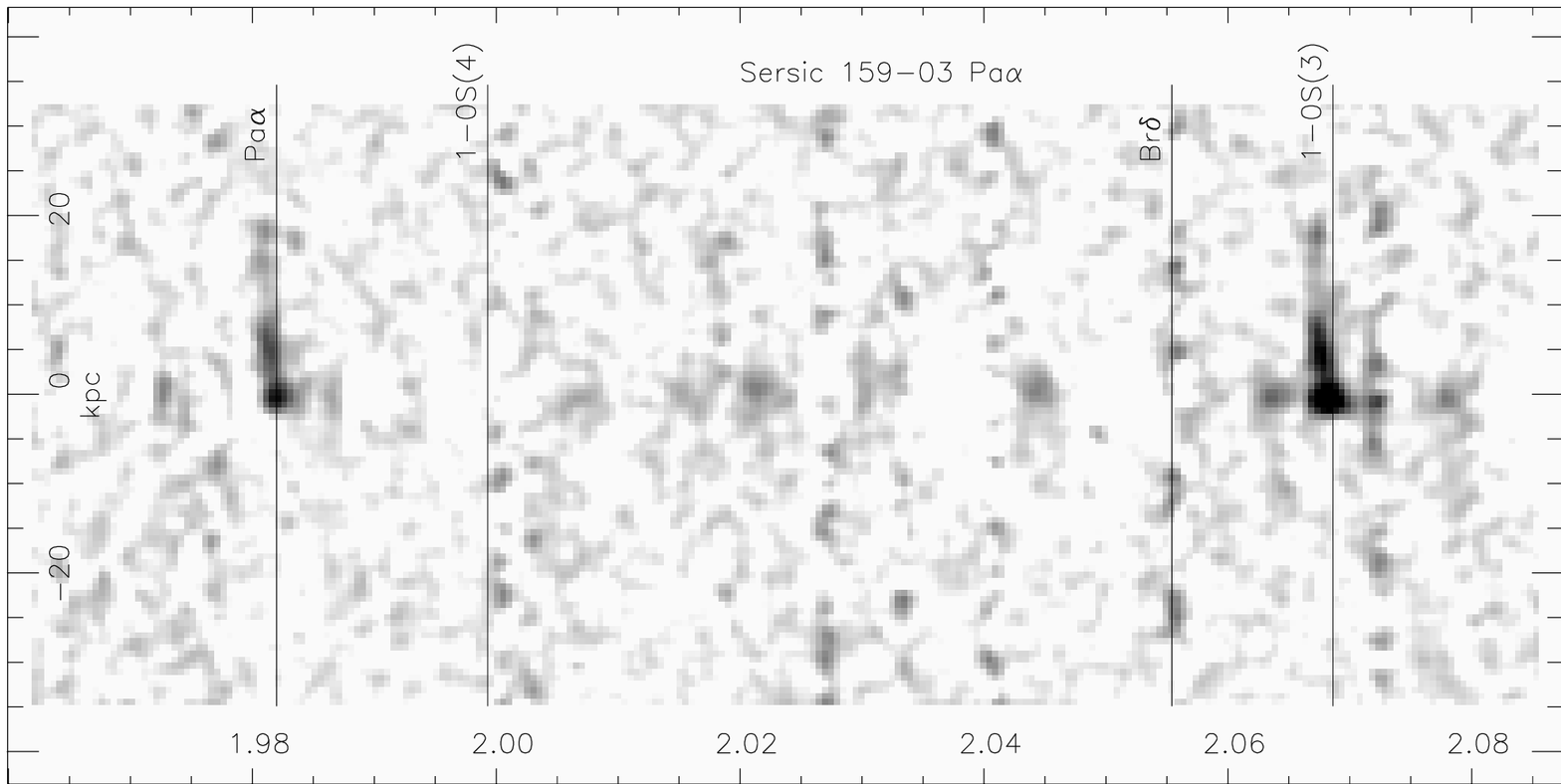}
\includegraphics[angle=90,width=17cm]{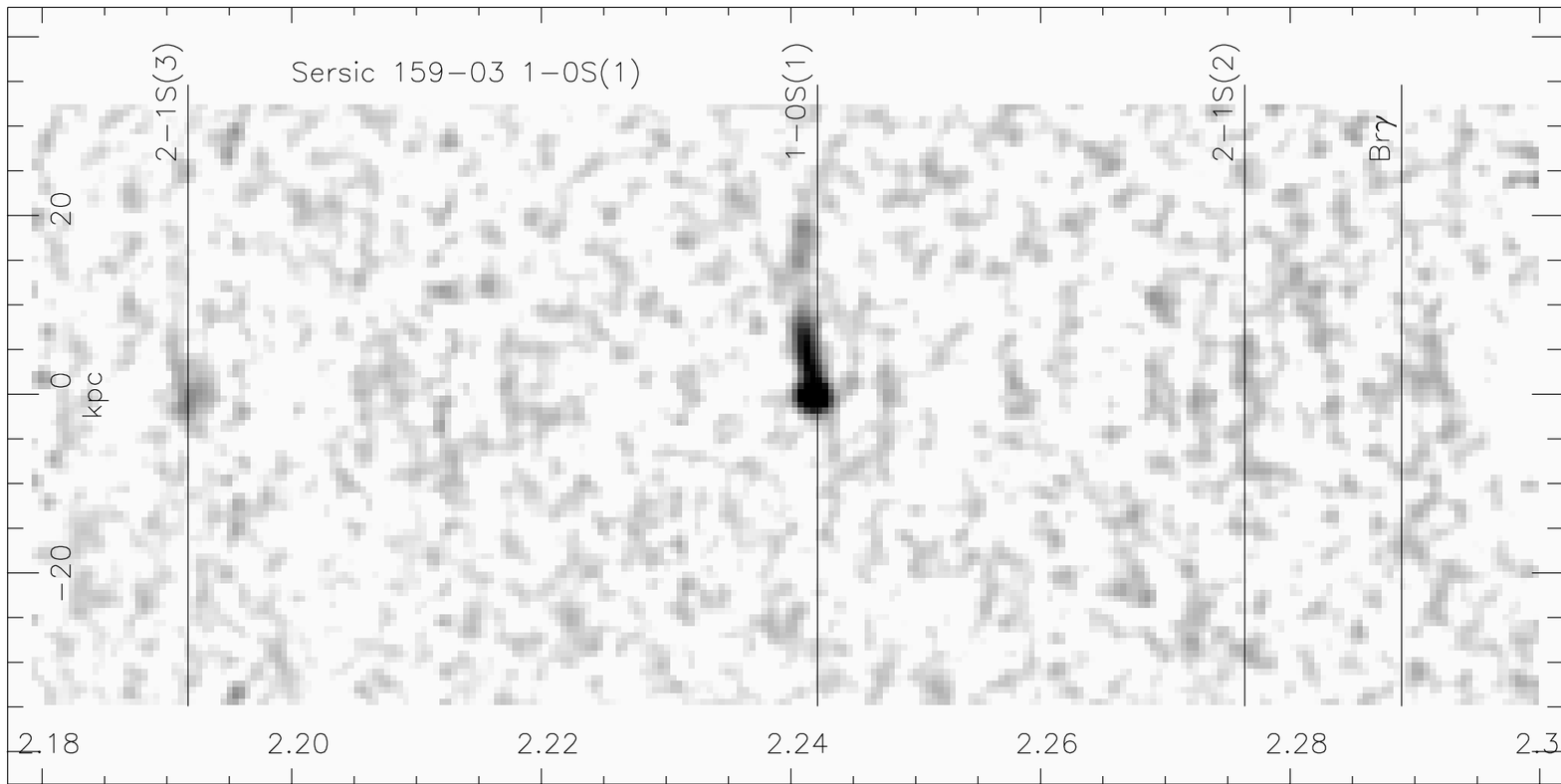}
\caption{Raw two-dimensional spectra of Sersic~159-03 in
the regions of Pa$\alpha$ and 1-0S(1) respectively.}
\label{figure:S159IR}
\end{figure*}

\begin{figure*}
\centering
\includegraphics[angle=90,width=17cm]{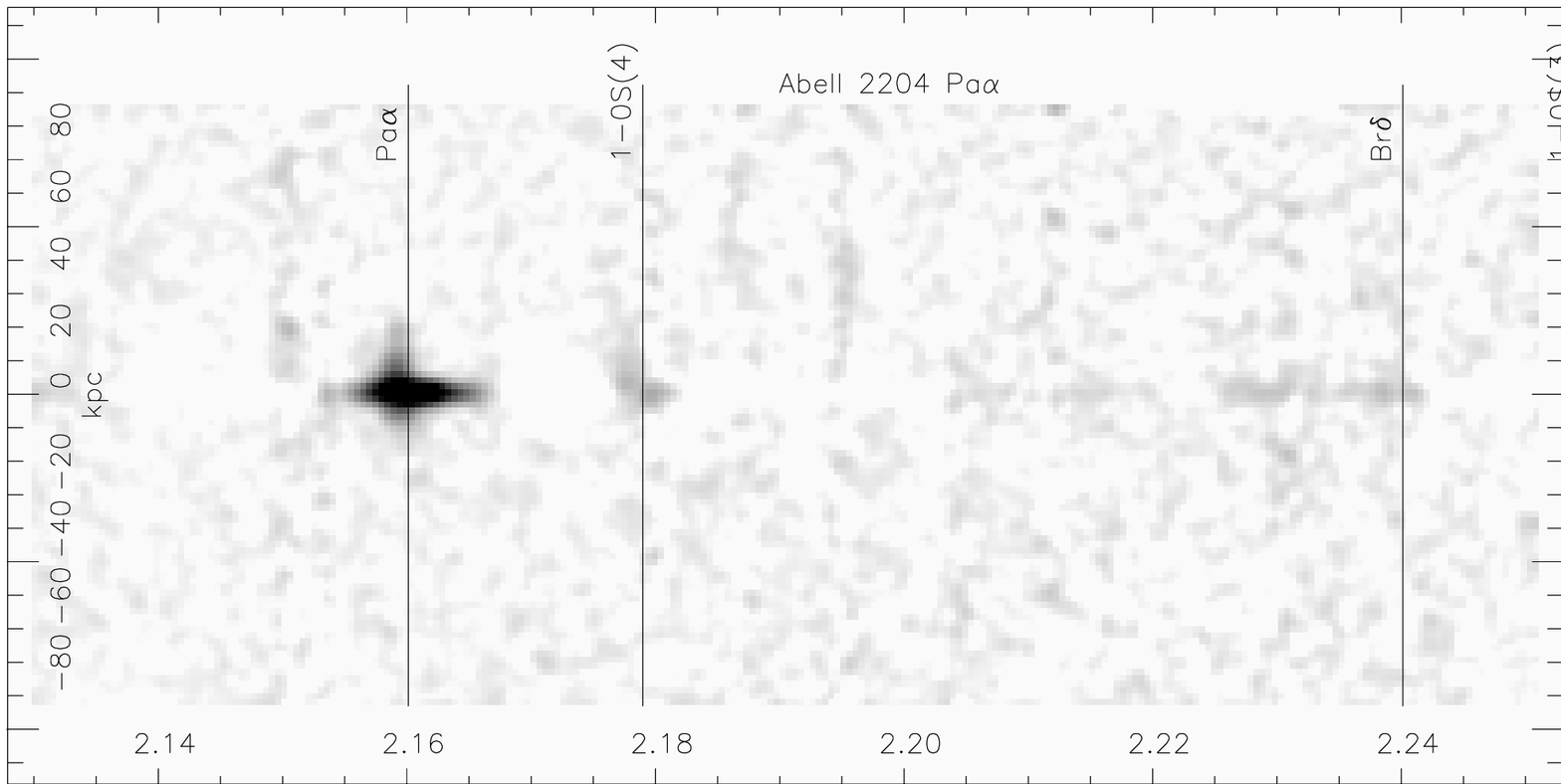}
\includegraphics[angle=90,width=17cm]{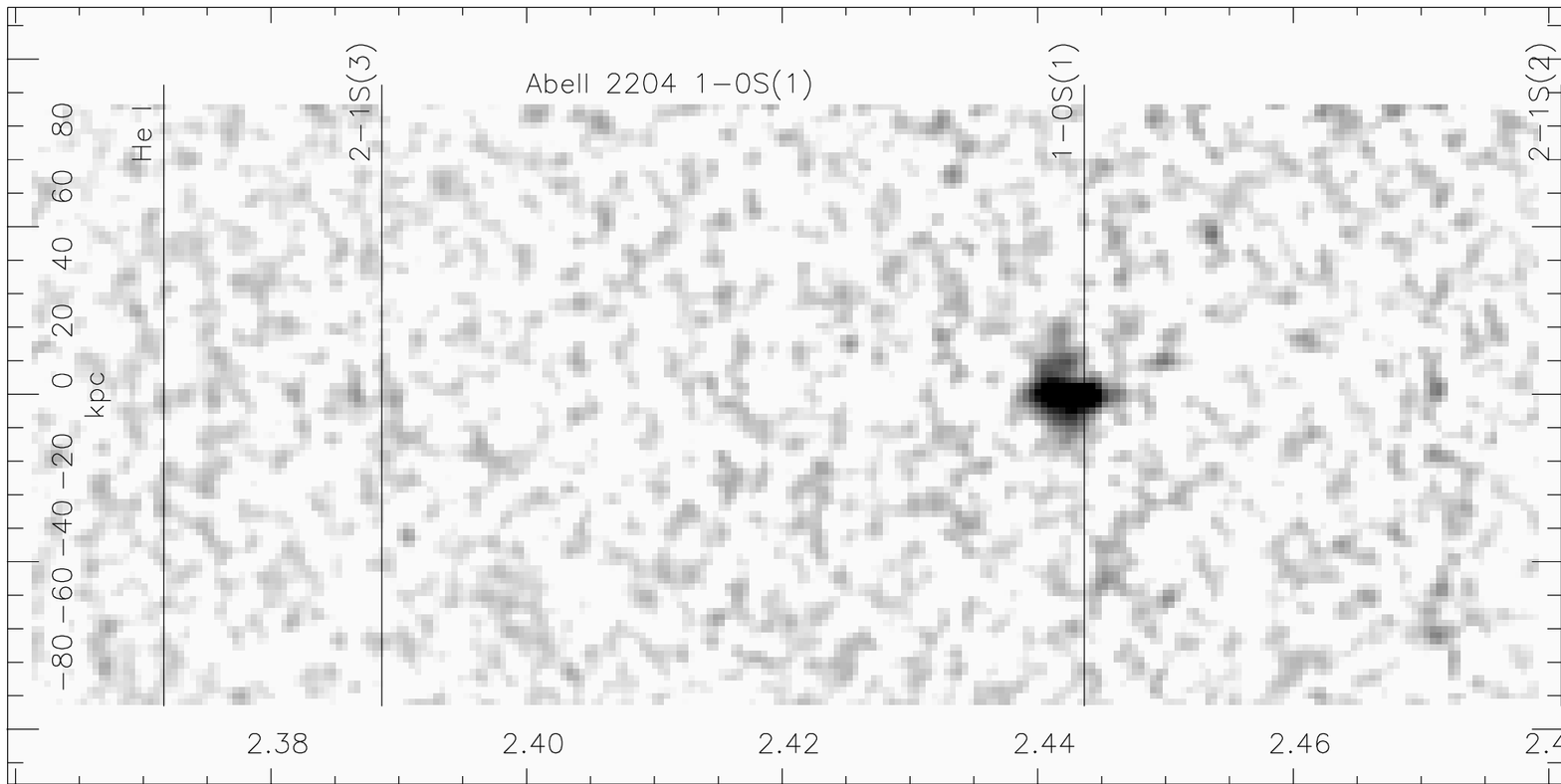}
\caption{Raw two-dimensional spectra of Abell~2204 in
the regions of Pa$\alpha$ and 1-0S(1) respectively.}
\label{figure:A2204IR}
\end{figure*}

These ``raw" two-dimensional spectra are shown in Figures \ref{figure:A2597IR}
through \ref{figure:A2204IR}.  In these 2d spectra we see the gaseous
emission lines, along with noisy vertical stripes at wavelengths of
high telluric absorption.  By Kirchoff's principle, the atmospheric
thermal emission at these wavelengths is maximum.  The photon noise
from the atmospheric emission is the chief source of noise in the
final spectra, and thus maximum at the position of the absorption
lines.  At the y-position of the maximum of the galactic continuum we
also see small features due to minor stellar absorption/emission
mismatches between target galaxy and standards.
\begin{figure*}
\centering
\resizebox{17cm}{!}{
\includegraphics[width=4cm]{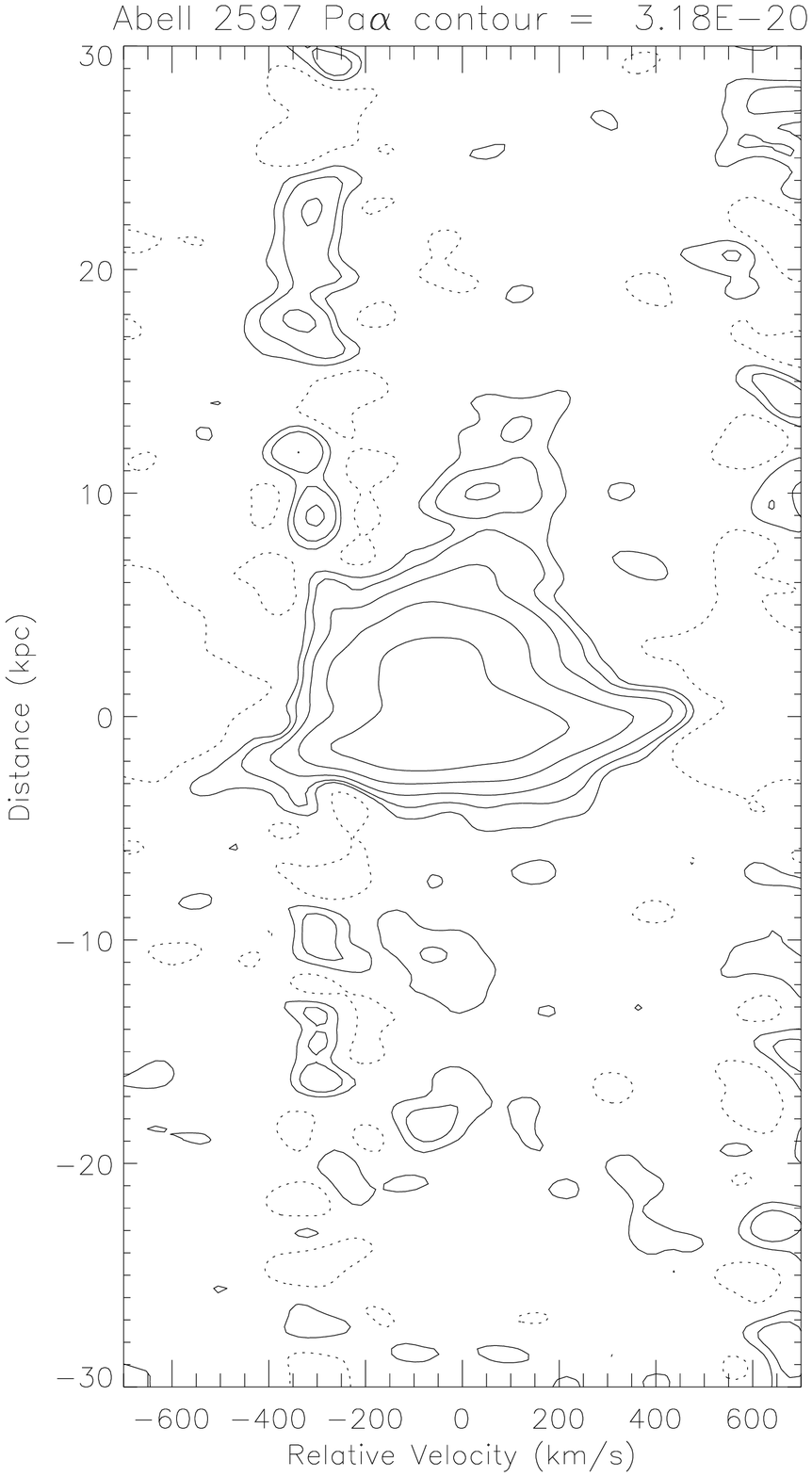}
\includegraphics[width=4cm]{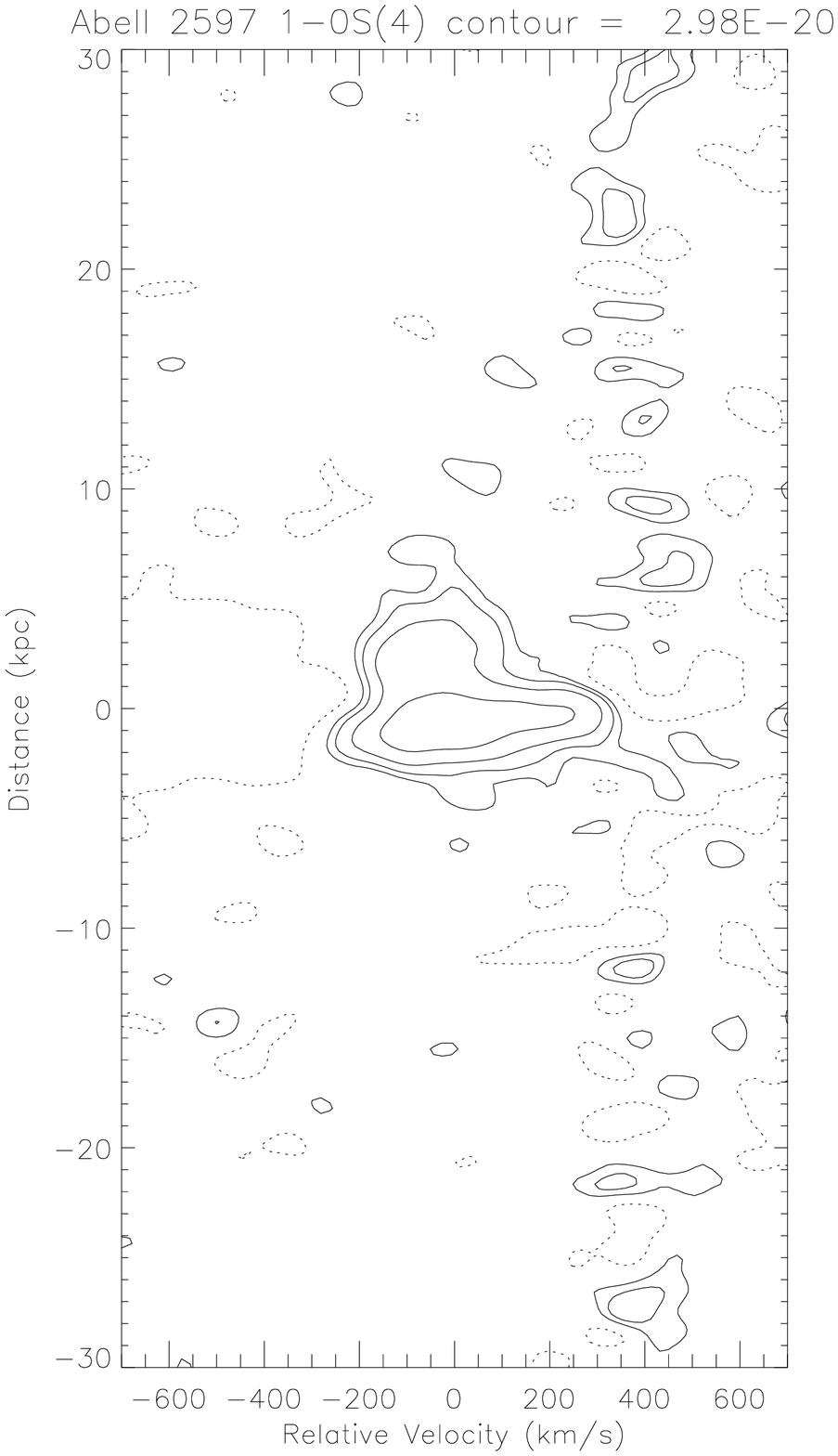}
\includegraphics[width=4cm]{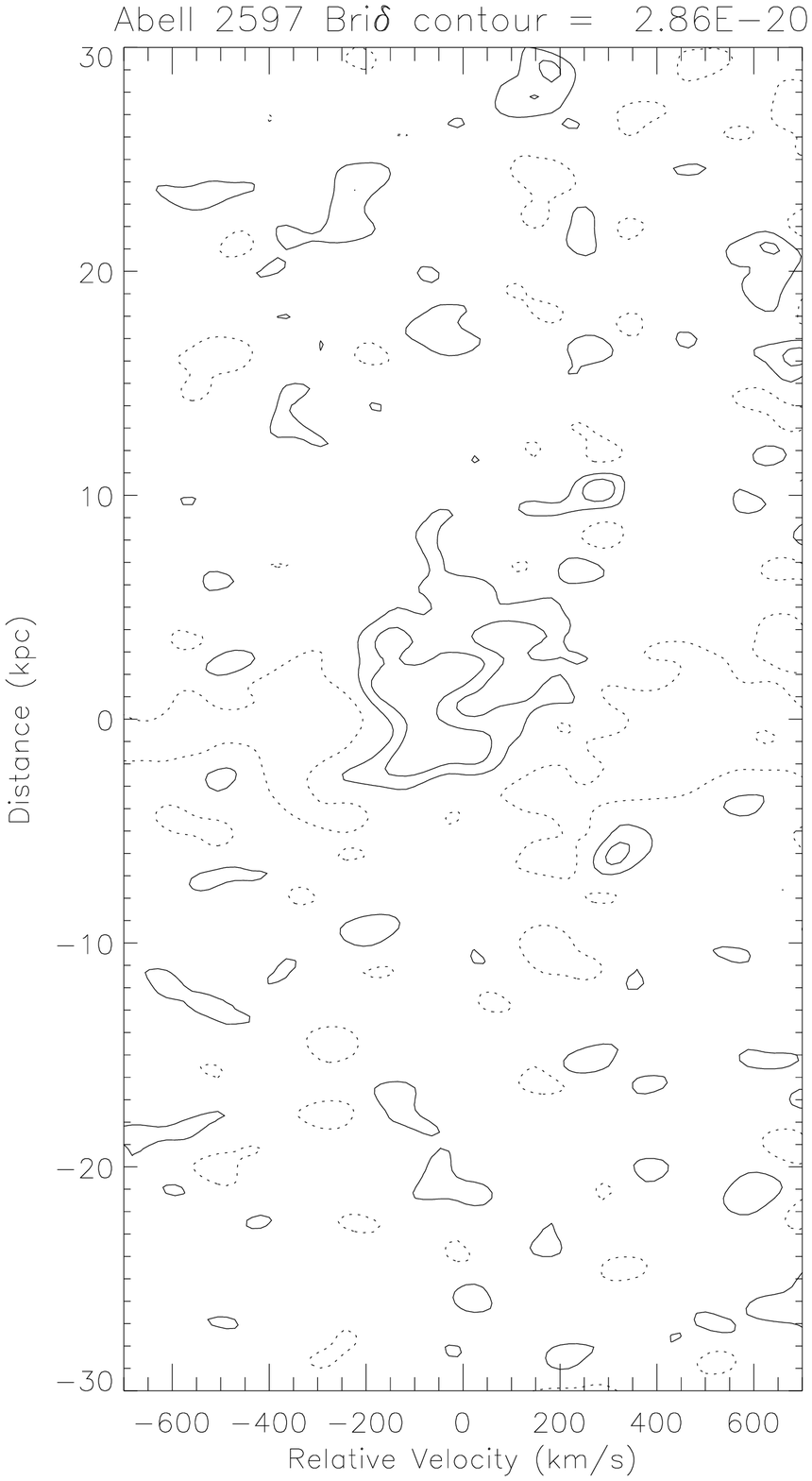}
\includegraphics[width=4cm]{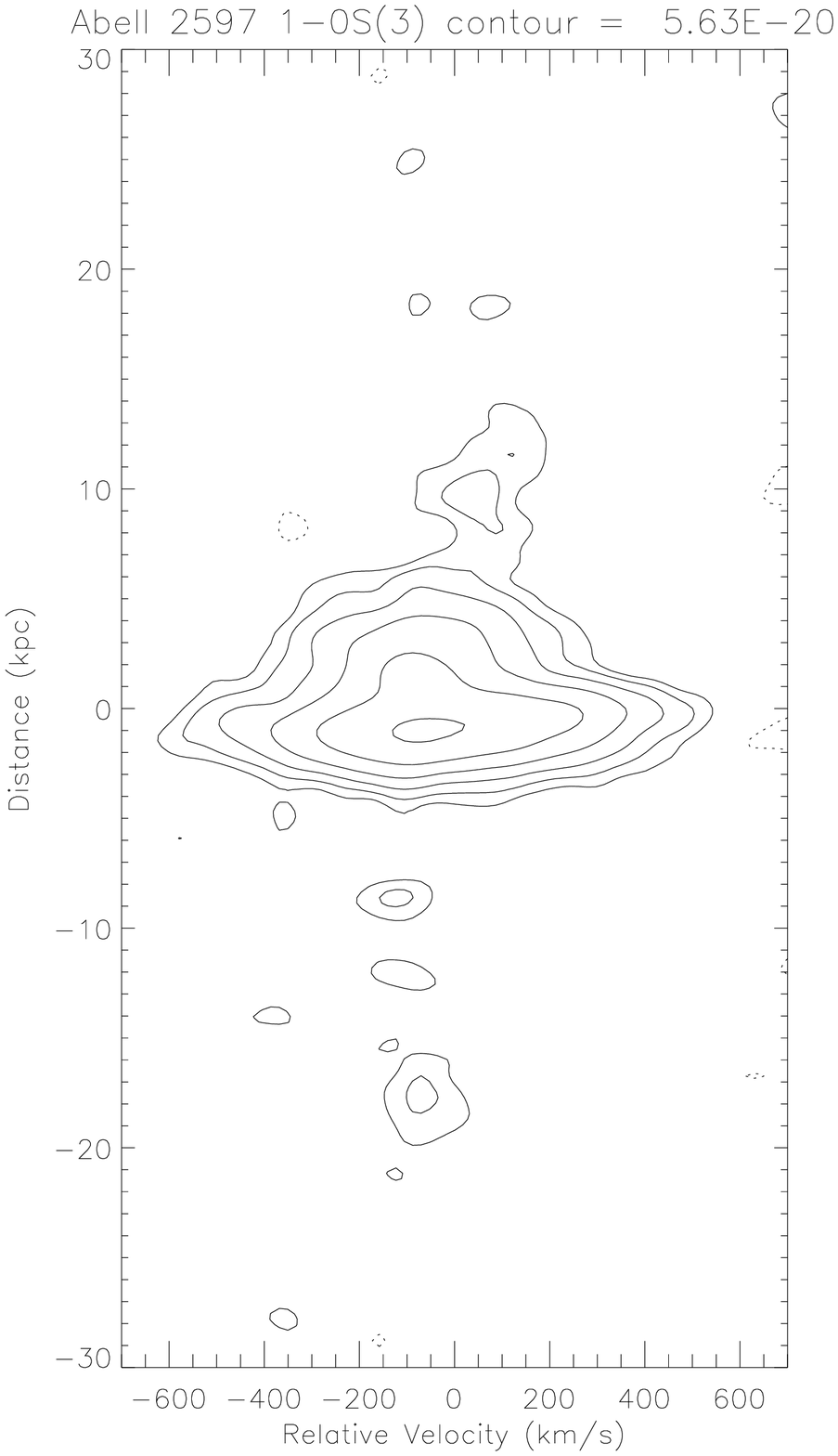}
}
\resizebox{17cm}{!}{
\includegraphics[width=4cm]{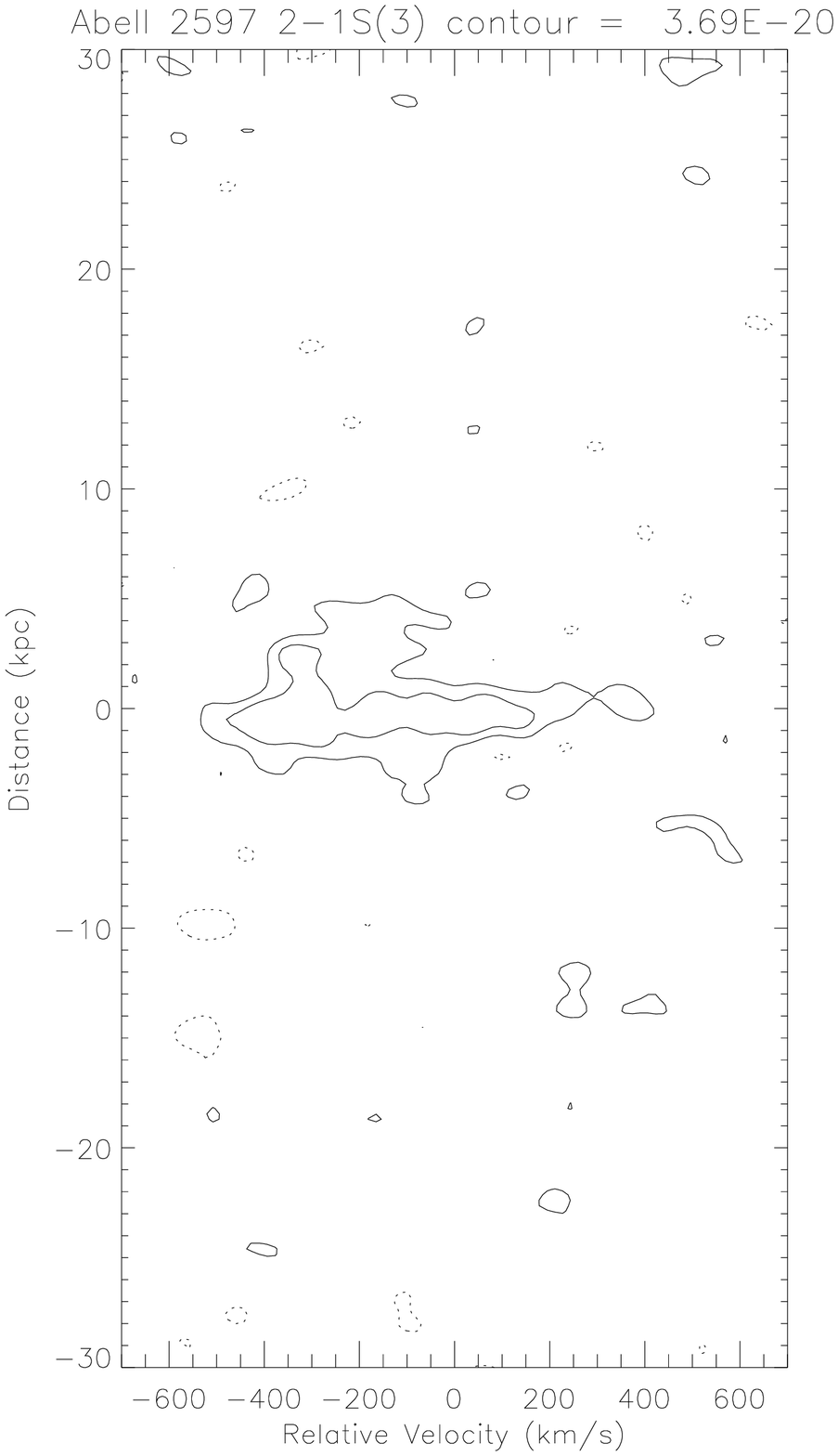}
\includegraphics[width=4cm]{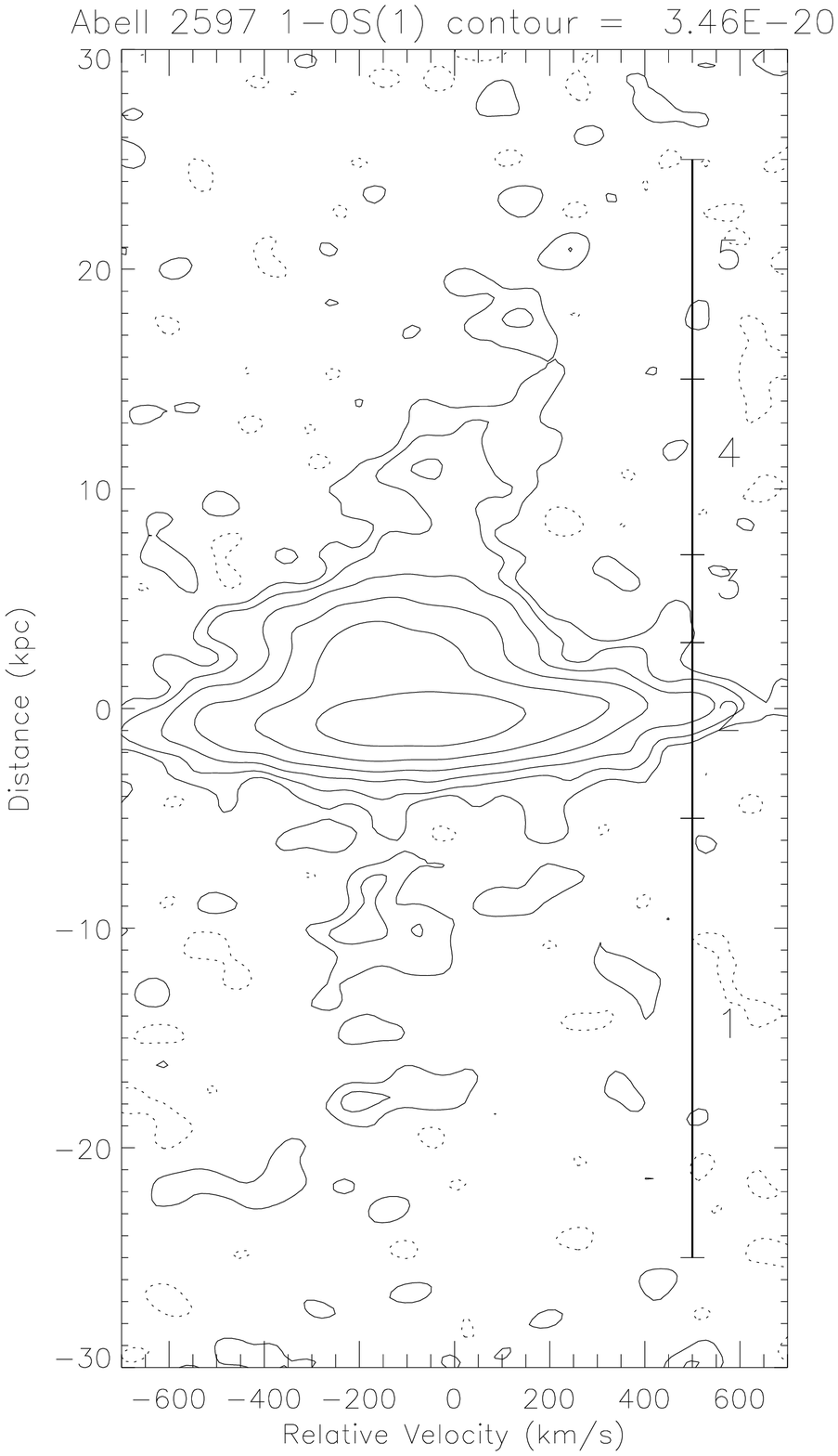}
\includegraphics[width=4cm]{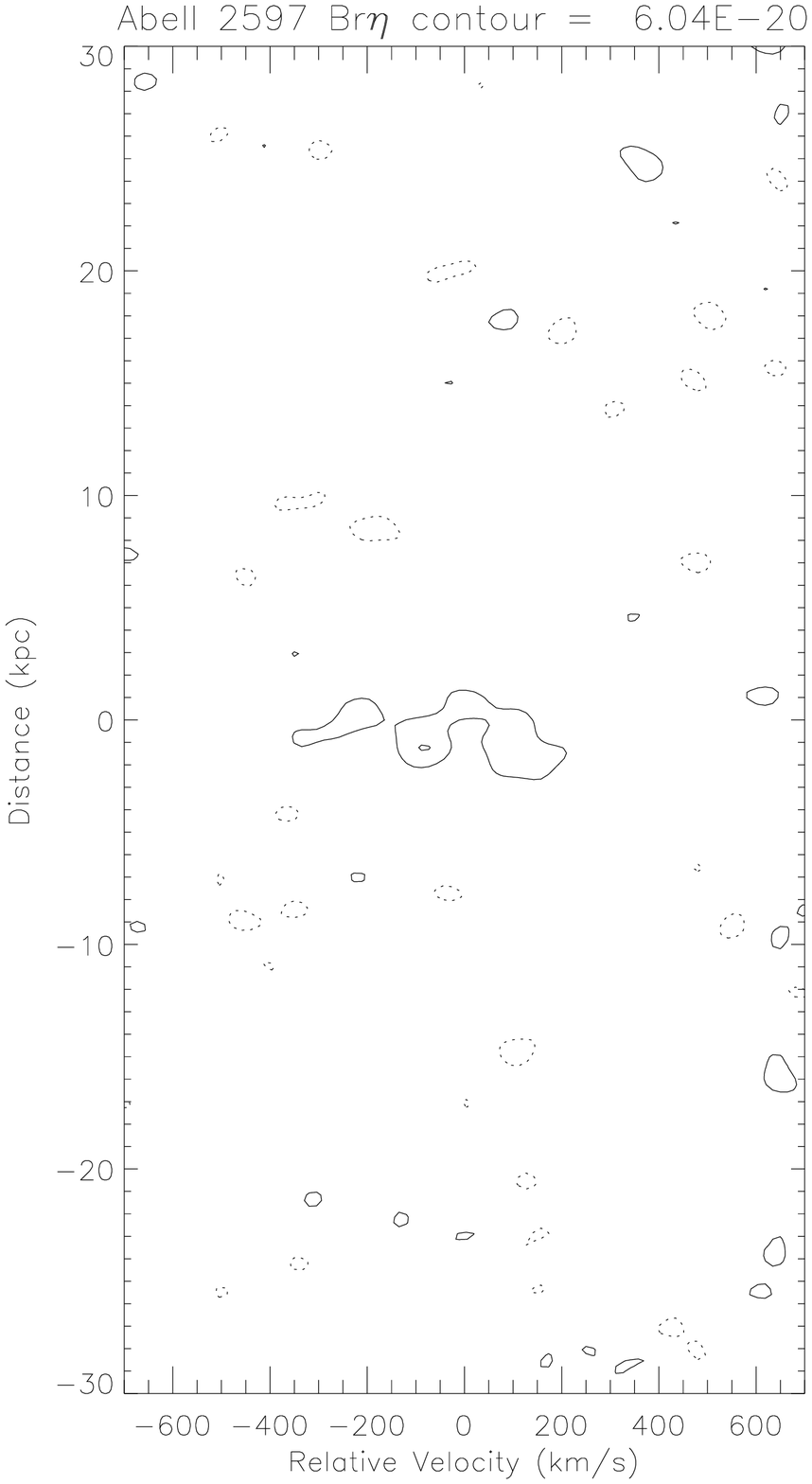}
\hbox to 2.5cm{}
}
\caption{Contour maps of calibrated stronger detected lines in Abell~2597.
Contours are [-1,1,2,4,8,16...] times the lowest level (in erg s$^{-1}$
cm $^{-2}$ arcsec$^{-2}$), which is
indicated in the figure title. The vertical
bar in the 1-0S(1) map marks the regions for which LTE line ratio
plots are given in Figure \ref{figure:A2597LTE}}
\label{figure:A2597contourIR}
\end{figure*}
\begin{figure*}
\centering
\resizebox{17cm}{!}{
\includegraphics[width=4cm]{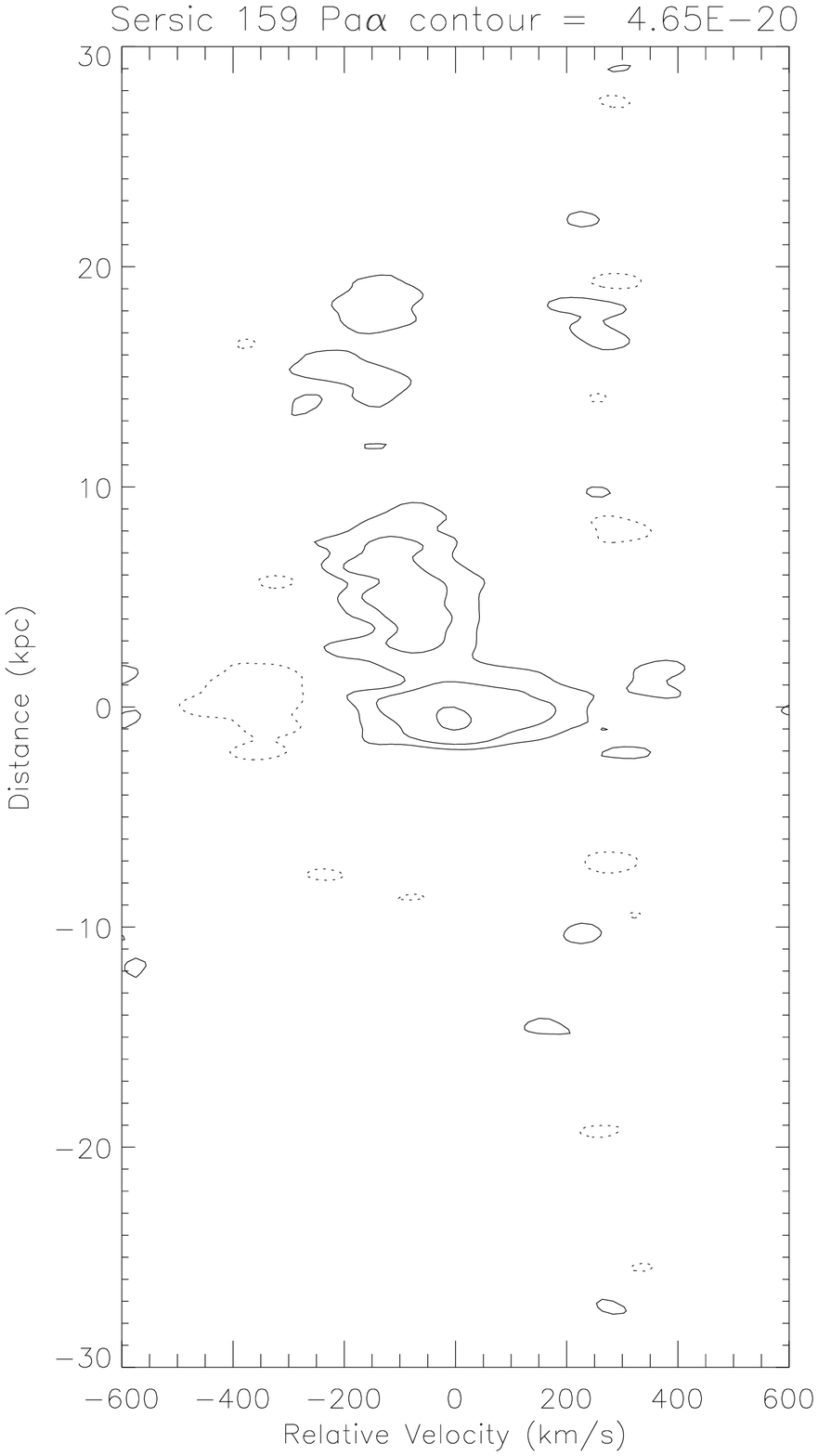}
\includegraphics[width=4cm]{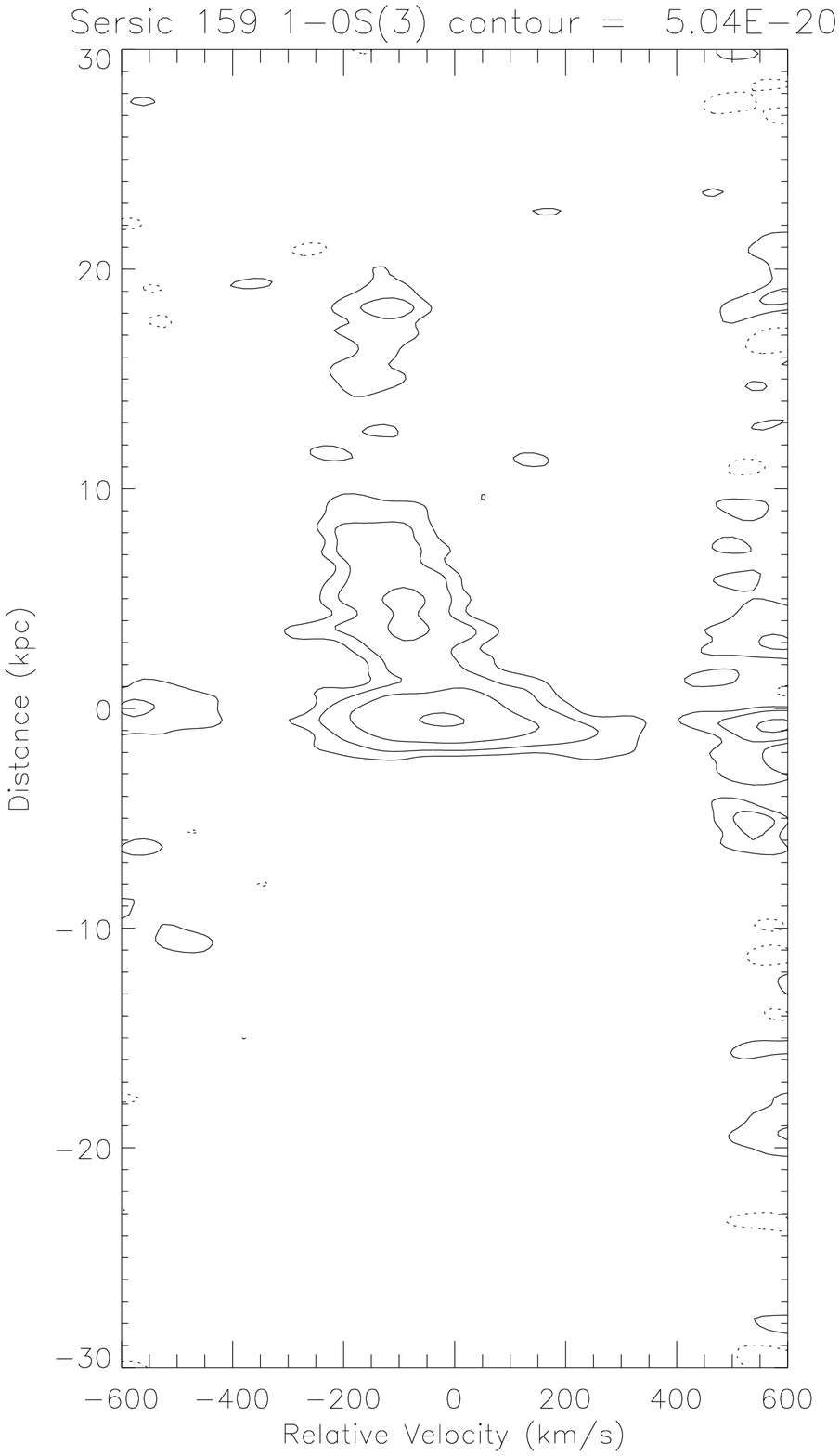}
\includegraphics[width=4cm]{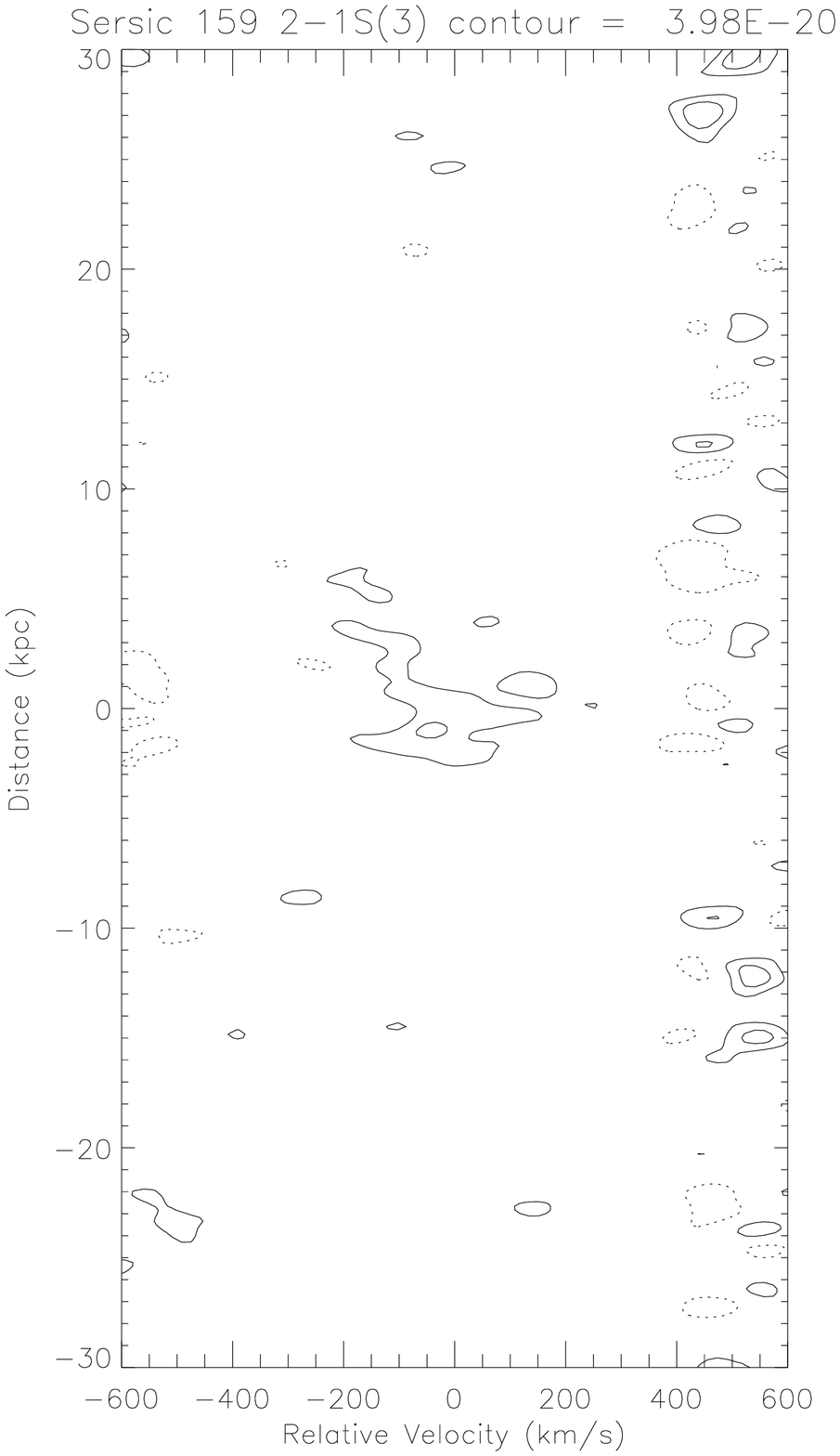}
\includegraphics[width=4cm]{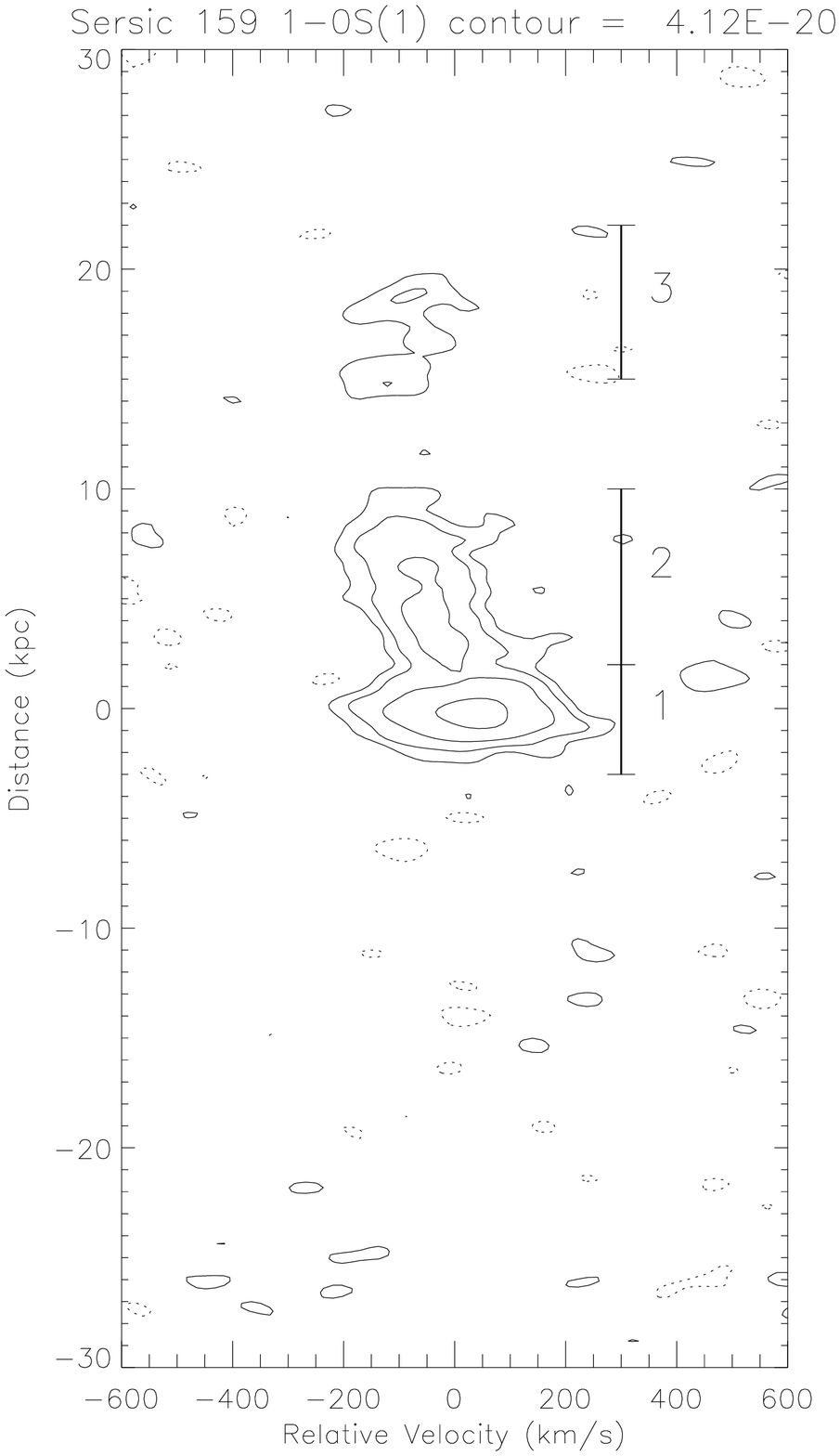}
}
\caption{Contour maps of calibrated stronger detected lines in Sersic~159-03}
\label{figure:S159contourIR}
\end{figure*}

\begin{figure*}
\centering
\resizebox{17cm}{!}{
\includegraphics[width=4cm]{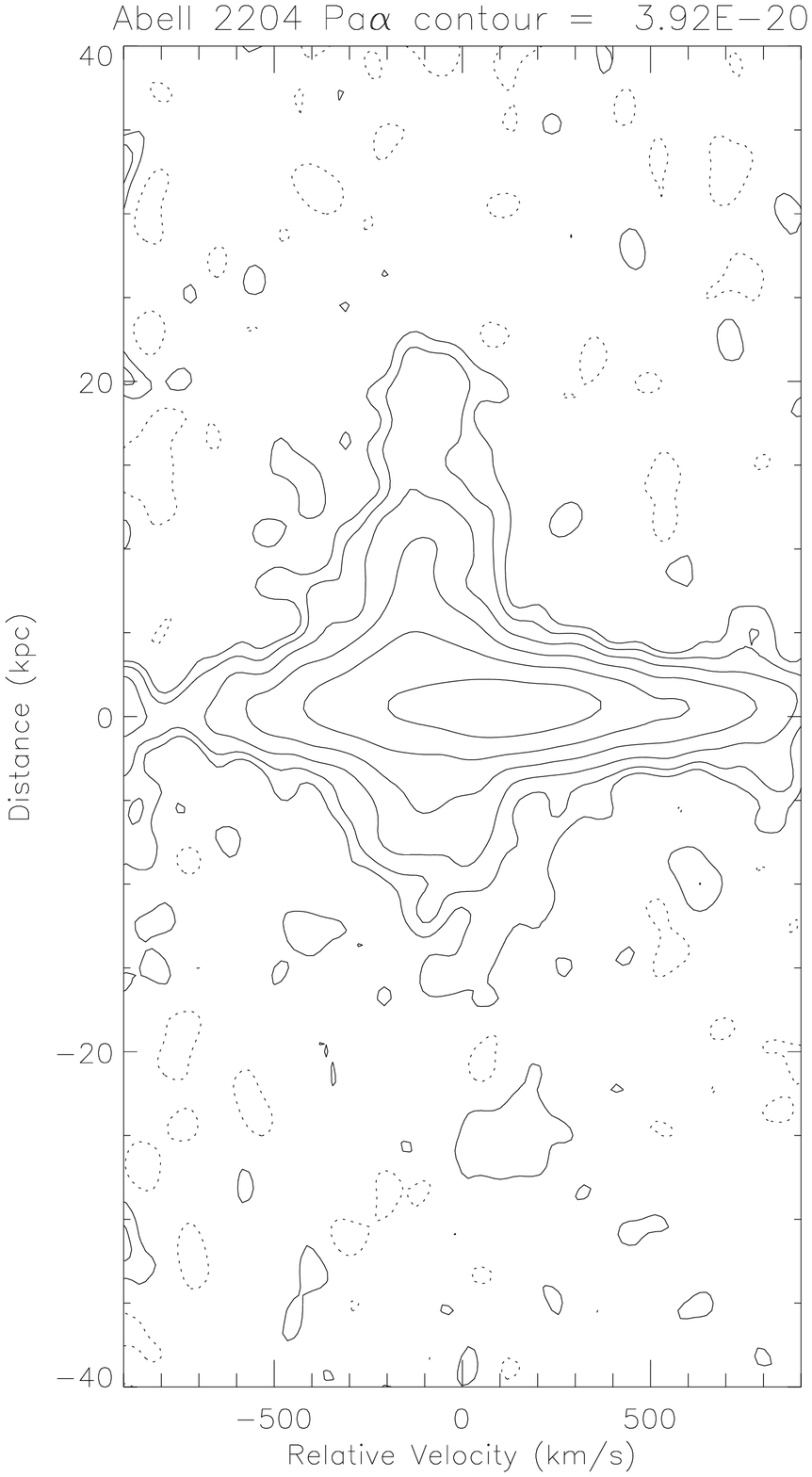}
\includegraphics[width=4cm]{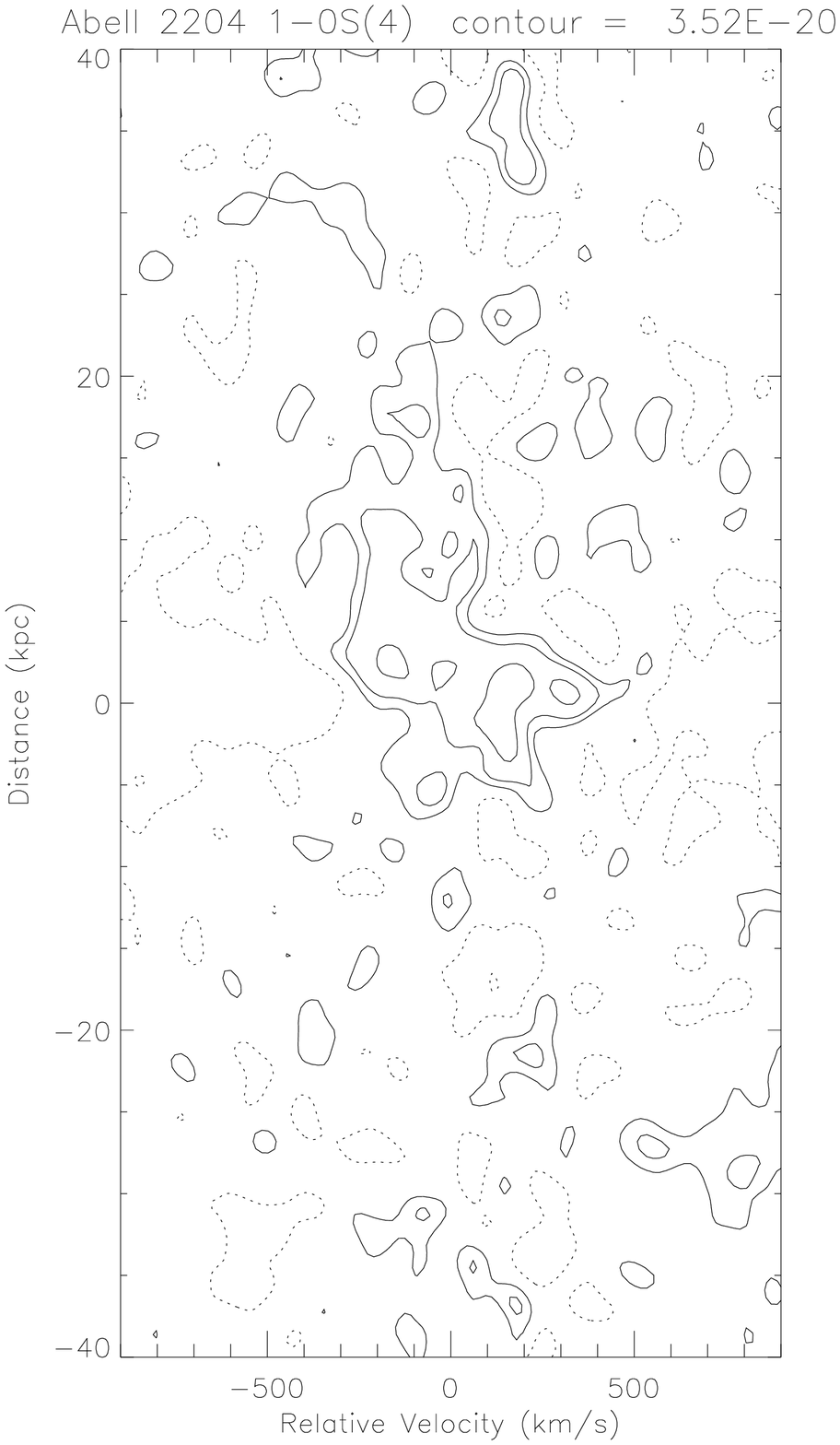}
\includegraphics[width=4cm]{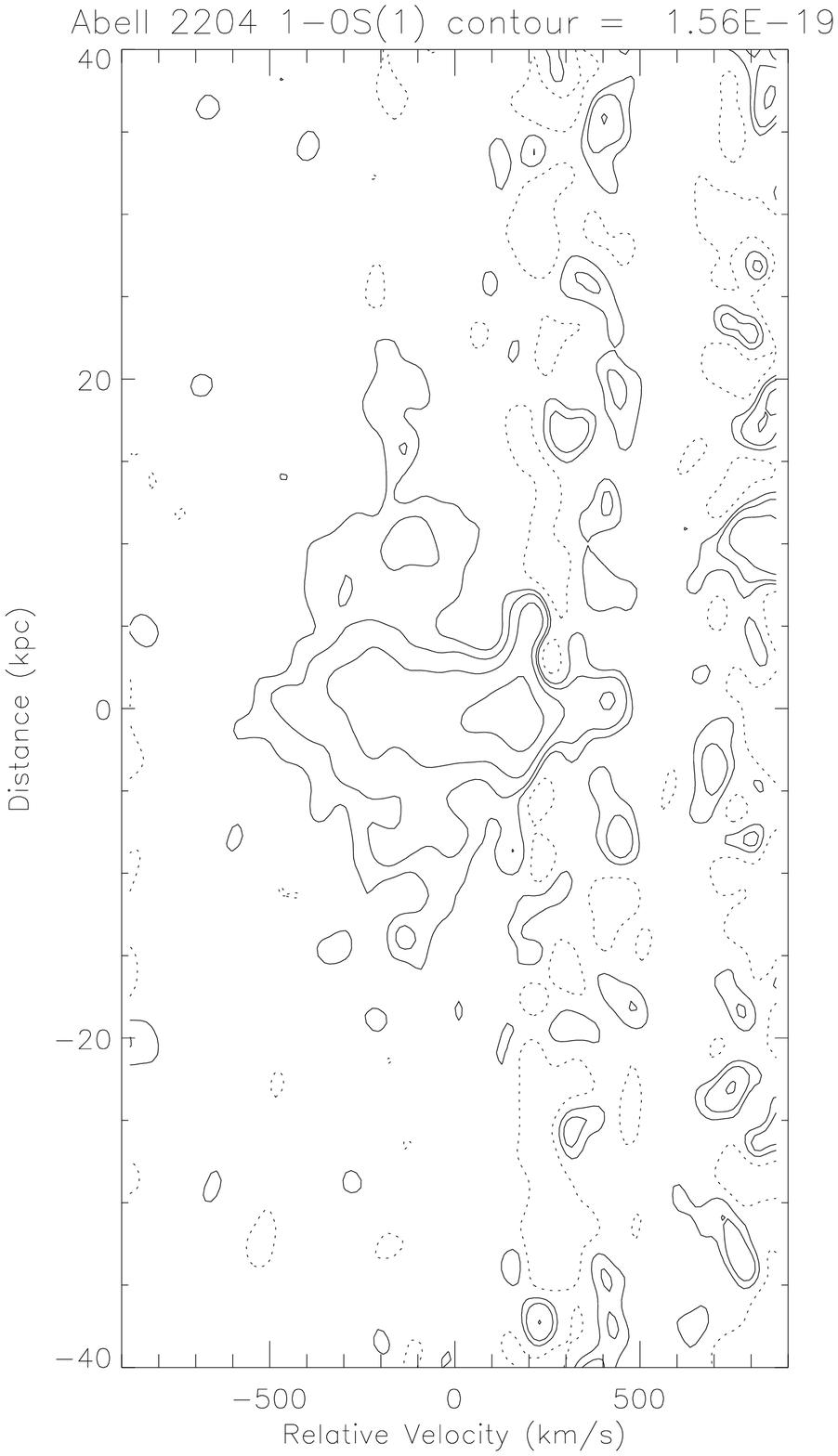}
\hbox to 2.5cm{}
}
\caption{Contour maps of calibrated stronger detected lines in Abell~2204}
\label{figure:A2204contourIR}
\end{figure*}

\begin{figure*}
\centering
\includegraphics[width=12cm]{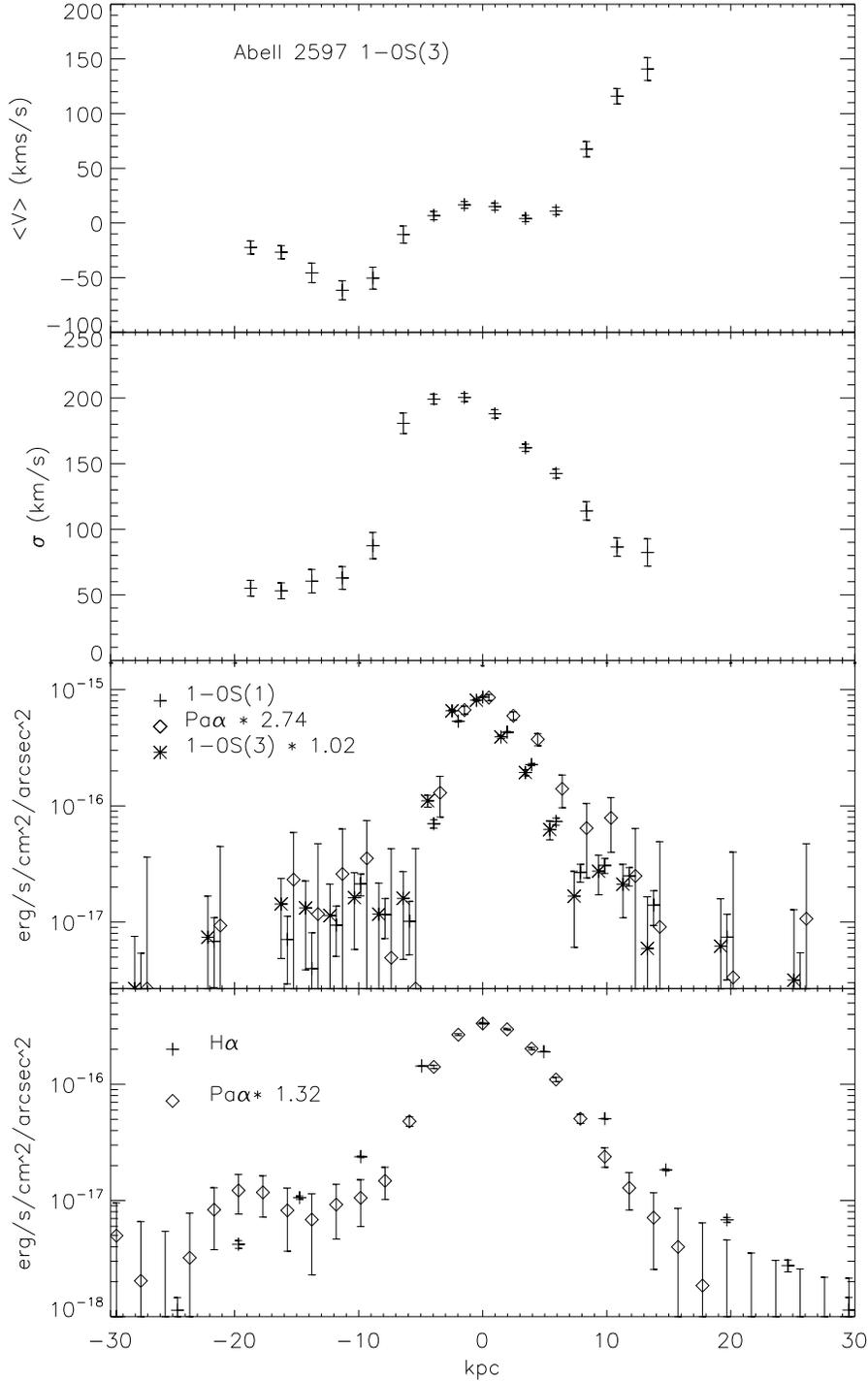}
\caption{Abell 2597 spectral data as a function of distance along the slit:
top and second panels give mean velocity and velocity dispersion taken from the 1-0S(3) line. %
The third panel gives the surface brightness of the IR lines.  The ordinate gives
the surface brightness of the 1-0S(1) line.  The other lines (Pa$\alpha$ and
1-0S(3)) have been scaled by the amounts indicated in the figure to match
1-0S(1) at the center. The points for all three lines are measured at the same positions, but
the \Pa points have been shifted 0.5 kpc to the right, and 1-0S(3) 0.5 kpc
to the left for clarity. The bottom panel gives the surface brightness of ionized lines.  The ordinate
gives the surface brightness of \Ha while \Pa has been scaled to match \Ha at the center.}
\label{figure:A2597sigma}
\end{figure*}

\begin{figure*}
\centering
\includegraphics[width=12cm]{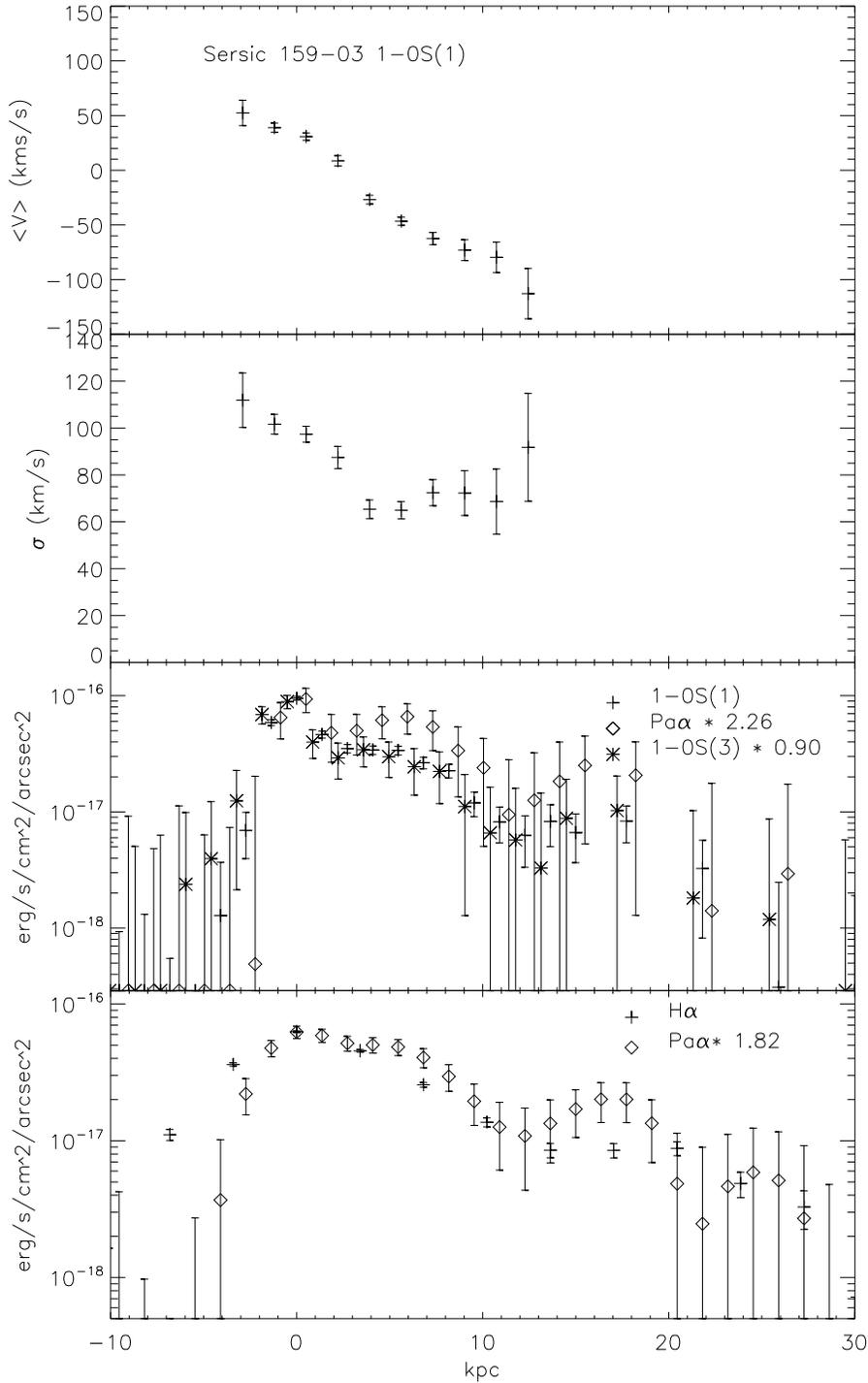}
\caption{Sersic~159-03 spectral data: 
top and second panels give  mean velocity and velocity dispersion taken from the 1-0S(1) line.%
The third panel gives the IR lines; the ordinate gives the
surface brightness of the 1-0S(1) line and the other lines have been
scaled by the amounts indicated to match 1-0S(1) at the center. The points for
the various lines have been shifted slightly in the x-direction as described
in the previous plot. The bottom panel gives \Ha and \Pa scaled to match \Ha at the center.}
\label{figure:S159sigma}
\end{figure*}

\begin{figure*}
\centering
\includegraphics[width=12cm]{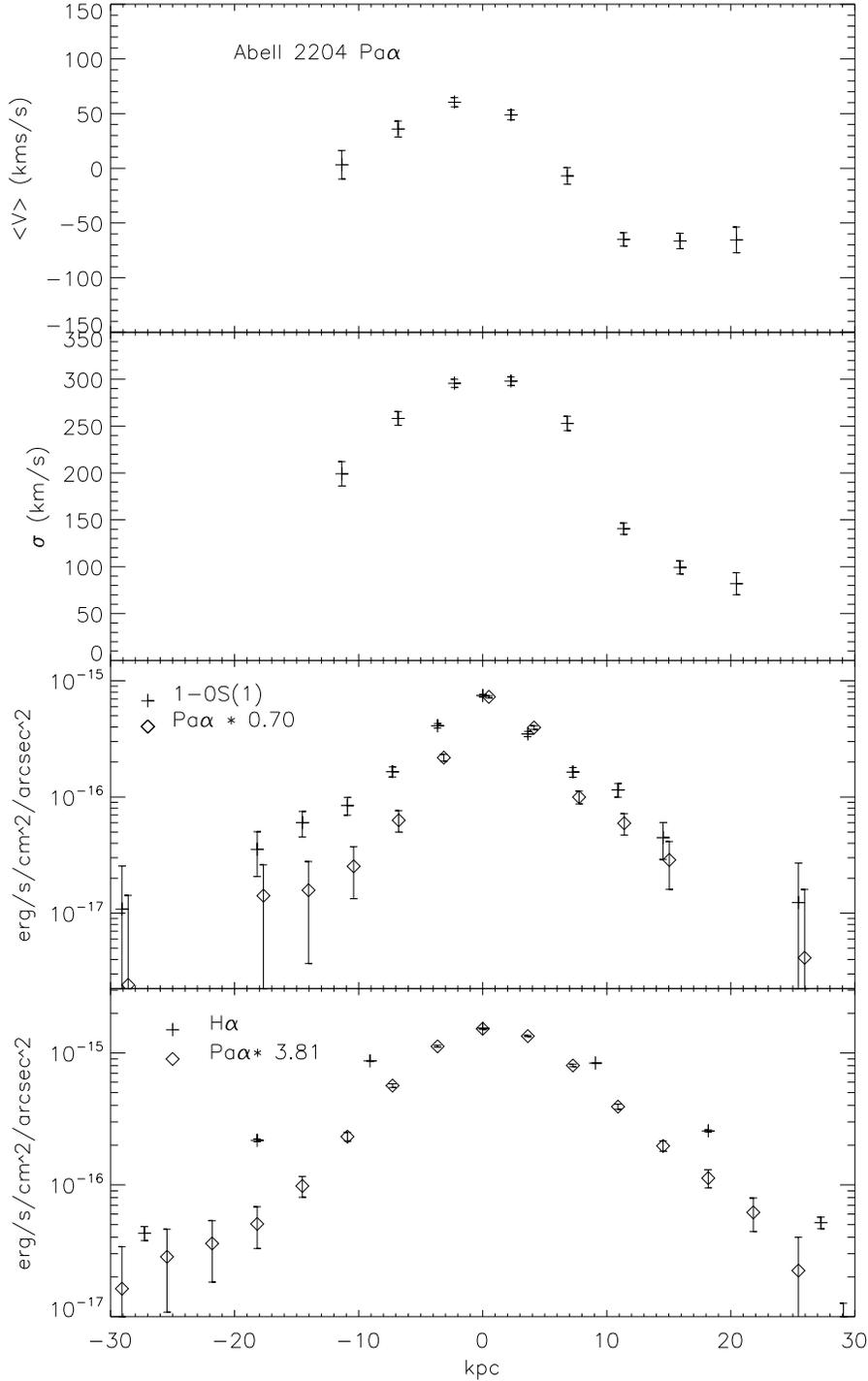}
\caption{Abell 2204 spectral data: 
top and second panels give mean velocity and velocity dispersion taken from the Pa alpha line.
The third panel gives 1-0S(1) and \Pa scaled to match 1-0S(1) at the center.
The bottom pannel gives \Ha and \Pa scaled to match \Ha at the center.}
\label{figure:A2204sigma}
\end{figure*}

\begin{figure*}
\centering
\resizebox{16cm}{!}{
\includegraphics[width=8cm]{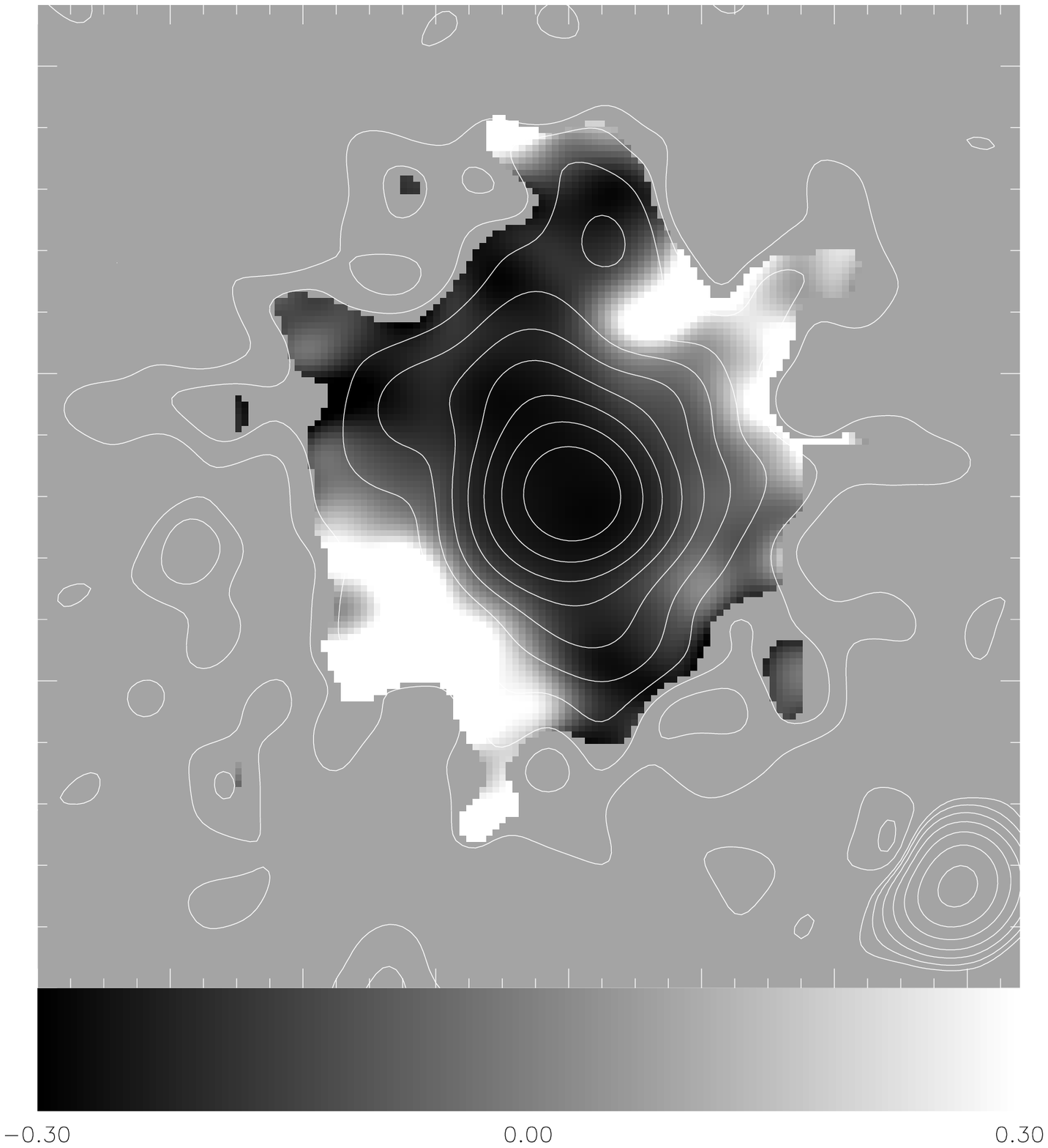}
\includegraphics[width=8cm]{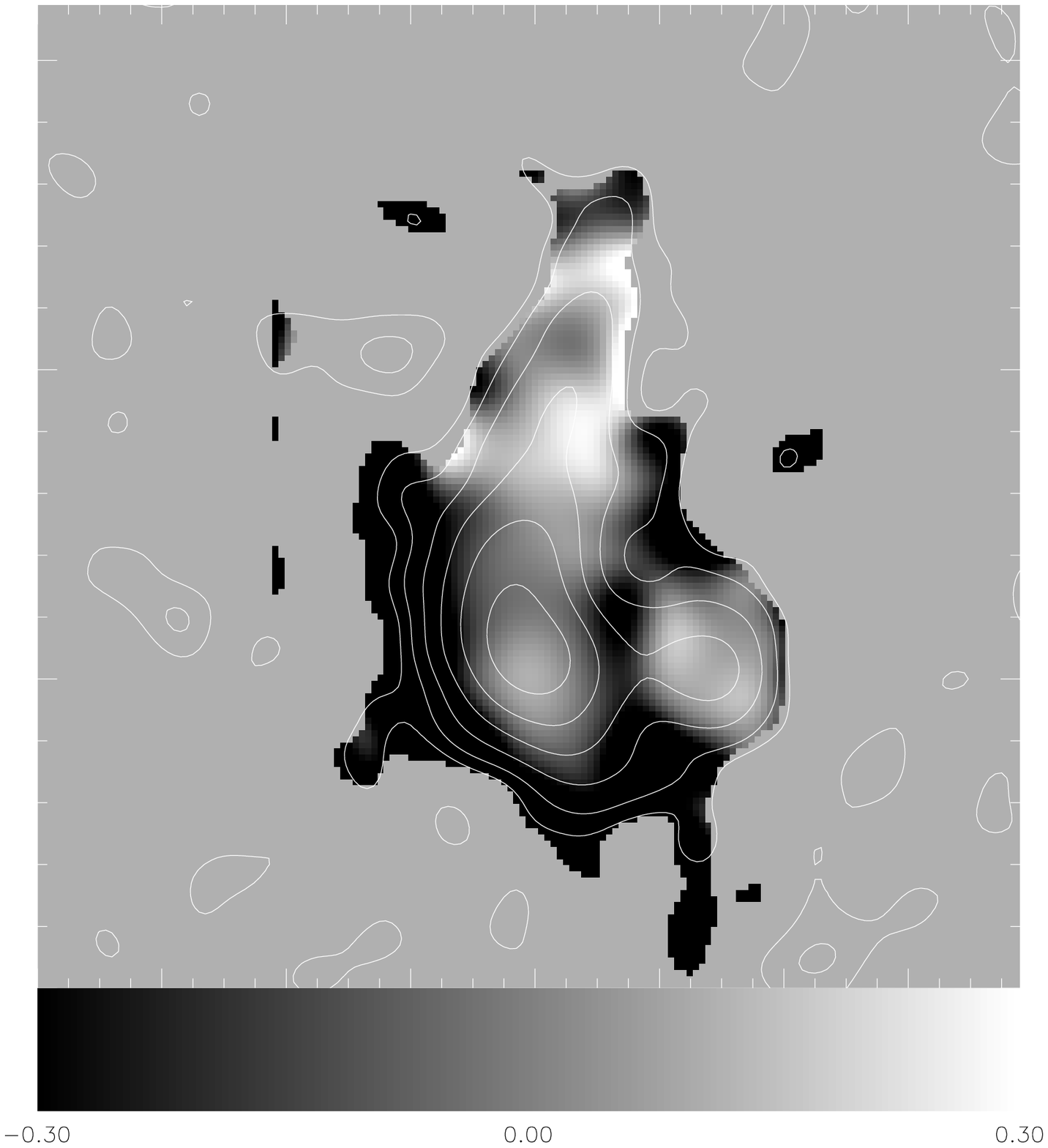}
}
\caption{(left) Greyscale representation of log([NII]/\halpha\ in the central region of Abell~2597 
with superimposed the contours from Figure \ref{figure:A2597optical}. The orientation
is the same as Figure \ref{figure:A2597optical}, while the field of view covers
60" in both coordinates. Where the signal/noise
was low the greyscale has been blanked to a uniform grey.  (right) log([NII]/\halpha\ for
Sersic~159-03 with the contours from Figure \ref{figure:S159optical}.  Again a 60" field is plotted.}
\label{figure:opticalratios}
\end{figure*}

\begin{figure*}
\centering
\includegraphics[width=12cm]{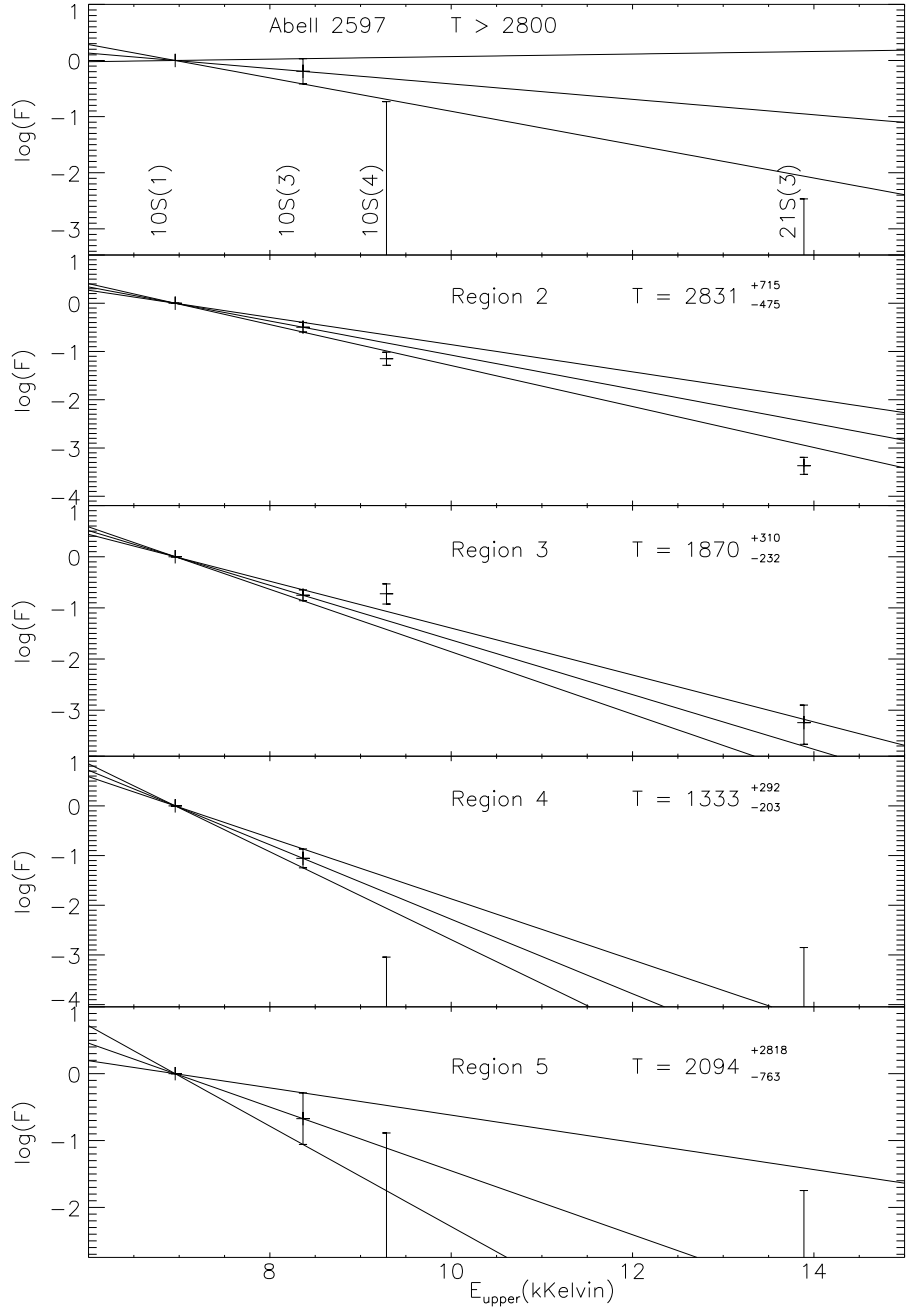}
\caption{LTE diagrams for Abell 2597.  Different regions along
the spectrometer slit, marked in Figure \ref{figure:A2597contourIR} are given
in different panels above. The abscissa marks the excitation
energy of the upper state of each transition (expressed in temperature
units), while the ordinate gives the flux of the corresponding line,
divided by its Einstein coefficient, $A$ and the statistical weight
of the upper state $g$.  Finally each flux is normalized to the flux
of the 1-0S(1) line.  In LTE, $F\propto gA\exp(-E_U/kT)$, so the
various transitions should lie along a straight line.}
\label{figure:A2597LTE}
\end{figure*}
\begin{figure*}
\centering
\includegraphics[width=12cm]{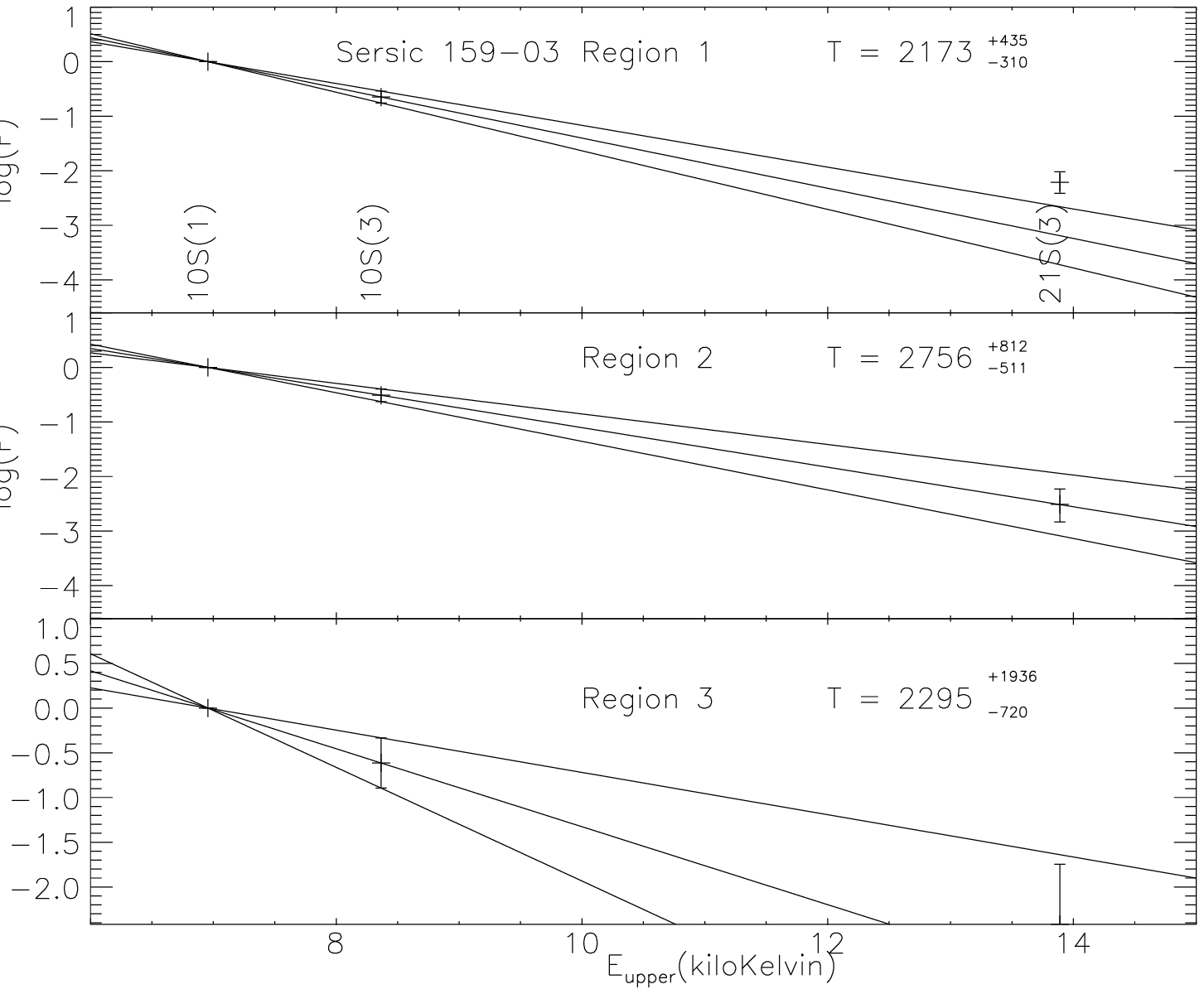}
\caption{LTE diagrams for Sersic~159-03.  The 1-0S(4) transition lies very
close to a sky line at this redshift, and is not given in the plot.}
\label{figure:S159LTE}
\end{figure*}

\section{Discussion}
\subsection{Extent, morphology and kinematics of the ``warm" gas phases}
\label{kinematics}
Figures \ref{figure:A2597optical} through \ref{figure:3C445optical} 
show that the warm HII phase ($T\sim 10,000$ K) can
be traced to radii beyond 20 kpc in all four cooling flows. In Abell~2204 
we see emission to 30 kpc and in Abell~2597 to beyond 40 kpc in smoothed 
images.   This last figure can be compared to HII structure seen in 
the same cluster to about 20 kpc in \cite{heckman89} and the bright
central features  associated with the central radio source at
$\sim 5$ kpc shown in \cite{koekemoer99}.  The increased maximum
radius of detection for this source corresponds to an increase 
in surface brightness sensitivity by a factor of 20 relative to \cite{heckman89}.

The four clusters have diverse morphologies: extended and roughly
circularly symmetric (Abell~2597), compact (Abell~2204), and arc-like
(Abell~1795 and Sersic~159-03).  Ignoring these differences, we have
plotted the radial surface brightness distribution, averaged over
annuli centered on the galaxy nucleus in figures
\ref{figure:A2597radial} and \ref{figure:A2204radial}.  The
averaging over a large number of pixels at larger radii improves
the signal/noise ratio at these radii.

Abell~2597 shows significant emission, distributed essentially
as an $r^{-3}$ power law to beyond 40 kpc, where it disappears
smoothly into the noise.  The ``arc-like" structure of Abell~1795,
when azmuthally averaged, shows similar behaviour, with perhaps 
a slightly flatter distribution with radius ($\propto r^{-2}$). 
Abell~2204 shows emission to a similarly
large radius ($\sim 35$ kpc) but drops more rapidly than a power law,
more like a gaussian, with a radius of $\sim$ 12 kpc.  In other words
the apparent size of the emission region seems to be a real physical
characteristic of the source, rather than a limit imposed by our
surface brightness sensitivity.  In Sersic 159-03
the circularly averaged emission can be traced to radii just
beyond 20 kpc and is fit approximately by an exponential with scale
sizes of $\sim 4$ kpc.  Thus the increased sensitivity afforded by the TTF system 
yields two contrasting results: in Abell~2597 and Abell~1795 we can trace the 
smooth decrease in surface brightness to double the previous limiting radius but find
no physical edge; in the other two clusters we can establish the existence of a
radius characteristic of the source physics.

Figures \ref{figure:A2597IR}-\ref{figure:A2204IR} show strikingly that
molecular emission can be traced to $\sim 25 $ kpc and that the 
\Pa and \Ht line profiles are virtually indistinguishable
both spatially and kinematically.  A comparison
of the \Pa lines and the 1-0S(1) and 1-0S(3) lines shows that,
taking differing signal/noise ratios and interfering sky lines
into account, every feature visible in one line is visible in
the others at the same velocity and position and essentially
the same intensity.  This implies directly that that the HII and
warm \Ht gas phases and their respective excitation sources
must be well mixed on scale sizes finer than 1 kpc, and the relative
mix is essentially constant over factors of 10 in radius.
The correspondence suggests further that the \Ht phase is present wherever 
the diffuse HII can be observed, i.e. to beyond 40 kpc in Abell~2597.

Consistent with our earlier work, the ratio of \Ht to HII line
strengths is far higher than that found in typical galactic star
forming regions and starburst galaxies (by HII lines we mean hydrogen
recombination lines from the Balmer and Paschen series, arising
in HII regions).  In fact the
combined luminosity in all the Near InfraRed (NIR) \Ht lines is
comparable to that in all the optical HII lines, although the
luminosity in \lya~ is probably an order of magnitude above this.

The velocity structure of the molecular gas in all three observed
clusters shows a division into two regions: a central area of radius
$\sim 5$ kpc with high velocity dispersion and a larger region of 5-25
kpc with much lower dispersion. The one-dimensional spectra presented
in previous works including our own are dominated by emission from
this central region.

In this region, where the gas is probably stirred by the AGN,
the velocity spreads -- here quoted as the line-of-sight rms velocity
dispersion $\sigma$ --  range from $\sim 120$ \kms for Sersic~159-03
through 200 \kms for Abell~2597 to 300 \kms for Abell~2204.  The areas
at larger radii show lower dispersion ($\lesssim 60$ km s$^{-1}$) and
apparent large scale rotation or shear motions of magnitude $\sim 100$
\kmsp.  These dispersion estimates are in fact upper limits set by the
velocity resolution of our wide slit spectroscopy and the
true values could be lower.  For Abell~2204 and
Abell~2597, \cite{edge01} finds line widths in the CO molecular
emission of 255 and 278 \kms respectively with a beam size of 22
arcsec.  These values seem lower than we would expect from our K-band
data, given that our central FWHP ``widths" are 700 and 470 \kms for
these two clusters, and that the K-band line emission is strongly
peaked toward the centres.  This would indicate that the cold
molecular gas responsible for the CO emission does not partake of the
violent motions near the galaxy center and is situated at radii beyond
$\sim 10$ kpc.

At the larger radii the gas velocity dispersions and rotational
velocities are much smaller than the typical stellar dispersions
velocities in BCGs: 400-600 \kmsp.  Thus the the molecular gas clouds
at large radii {\it are not kinematically supported} nor, given their
high density, can they be supported by pressure gradients in the
surrounding hot gas.  The low velocities that we see are dynamically
consistent only if this gas has been formed from a pressure supported
phase in the last 10$^7$ years (the time necessary for gravity to
accelerate them to above 100 \kmsp) or an as yet unknown mechanism
supports them. \cite{daines94} have suggested magnetic support of the
clouds. This seems unlikely to us however, because of the Parker
instability \citep{parker68} generally present in gas supported against
gravity by magnetic fields.

From our \halpha\ fluxes, with corrections for dust absorption (see below)
we can calculate the total mass of warm ionized gas in the clusters
centres.  Because of the relatively fast drop of brightness with
radius we do not, however, expect these estimates to be much larger
than older data obtained at higher surface brightness limits.

For Abell~2597, assuming $n_e\sim 200$ cm$^{-3}$ \citep{heckman89} and
$A_{H\alpha}\sim 1$ we find an HII mass of $\sim 3\times 10^7$ M\subsun.
The mass in {\it warm} \Ht can be estimated from $M_{H_2} = 2 m_H
L_1/(\epsilon_1 A_1 F_1)$ where $L_1$ is the luminosity in the 1-0S(1)
line, $\epsilon_1$ is the energy of the photon from this transition,
$A_1$ is the Einstein coefficient for this transition, and $F_1$ is
the fraction of all \Ht molecules to be found in the upper state for
this line.  For LTE conditions at $T\sim$ 2000 K (c.f. Section~\ref{LTE}), 
$F_1$ is $\sim .015$.  With the assumption that the ratios of 1-0S(1) to \Pa and \Pa
to \halpha\  vary globally the same way as they do along our slit
measurements, we find $L_1\sim  10^{42}$ erg s$^{-1}$ for both
Abell~2597 and Abell~2204, with a corresponding $M_{H_2}$ of $\sim 4\
10^5$ M\subsun.  This is much smaller than the masses of {\it cold}
\Ht found by \cite{edge01} to be $\sim 10^{9-10}$ M\subsun.  The \Ht
masses calculated in that article are based on ``standard" ratios
of H$_2$/CO and \cite{israel97} shows that these may in fact be serious
underestimates in cases where the molecular gas is exposed to
a bright UV/X-ray radiation field, which dissociates the CO molecule.
The kinetic energy in the warm molecular gas in the ``violent" central
region is of order $10^{54}$ ergs and if it is dissipated in the
central dynamic time of $\sim 2\times 10^7$ years the corresponding kinetic
luminosity is of order $10^{39}$ erg s$^{-1}$, which is small relative
to other luminosities associated with the AGN and cooling flow.  If,
contrary to our inference above, the cold gas also partakes of this
rapid turbulent motion, its kinetic luminosity would of course be much
larger and would become important to the energetics of the system as a
whole.

The problem of the support of the molecular gas takes on its most
serious form for the cold ``CO" gas.  The gravitational potential
energy of this gas,$\Phi_{cold}$ can be estimated as $M_{cold} \sigma_*^2$,
where $\sigma_*$ is the 3-dimensional velocity dispersion of the stars
$\sim 600$ \kmsp; $\phi_{cold}\sim 3\times  10^{58}$ erg.  This is the
same order of magnitude as the {\bf total} thermal energy of the
X-ray gas within 15 kpc of the galaxy center (\citealt{kaastra04}) and
similarly the entire magnetic energy in this region, assuming $B\sim 10 \mu$G
(e.g. \cite{taylor02}).

\subsection{Line ratios and thermal structure of the warm gas phases}
Figures \ref{figure:A2597IR} to \ref{figure:A2204contourIR} illustrate
the similarity of the surface brightness profiles for the NIR HII and
\Ht lines.  To allow a more direct comparison of the line ratios of
the strongest lines (1-0S(1), 1-0S(3) and \Pap, we have integrated
them in velocity and plotted them as a function of slit position in
the third panels of
Figures \ref{figure:A2597sigma} to \ref{figure:A2204sigma}.  In each
case the spectra have been plotted on a logarithmic intensity scale, and scaled
to the flux of 1-0S(1) at the center.  In the fourth (d) panel of
these figures we have plotted in a similar fashion the intensities of
\halpha, [NII] and \Pa, where the {\it optical} lines have been
evaluated from the TTF images along the position of the {\it NIR}
slit.  In estimating the absolute IR/optical ratios, some
uncertainties in this procedure are inevitable: the different
observing and calibration techniques may introduce errors of $\sim
30\%$, and errors in the exactly alignment of the slit on the images
may be of order 1 arcsec.  Experiments with moving the positions or
position angles of the slit on the image did not produce significant
changes in the shapes of these plots.  Additionally, the narrow
bandpass (12\AA) of the TTF filters will miss some optical flux at the
central high velocity width position.  Assuming a gaussian profile
with the dispersions plotted in the Figures, we find about 40\% of the
\halpha\ flux at the center of Abell~2204 would be lost ($\sigma$=300
\kms), 20\% at the center of Abell~2597 ($\sigma$=200\kms), and a
negligible loss in Sersic~159-03.

\subsubsection{Dust extinction}
Under the assumption of Case-B ionization equilibrium in the HII
regions, the P$\alpha$/H$\alpha$ ratio is $\sim 0.12$
(\citealt{osterbrock74}).  All three of the objects with ISAAC spectra
show larger ratios at the peak of their surface brightness profiles:
0.76 for Abell~2597, 0.55 for Sersic~159-03 and 0.24 for
Abell~2204. For the first two clusters this implies extinction of
$A_{H\alpha}\sim 1.6-1.8$ mag with calibration uncertainties of $\sim
0.5$ mag, and reddening of $E(B-V)\sim 0.6$.  For Abell~2204, after
correction for \halpha\ lost outside the filter bandpass, the
estimated extinction is small, $<1$ mag. In determining these numbers,
we have assumed that the dust is in a screen in front of the gas,
whereas in reality the dust could well be mixed-in. Nevertheless,the
numbers give an indication of the amount of dust in or around the
emission-line regions.

In the case of Abell~2597 this ratio decreases smoothly by a factor
of $2-3$ as $r$ increases to $\pm 15$ kpc from the centre of the nebulosity, 
corresponding to a drop in extinction to $\sim 1$ mag. We see a similar
trend for Abell~2204 probably due to a reduction in loss outside the
\halpha\ filter i.e. showing little evidence for extinction at any radius.

The situation is more complicated for Sersic~159-03. Along the ``tail'' of
emission north of the peak surface brightness, the line ratio barely
varies significantly, possibly the P$\alpha$/H$\alpha$ ratio increases
slightly relative to the peak, implying an increase of 0.2-0.3 in
E(B-V) at most. To the south, the surface brightness drops
sharply. Although Figure \ref{figure:S159sigma}d shows a sharper 
drop in \Pa than
\halpha\ indicating a decrease in reddening, differences in the
seeing and pixel size between the two data sets may cause this. These
systematic effects limit our ability to draw firm conclusions about
the effect of reddening in this region.

What is clear from the above is that many line emitting clouds up to
20 kpc of the centres of these clusters contain dust and the reddening
effect of the dust does not vary dramatically across the nebulosity,
apart from at their very centres.  \cite{koekemoer99} have already
shown that there is a probable dust lane crossing the nucleus of
Abell~2597, this is the obvious explanation for the excess reddening
we see at the centre for this source.

\subsubsection{ [NII]/\halpha\ line ratios}
The ratio of the forbidden [NII] line to the permitted \halpha\ line 
can be taken as a crude measure of ionization parameter $\Xi\equiv n_{photon}/n_H$;
larger values of [NII]/\halpha\ correspond to larger values of $\Xi$.
\cite{heckman89} have shown that from cluster to cluster, various ionization measures 
correlate with each other, and that ionization correlates inversely with \halpha\ luminosity.
\cite{crawford92}, \cite{crawford95} and \cite{crawford99} have shown that these correlations
are quite general in cooling flows, but that the line ratios vary continuously
rather than forming two distinct classes as claimed by \cite{heckman89}.  
For the two clusters where we have good TTF maps of these lines, the line ratios
are displayed in Figure \ref{figure:opticalratios}.  

We find that between these two clusters, the relation between [NII]/\halpha\ varies
in the sense described in the quoted papers, but that locally within each cluster,
this is not true, and that the two clusters appear quite different.  In Abell~2597
there is a some tendency for this ratio to increase where the surface brightness
of \halpha\ decreases, but the effect is only strong at the very edges of the
nebulae.  Log([NII]/\halpha) is in fact very nearly constant at a low value $\sim -0.2$ typical
of the high luminosity ``Type II" of \cite{heckman89} along a NE-SW axis,
more or less parallel with the radio jet axis, rising to the ``Type I" value
of $\sim +0.3$ on the NW and SE perimeters.  The source average is dominated
by the central, lower value.  Sersic~159-03 shows the opposite behaviour,
the brighter central regions show a high value of log([NII]/\halpha), while
the perimeter shows lower values; the high value dominates the average.

We note that the ionization level is not simply related to distance from
the galactic nucleus.  In Abell~2597 the line ratio varies little when
the radius varies by a factor of more than 10.  This fact emphasizes,
as other authors have noted, that the source of excitation is almost 
certainly not the AGN but rather of an origin local to the ionized gas.  
The difference between the two clusters emphasizes that the interpretation
of  these line ratios is not simple, as they depend on metallicity,
ionization parameter, and density.

\subsubsection {Comparison between H$_2$ lines and Pa$\alpha$}
Using our ISAAC spectroscopy, we are able to compare the spatially
resolved properties of the P$\alpha$ and H$_2$1-0 lines. Previous work
has concentrated on the integrated properties of the highest surface
brightness IR lines. These spectra are invariably dominated by
emission from the very centres of the galaxies (which harbour strong
radio sources) and likely to be influenced by the effect of AGN
emission.

We can compare three main spatially-varying parameters of the lines:
Their surface brightness profiles, their velocity dispersions and
their peculiar velocities relative to the centre of the
galaxy. Figures 6-14 show that the atomic and molecular hydrogen
lines behave very similarly as a function of position. 

The velocity structure in all the lines looks similar.  As noted in
Section~\ref{kinematics} the central high surface brightness regions
have velocity dispersions of several hundred \kms, while at larger
radii the dispersion in all lines drops to approximately 60 \kms, 
the resolution of
the spectrograph, and probably lower.  The molecular and atomic lines
show similar variations in peculiar velocity as a function of
position, showing that these gas phases are closely linked
kinematically, even though their dynamical properties, e.g. the ratio
of gravitational to drag forces, are quite different because of the
large difference in density: $n_e < 10^3$ cm$^{-3}$ in the ionized
phase \cite{heckman89} versus $n_e\sim > 10^5$ cm$^{-3}$ in the
molecular phase (c.f. Section \ref{LTE}).  The surface brightness
profiles of the atomic and molecular phases also follow broadly
similar patterns. In Figures \ref{figure:A2597sigma}b to
\ref{figure:A2204sigma}b the brightness profiles integrated across the
slit are normalised to the brightest point in the 1-0S(1) profile. For
Abell~2597 the ratio \Pap/1-0S(1) appears a factor two lower at the
peak of the profile than in the rest, with molecular and atomic
emission otherwise tracing each other. For Abell~2204 the ratio
decreases by no more than a factor of two from the centre to the edge
of the detected nebulosity. For Sersic~159-03 the ratio appears
constant across the northern tail, with a decrease of a factor of 2 in
the relative strength of P$\alpha$ at the peak of the profile, similar
to that seen in Abell~2597. The current data do not allow us to trace
the atomic emission to the south, but the non-detection is consistent
with the ratio found for the north.

We have earlier noted (\citealt{jaffe01}) that this ionized/molecular line ratio
is relatively constant from cluster to cluster, and is typically at
least a factor of ten lower than that usually seen in photo-dissociation regions 
(PDRs) around star-forming regions.  The current observations show that this
ratio does in fact vary somewhat, by factors of $\sim 2$ from cluster to cluster
and within clusters.  We will discuss in Section~\ref{photons} some of the possible
reasons for this variation, but our principle remark is that the excitation sources
of the HII and warm \Ht phases must be very closely linked in order that
the line ratios and velocities remain so close to each other despite changes
in radius factors $> 10$, surface brightness by factors $>100$ and local 
velocity dispersion by factors $> 4$.

If we accept the hypothesis that the ratios of HII and \Ht lines is
essentially constant over the area that we see HII line emission, 
we can calculate total galaxy luminosities in the \Ht lines.
As given in Section \ref{kinematics}, for the 1-0S(1) line 
in Abell~2597 this yields $L_1 \sim 2\times 10^{42}$
erg s$^{-1}$, for Sersic~159-03 $L_1\sim 2\times 10^{41}$ erg s$^{-1}$ and
for Abell~2204 $L_1 \sim 4\times 10^{41}$ erg s$^{-1}$ with uncertainties
of order a factor of 2.  The luminosities
in the H$_2$ 1-0S(3) are similar, and detailed models of excitation
of \Ht gas in many circumstances (e.g. \citealp{black87, sternberg89})
generally find that the luminosity in the strongest single rotational/vibrational 
line is generally $\sim 1-2$\% of the total molecular NIR emission due to \Hp.  
This implies a typical total \Ht NIR luminosity of $10^{43}-10^{44}$ erg s$^{-1}$.
This is up to two orders of magnitude greater than the \halpha\ luminosity
and may be greater than the total optical line luminosity (assuming
that \lya\ luminosity is $\sim 10$ times that of \halpha).

\cite{kaastra04} quote total X-ray fluxes (0.1-2.4 keV) for Sersic~159-03
and Abell~1795 that are approximately 2000 times larger than the \halpha\
fluxes quoted in Table 1.  With the conversion factors used in the
previous paragraph the total NIR luminosity for these galaxies
is $\sim 5$\% of the total X-ray luminosity at all radii, and a
larger fraction of that arising in the central $\sim 20$ kpc.  For
example in the Chandra data of Abell~1795 (\citealt{ettori02}) only
about 4\% of the total X-ray emission in the 0.5-7 keV range arises
in the central 17 kpc.  This suggests that the NIR and X-ray luminosities
in this central region are similar.

\subsubsection{\Ht line ratios and LTE calculations}\label{LTE}
In Figure~\ref{figure:A2597LTE} we show the relative strengths
of the 1-0S(1),S(3),S(4) and 2-1S(3) lines averaged within 
five spatial regions marked
on Figure~\ref{figure:A2597contourIR} and displayed in the
form of $log(F/gA)$ versus $E_{upper}$  where $F$ is the line flux,
$g$ the statistical weight of the upper state, $A$ the Einstein coefficient
of the upper state, and $E_{upper}$ the energy of the upper state of
the transition expressed in temperature units.  
In local thermodynamic equilibrium (LTE), the resulting points 
should lie on a straight line with slope
$1/T_{LTE}$.  For Sersic~159-03 the same plots are given in 
Figure~\ref{figure:S159LTE}, but the 1-0S(4) line is not included because
its wavelength falls near that of a bright sky line.  The values and
uncertainties of $T_{LTE}$ printed in each plot are derived solely
from the 1-0S(1) and S(3) lines, which have by far the best signal/noise ratios.

Looking first only at the temperature estimates from these two lines we
see in Abell~2597 a gradient of decreasing temperature from the
the southern region {1} to region {4} at a height of $\sim 10$ kpc.
While it is tempting to conclude that the temperature decreases with
distance from the nucleus the estimated average value for the
southern region {1} is actually higher than that for region {2}.
The temperature uncertainty in region {1} is, however, quite large due
to the low signal level there. Comparison of the error bounds 
indicates that while $T_1$ could be the same as $T_2$, it cannot be not significantly lower, 
e.g. not as low as $T_3$.  The weighted best LTE temperature for all five regions
together is $\sim 2600$ K, somewhat higher than the central
value of 2160 K given by \cite{wilman02} for this cluster, but within the range typical for
cooling flows.  In Sersic~159-03 there is an indication of increasing temperature with distance from
the nucleus, but the significance is low.  The global average temperature along the slit
is $\sim 2400 $ K.

In none of the regions is there significant evidence for line ratio deviations from the
LTE curve.  There are several regions where the signal at the 2-1S(3) line, which has
an excitation energy of 13890 K, is significant, namely regions {2} and {3} of
Abell~2597 and region {1} and {2} of Sersic~159-03.  In these cases the positions
of this line near the LTE line imply densities of $n\sim 10^7$ cm$^{-3}$ in
order for there to be enough energetic collisions to keep the upper level populated.

\subsection{Source of molecular excitation}\label{photons}
The problem of the source of excitation of the ionized gas, and to a
lesser extent that of the molecular gas, has previously been discussed by many
authors, including ourselves.  The new contributions of the data reported
here consist firstly of the demonstration of the very close spatial and kinematic correlation 
of the HII and \Ht lines and secondly of the extension of these correlations
to large radii with respect to the central galaxy.

The strong correlation between the emission characteristics of the two
gas phases argues that they are excited by a single physical mechanism
and not, for example, that stellar photoionization excites the HII
region, while heat conduction from the X-ray medium excites the \Ht
emission.  Shock excitation provides such a natural connection because
the various gas phases represent a time sequence in the shock heating
and cooling history, but optical spectra in cooling flows at the
[OIII] $\lambda$ 4363 line (\citealt{voit97} and more stringently
\cite{baker04}) are not consistent with shock models, and we will not
discuss these further.

A more likely explanation is excitation of both phases by relatively hard
radiation.  In  starforming photodissociation regions (PDRs) excitation
by young stars causes HII line emission due to absorption of Lyman continuum
photons and subsequent recombination.  In these regions,
\Ht lines arise from the absorption of Werner and Lyman lines by \Ht at longer
wavelengths near 1000 \AA\ and radiative or collisional de-excitation of the
molecules (\citealp{black87, sternberg89}).  Photons beyond the Lyman limit
have a low cross section for absorption in the HII regions, but can be
absorbed in dense \Ht regions with the release of energetic photoelectrons
which lose their energy primarily by spinning up \Ht molecules 
(\citealp{tine97,maloney96}).  This
efficient conversion of hard photons to molecular vibration-rotation lines
should lead to higher molecular/ionic line ratios than those from
softer sources.  In fact, in \Ht regions surrounding planetary nebulae, the
molecular/ionic line ratios are similar to those in cooling flows (Jaffe
and Mellema, in preparation).  We have not yet carried out detailed modelling
of these line ratios in cooling flows, because the large pressure
difference between the two phases implies that a full dynamic model, including
outflows is probably necessary.

The assumption of a hard ionizing spectrum is attractive because it is consistent
with detailed analyses of the optical spectra (\citealp{voit97,crawford92,baker04}) which indicate the presence of radiation from a source harder or hotter
than O-stars (e.g. white dwarfs or Wolf-Rayet stars).  
\cite{odea04} find evidence of copious FUV radiation in the
regions of Abell~1795 and Abell~2597 that show molecular emission in our
maps, although the radiation detected in the HST/STIS observations is not
``hard" compared to Lyman continuum.  The source of these hard photons
is still unknown.  \cite{crawford92} suggest, for example, radiation
from a non-equilibrium ``mixing-layer" of X-ray and molecular gas.
Our observations re-enforce the need for hard radiation, but do not
give direct evidence for its source.  They do however lead to several
interesting constraints on the spectrum and spatial distribution of the
exciting photons.

The cooling rate of dense \Ht gas at LTE from rotational/vibrational
radiation alone is quite high and the gas is quite difficult to 
keep warm (e.g. \citealt{crawford92}).  The cooling rate per molecule from
this radiation is about $3\times 10^{-17}$ erg s$^{-1}$ near 2000 K.
The heating rate is $F_E\sigma_h$  where $F_E$ is the exciting
flux (erg s$^{-1}$ cm$^{-2}$) at energy $E$, and $\sigma_h$ the 
cross section for absorbing photons of energy $E$,
which is roughly $\sigma_h\sim 5\times 10^{-18} (E/20 {\rm eV})^{-3}$
cm$^{2}$.  If we assume 100\% efficiency of converting
hard photons to NIR emission, and a uniform
distribution of the heating sources within a radius
$r\sim 20$ kpc, we can estimate $F_E\sim L_{NIR}/4\pi r^2$.
For $L_{NIR}\sim 3\times 10^{43}$ erg s$^{-1}$, 
$F_E\sim 10^{-3}$ erg s$^{-1}$ cm$^{-2}$.  Thus even
for photons with energies just above 1 Rydberg (say 20 eV) the
heating rate per molecule is $\sim 5\times 10^{-21}$ erg s$^{-1}$ or
four orders of magnitude below the cooling rate.

Thus if photons beyond the Lyman limit are the heating source, we must
conclude both that these photons are relatively soft (EUV rather than
soft X-ray) and that the surface filling factor of the photon sources
is very low.  In other words, the sources of photons must be clumpy
and lie very close to the molecular gas, so that the {\it local} value
of $F_E$ is much larger than its {\it average} value. Because of the
difficulty of detecting EUV photons, there is currently little direct
evidence for such emission in clusters and existing observations are
limited. \cite{Oegerle01} detected O {\rm VI} emission from A2597 in
{\it FUSE} observations, indicating the presence of gas at $\sim
10^5$K., whereas \cite{lecav04} failed to detect the same lines in
Abell~1795. Although a {\it soft excess} of EUV and soft X-rays has now
been detected in several clusters (e.g. \citealt{kaastra03}), this is
generally detected across the entire cluster and only in the case of
Sersic~159-03 shows any sign of peaking close to the cluster centre. The
large-scale of the emission contrasts with the highly clumped nature
of any EUV heating source indicated by our data.

The CO observations of \cite{edge01} support the idea of clumpy, low
surface filling factor, molecular gas; for several clusters the
absorbing column calculated from an assumed smooth distribution
exceeds that found in X-ray observations.  For the clusters Abell~1068
and Abell~1835, for example, \cite{edge01} find values of
$I_{CO(1-0)}\equiv \int T_a\ dV$ of approximately 3 and 2 K km
s$^{-1}$ respectively.  \cite{israel97} shows that the ratio of the
column density $N_{H_2}$ (cm$^{-2}$) to $I_{CO(1-0)}$ is typically
$\sim 3\times 10^{21}$ in external galaxies.  Thus the CO emission
fluxes are equivalent to column densities $N_{H_2}$ of 9 and 6 $\times
10^{21}$ cm$^{-2}$.  The estimated X-ray {\it absorption} column
densities are, in contrast, about $2\times 10^{20}$ for both clusters
(\citealt{wise04} and \citealt{peterson01}).  The molecular gas implied by
the CO observations would have been visible in the X-ray observations
had it smoothly covered the regions of X-ray emission; the absence of
this absorption implies a small covering factor for the molecular gas.

The luminosity needed to heat the
\Ht gas and that missing from the relatively cool
gas in cooling flow models (\citealt{peterson03}) are
both of order $\sim 10$\% of the total central
X-ray flux.  This prompts us to speculate that
the \Ht medium in fact absorbs and hides most of the
softer radiation from the cooling flow.  While
the X-ray spectra are not consistent with a foreground
screen of absorbing gas covering all the emitting gas,
they might still be consistent with clumpy selective
absorption.  Here we postulate that the cooler regions
of the X-ray gas, with $T<T_{ambient}/3$ (\citealt{peterson03})
consist of small clouds each surrounded by even
colder, dense, \Ht gas.  
Because many other processes in the central regions of
rich clusters have similar radiative or kinematic 
luminosities, however, the issue of the \Ht excitation source
and/or the heating of the cooler X-ray gas is still
wide open.

\section{Summary and Conclusions}
With narrow-band optical imaging of seven clusters we have shown that
H$\alpha$ and [NII] emission can be traced to beyond 20 kpc from the
centre in four clusters previously classified as cooling
flows. Although our images of these sources are deeper than those
already in the literature, they show very similar morphologies to
the previous data. A
typical non-cooling flow radio galaxy at a similar redshift to these
clusters shows only nuclear H$\alpha$ emission when observed to the
same surface brightness limit.

There are clear variations in H$\alpha$ morphology from source to
source. Two sources are filamentary (Abell~1795, Ser159-03) and two are
roughly circularly symmetric (Abell~2204, Abell~2597). Abell~2597 shows an
azimuthally averaged surface brightness profile which follows an
$R^{-3}$ distribution to beyond 40 kpc (limited by the signal-to-noise
in our data). The profiles of the other three clusters drop more
rapidly with radius than this. In particular the filaments in Abell~1795 and
Ser159-03 appear particularly strongly confined (hence the greater
depth of our images show little difference from previously published
images). We are therefore seeing a physical limit on the size of the
nebulosity, rather than a limit imposed by the signal-to-noise of our
images.

Our spectra show that for each source the \Ht emission follows the \Pa
emission closely in spatial distribution, surface brightness and
dynamics across the nebulosity. Clearly all phases of the nebulosity
and the source of excitation must be very intimately mixed on scales
much smaller than our spatial and spectral resolution. Given the lack
of dramatic variation in molecular to atomic emission as a function of
radius, the ratio of physical parameters leading to the line ratios
vary little over scales of 2 to 20 kpc from the centre of the
cluster. The consequence of this is that we expect to see the
molecular gas follow the atomic gas out to well beyond 40 kpc in
systems such as Abell~2597. This has a minimal consequence on the total
$K-$band \Ht luminosity because of the rapid dropoff of surface
brightness with radius as noted above.

Our detailed and sensitive $K-$band long slit spectroscopy of three of
these sources confirms the results of \cite{jaffe01}
that the P$\alpha$ to H$_2$ line ratios differ radically from
those found in starburst galaxies. They are indicative of excitation
and heating by a spectrum harder than O-stars (or more particularly a
``bluer'' EUV-FUV colour). This data does not allow us to determine
whether this is caused by hotter stars, a substantial soft tail 
to the X-ray spectrum of the ICM, or
some other means. The hardness of the spectrum can be independent of
the ionization parameter in the gas.

The difficulty of keeping molecular gas warm by any plausible
radiative excitation suggests that in this case the heating sources must
be very inhomogeneous distributed, and in close association with
the gas clouds.  This inhomogeneity of the gas is supported by
a comparison of CO emission and X-ray absorption measurements
in at least 2 clusters.

The velocity dispersion curves of the molecular and ionized gas are
sharply peaked within 8 kpc of the centres of Abell~2204 and 2597 with
dispersions typically peaking at $\sigma \sim 200-300$ km
s$^{-1}$. All previously published $K-$band spectra of cooling flows
are of emission from this region. The motions in this region are
probably dominated by the energetics of the AGN. In the case of
Ser159-03, our $K-$band slit traced the large-scale emission filament
and did not cross the nucleus of the galaxy and avoids this high
velocity dispersion region. Beyond 8 kpc the velocity dispersion drops
below the instrumental resolution of approximately 50 km s$^{-1}$.

The mean velocity of the gas relative to the centres of each system
show values of typically $\pm 100 $ km s$^{-1}$ and in Abell~2597 and
Ser159-03 show evidence for coherent shear or rotation over 20-30
kpc. These velocities are much smaller than the typical stellar
velocity dispersions in brightest cluster galaxies and are therefore
insufficient to support the molecular gas against the gravitational
field of the central galaxy. Moreover, the density of the molecular
gas indicates that it cannot be pressure supported in equilibrium with
the hot ICM. Either the molecular gas is continually replenished from
a pressure-supported reservoir of gas on timescales of 10 Myr or is
required to be supported by some other mechanism. The kinematics of
the molecular gas at large radii seems to exclude an origin for the
gas either in material stripped from infalling galaxies or material
pushed out of the centre of the system by the action of the central
radio source.

\section{Acknowledgements}
The authors wish to thank the Anglo-Australian Observatory and
the staff of the Anglo-Australian Telescope, in particular
J. Bland-Hawthorn, for their support of the AAT observations.
We thank Martin Hardcastle for help with the X-ray data.
The ISAAC observations were taken at the Very Large Telescope
(VLT) facility of the Europoean Southern Observatory (ESO)
as part of project 71.A-0239.  We also thank the ESO staff
in Garching and at Cerro Paranal for their support.
\bibliographystyle{mn2e}
\bibliography{MF067rv2}
\label{lastpage}
\end{document}